\documentclass[12pt]{article}
\pdfoutput=1
\usepackage{yfonts}
\usepackage{color}
\usepackage{mhchem}
\usepackage{xcolor}
\usepackage{cite}
\usepackage{hyperref}
\hypersetup{colorlinks=true,linkcolor=red,anchorcolor=black,citecolor=green}
\usepackage[toc,page]{appendix}
\usepackage{amsfonts}
\usepackage{bbold}
\usepackage{textcomp}
\usepackage[DIV13]{typearea}
\usepackage{amsmath, amsthm, amssymb, mathtools,empheq,latexsym,dsfont}
\usepackage{bbm}
\usepackage{slashed, simplewick}
\usepackage[utf8]{inputenc}
\usepackage{graphicx,placeins}
\usepackage{makeidx}
\usepackage[font=small,labelfont=bf]{caption}
\usepackage{nicefrac}
\usepackage{subfigure}
\usepackage{array, bigdelim,multirow,multicol}
\usepackage[integrals]{wasysym}
\usepackage{fancybox}
\usepackage{bm}
\usepackage{float}
\usepackage{rotating}
\usepackage{colortbl}
\usepackage{booktabs}
\usepackage[top=2cm,textwidth=16.6cm,textheight=22.75cm]{geometry}
\usepackage{doi}
\graphicspath{{immagini/}}

\definecolor{Gray}{gray}{0.92}
\newcommand{\nn}{\nonumber}
\newcommand{\be}{\begin{equation}}
\newcommand{\ee}{\end{equation}}
\newcommand{\bea}{\begin{eqnarray}}
\newcommand{\eea}{\end{eqnarray}}

\makeatletter
\renewcommand*{\@fnsymbol}[1]{\ensuremath{\ifcase#1\or *\or  \mathsection\or \ddagger\or
\dagger\or \mathparagraph\or \|\or **\or \dagger\dagger
\or \ddagger\ddagger \else\@ctrerr\fi}}
\makeatother

\makeatletter
\@addtoreset{equation}{section}
\makeatother

\begin{document}
 \unitlength = 1mm

\setlength{\extrarowheight}{0.2 cm}

\title{
\begin{flushright}
\hfill\mbox{{\small\tt USTC-ICTS/PCFT-20-28}}\\[5mm]
\begin{minipage}{0.2\linewidth}
\normalsize
\end{minipage}
\end{flushright}
 {\Large\bf Automorphic Forms
and Fermion Masses}\\[0.2cm]}
\date{}

\author{
Gui-Jun~Ding$^{1,2}$
\thanks{E-mail: {\tt dinggj@ustc.edu.cn}}
\
Ferruccio~Feruglio$^{3}$
\thanks{E-mail: {\tt feruglio@pd.infn.it}}
\ and
Xiang-Gan~Liu$^{1,2}$
\thanks{E-mail: {\tt hepliuxg@mail.ustc.edu.cn}}
\
\\*[20pt]
\centerline{
\begin{minipage}{\linewidth}
\begin{center}
$^1${\small Peng Huanwu Center for Fundamental Theory, Hefei, Anhui 230026, China} \\[2mm]
$^2${\small
Interdisciplinary Center for Theoretical Study and  Department of Modern Physics,\\
University of Science and Technology of China, Hefei, Anhui 230026, China}\\[2mm]
$^3${\small
 Dipartimento di Fisica e Astronomia `G.~Galilei', Universit\`a di Padova\\
INFN, Sezione di Padova, Via Marzolo~8, I-35131 Padua, Italy}\\
\end{center}
\end{minipage}}
\\[8mm]}
\maketitle
\thispagestyle{empty}

\centerline{\large\bf Abstract}
\begin{quote}
\indent
We extend the framework of modular invariant supersymmetric theories to encompass invariance under more general discrete groups $\Gamma$, that allow the presence of several moduli and make connection with the theory of automorphic forms. Moduli span a coset space $G/K$, where $G$ is a Lie group and $K$ is a compact subgroup of $G$, modded out by $\Gamma$. For a general choice of $G$, $K$, $\Gamma$ and a generic matter content, we explicitly construct a minimal K\"ahler potential and a general superpotential, for both rigid and local ${\cal N}=1$ supersymmetric theories. We also specialize our construction to the case $G=Sp(2g,\mathbb{R})$, $K=U(g)$ and $\Gamma=Sp(2g,\mathbb{Z})$, whose automorphic forms are Siegel modular forms. We show how our general theory can be consistently restricted to multi-dimensional regions of the moduli space enjoying residual symmetries. After choosing $g=2$, we present several examples of models for lepton and quark masses where Yukawa couplings are Siegel modular forms of level 2.
\end{quote}

\newpage

\section{Introduction}
The description of the flavour sector of particle physics requires up to 22 independent measurable parameters if neutrinos are Majorana particles. An extraordinary experimental activity has supplied not only an accurate determination of most of them, but also many redundant cross-checks, providing one of the most robust pillars of the Standard Model. The discovery of neutrino oscillations, and the related impressive recent experimental progress has brought also the lepton sector into a precision era, with many combinations of masses and mixing parameters measured with an accuracy approaching the percent level. The need of a large number of flavour input parameters and the presence of significant qualitative differences between the quark and the lepton sector constitute one of the major mystery of particle physics \cite{Weinberg:1977hb}, which has motivated an intense activity on the theoretical side to solve or mitigate the flavour puzzle.

In a bottom-up perspective, most of the attempts have been based on hypothetical flavour symmetries. The great relevance of symmetry considerations in the description of gauge interactions and of the electroweak symmetry breaking sector has certainly played a main role in guiding our efforts towards the solution of this puzzle. Hence, it is frustrating to admit that so far such a powerful tool has not delivered the expected result. While there are many working models with some degree of predictability, we still lack both a baseline theory and a
fundamental principle. A common disappointing aspect of models based on flavour symmetries is that these have to be broken and probably realized far from the symmetric phase.  As a result, most of the existing constructions are based on ingenious but very complicated symmetry breaking sectors, that include many scalar multiplets, whose vacuum expectation values (VEVs) have to be selectively coupled to different matter species and suitably oriented in flavour space to achieve the desired result.

Looking at the top-down approach offered by string theory, Yukawa couplings
are not independent input parameters, but rather field dependent quantities.
Such a dependence arises when the background over which the string propagates
is chosen. Contact with our four-dimensional universe requires several compact extra-dimensions, and a significant part of this background is of geometrical origin. The components of the metric tensor along the extra spatial dimensions, possibly combined with other types of background, give rise to moduli, scalar fields taking values on a definite moduli space. Yukawa couplings depend on moduli and their observed value is settled once moduli acquire their VEVs. The moduli space is often a symmetric space of the type $G/K$, $G$ being some noncompact continuous group and  $K$ a maximal compact subgroup of $G$~\cite{Maharana:1992my}. A discrete subgroup $\Gamma$ of $G$, the duality group, realizes its natural action on the moduli space $G/K$~\cite{Giveon:1994fu}. Hence such a framework is naturally equipped
with a symmetry $\Gamma$ and a symmetry breaking sector spanning $G/K$.
Moreover such a symmetry is always broken, since $\Gamma$ is non-linearly realized on  $G/K$. Matter fields generally have non-trivial transformation properties under $\Gamma$. Thus it is very tempting to interpret $\Gamma$ as (part of) the flavour symmetry of the theory.

An important part of the theoretical activity has indeed been devoted
to the study of Yukawa couplings in realistic string theory compactifications~\cite{Ibanez:1986ka,Casas:1991ac,Lebedev:2001qg,Kobayashi:2003vi,Brax:1994kv,Binetruy:1995nt,Dudas:1995eq,Dudas:1996aa,Leontaris:1997vw,Dent:2001cc,Dent:2001mn}
and their modular properties~\cite{Hamidi:1986vh,Dixon:1986qv,Lauer:1989ax,Lauer:1990tm,Erler:1991nr,Cremades:2003qj,Blumenhagen:2005mu,Abel:2006yk,Blumenhagen:2006ci,Marchesano:2007de,Antoniadis:2009bg,Kobayashi:2016ovu,Kobayashi:2020hoc,Cremades:2004wa,Abe:2009vi,Kikuchi:2020frp,Kikuchi:2020nxn}.
Nevertheless, the above considerations suggest a complementary analysis
where the flexibility of the bottom-up approach can be combined with the clues coming from string theory. A first step in this direction has been taken with the proposal of ref.~\cite{Feruglio:2017spp} where the role
of modular invariance as principle governing the lepton flavour sector has been advocated. In terms of the previous data, this corresponds to the choice: $G=SL(2,\mathbb{R})$, $K=SO(2)$ and $\Gamma=SL(2,\mathbb{Z})$. The moduli space $G/K$ is the upper half plane and Yukawa couplings are
classical modular forms. Model building~\cite{Feruglio:2019ktm} relies on the classification of modular forms under the principal congruence subgroups of $SL(2,\mathbb{Z})$.

In this paper we would like to widen this viewpoint and explore more general possibilities for the groups $G$, $K$ and $\Gamma$. While aiming at the mathematical consistency of our construction, we work in a bottom-up perspective and our setup does not necessarily correspond to a specific string theory realization. As we explain in section~\ref{S2}, in case of a generic triplet $(G, K, \Gamma)$, Yukawa couplings become automorphic forms and model building requires the classification of these forms under subgroups of $\Gamma$. In section~\ref{SS} we develop the formalism of global and local ${\cal N}=1$ automorphic supersymmetric theories. This generalizes and extends the classical works of Ferrara and collaborators of the late eighties on modular invariant supersymmetric theories~\cite{Ferrara:1989bc,Ferrara:1989qb}. After defining the general context, in section~\ref{S3} we move to a specific realization representing the most direct extension of modular invariance to a theory including several moduli. Such a construction is based on the choice $G=Sp(2g,\mathbb{R})$, $K=U(g)$ and $\Gamma=Sp(2g,\mathbb{Z})$. The moduli space $Sp(2g,\mathbb{R})/U(g)$ is the Siegel upper half plane and the automorphic forms are related to Siegel modular forms. In the second half of our paper we focus on the case $G=Sp(4,\mathbb{R})$, the simplest non-trivial extension of $G=SL(2,\mathbb{R})$. We present in detail properties of the Siegel modular group $\Gamma=Sp(4,\mathbb{Z})$, the construction of the fundamental region of $Sp(4,\mathbb{Z})$ in $G/K$ and its fixed points, the explicit form of Siegel modular forms at genus two and their restrictions on invariant loci of the moduli space. Finally we present examples of Siegel modular invariant models for lepton and quark masses.
\section{Automorphic Forms}
\label{S2}
Automorphic forms~\cite{Borel,Borel2} can be roughly regarded as generalization of periodic functions:
\be
f(x+a)=f(x)~~~,
\ee
where $a$ is a real constant and the above relation holds for any real number $x\in\mathbb{R}$. This can be generalized by replacing the set $\mathbb{R}$ by a coset space $G/K$, where $G$ is a continuous group, and $K$ a subgroup of $G$, and by substituting the discrete translation group $\mathbb{Z}$ with some discrete subgroup of $G$.

As an introduction to the subject, it is useful to recall how automorphic forms are related to classical modular forms and to modular invariant supersymmetric theories. In ${\cal N}=1$ supersymmetric theories, modular transformations on the modulus $\tau$ and on matter multiplets $\varphi^{(I)}$ are defined by
\be
\tau\to\gamma \tau=\frac{a\tau+b}{c\tau+d}~~~,~~~~~~~~~~~\varphi^{(I)}\to (c\tau+d)^{k_I}\rho_I(\gamma)~\varphi^{(I)}~~~,
\label{mod}
\ee
where $\gamma\in\Gamma=SL(2,\mathbb{Z})$, the homogeneous modular group, and $k_I$ is a real integer~\footnote{Extensions to non-integer weights $k_I$ are also possible~\cite{Liu:2020msy}.}. The modulus $\tau$ varies in the upper half plane ${\cal H}$~\footnote{The modulus transforms under the inhomogeneous part of $\Gamma$, $\overline{\Gamma}=SL(2,\mathbb{Z})/\{\pm \mathbb{1}\}$.}. The parameters $a$, $b$, $c$ and $d$ are integers obeying $ad-bc=1$. The matrix $\rho_I(\gamma)$ defines a unitary representation of a finite modular group $\Gamma/G_d$, $G_d$ being a normal subgroup of $\Gamma$ of finite index. Invariance under the transformations in eq.~\eqref{mod} requires that Yukawa couplings among matter fields $\varphi^{(I)}$ are classical modular forms $Y(\tau)$, which are the building blocks of the theory. Classical modular forms are holomorphic functions of $\tau$ transforming as:
\be
Y(\gamma\tau)=(c\tau+d)^k Y(\tau),~~~~~~~~~\gamma\in G_d~~~.
\label{cmf}
\ee
They form a linear space of finite dimension, thus constraining the possible Yukawa interactions. We see that the data needed to specify $Y(\tau)$, up to holomorphic requirements, are the moduli space ${\cal H}$, the group $G_d$ and the automorphy factor $(c\tau+d)^k$.

To make connection with the theory of automorphic forms~\cite{Borel}, it is useful to consider classical modular forms under a different perspective. For completeness we briefly summarize here such a viewpoint, standard in the mathematical literature~\cite{Fleig:2015vky}.
\begin{itemize}
\item[$1.$]
First of all, we can equivalently define the moduli space ${\cal H}$ as the quotient space $G/K$, where $G=SL(2,\mathbb{R})$ and $K=SO(2)$ is a maximal compact subgroup of $G$. Indeed any $\tau=x+i y$ $(y>0)$
can be represented as the action of $\textsf{g}\in SL(2,\mathbb{R})$ on the fixed complex number $\tau_0=i$. This action is defined in the sense of eq.~\eqref{mod}, that is:
\be
\textsf{g}~\tau_0=
\left(
\begin{array}{cc}
a& b\\
c&d
\end{array}
\right)\cdot\tau_0=
\frac{a\tau_0+b}{c\tau_0+d},~~~~~~(ad-bc=1)~~~.
\ee
We can uniquely decompose the generic element of $SL(2,\mathbb{R})$ as:
\be
SL(2,\mathbb{R})\ni \textsf{g}=\left(
\begin{array}{cc}
\sqrt{y}& x/\sqrt{y}\\
0&1/\sqrt{y}
\end{array}
\right)
~\textsf{k}\,,~~~~~~y>0~~~,~~~~~~~\textsf{k}=\left(\begin{matrix}
\cos\theta& -\sin\theta\\
\sin\theta&\cos\theta
\end{matrix}\right)~~,
\ee
where $\textsf{k}$ belongs to $K=SO(2)$, while the first factor in the decomposition of $\textsf{g}$ is an element of the coset $SL(2,\mathbb{R})/SO(2)$.
Since $\tau_0=i$ is left invariant by the action of $K=SO(2)$~\footnote{~$\textsf{k}~i=\left(\begin{smallmatrix}
\cos\theta& -\sin\theta\\
\sin\theta&\cos\theta
\end{smallmatrix}\right)
\cdot i=i$.}, we have:
\be
\textsf{g}~\tau_0=\left(
\begin{array}{cc}
\sqrt{y}& x/\sqrt{y}\\
0&1/\sqrt{y}
\end{array}
\right)\cdot i=x+i~y=\tau~~~.
\label{gi}
\ee
This shows the one-to-one correspondence between the elements of ${\cal H}$ and those of $SL(2,\mathbb{R})/SO(2)$.
\item[$2.$]
Second, we can relate a modular form $Y(\tau)$ to a periodic function $\Psi(\textsf{g})$ under the action of the discrete group $G_d$. We define:
\be
\label{ex1}
\Psi(\textsf{g})=j(\textsf{g},\tau_0)^{-1}~Y(\textsf{g}~\tau_0)~~~,
\ee
where $j(\textsf{g},\tau)$ is the automorphy factor:
\be
\label{ex2}
j(\textsf{g},\tau)=(c\tau+d)^k~~~,
\ee
satisfying the so-called cocycle condition:
\be
\label{eq:cocycle}j(\textsf{g}_1\textsf{g}_2,\tau)=j(\textsf{g}_1,\textsf{g}_2\tau)j(\textsf{g}_2,\tau)~~~.
\ee
The function $\Psi(\textsf{g})$ is called automorphic form and satisfies:
\bea
\Psi(\gamma \textsf{g})&=&\Psi(\textsf{g})\,,~~~~~~~~~~~~~~~~~~\gamma\in G_d~~\,,\nn\\
\label{autom0}
\Psi(\textsf{g}~\textsf{k})&=&j(\textsf{k},\tau_0)^{-1}~\Psi(\textsf{g})\,,~~~~~~~~\textsf{k}\in K~~.
\eea
The second relation in eq.~\eqref{autom0} follows from the invariance of $Y(\tau)$ under the action of the group $K$. Conversely, we can also start from $\Psi(\textsf{g})$ satisfying $(\ref{autom0})$ and define $Y(\tau)$ through:
\be
Y(\tau)=j(\textsf{g},\tau_0)~\Psi(\textsf{g})~~~.
\ee
From eqs.~\eqref{gi}, \eqref{eq:cocycle} and \eqref{autom0} we see that
\be
Y(\gamma\tau)=j(\gamma,\tau)~Y(\tau)~~~,
\label{af}
\ee
which reproduces the property (\ref{cmf}).
\item[$3.$]
Third, being holomorphic, the classical modular form $Y(\tau)$ satisfies the obvious differential relation $\partial Y(\tau)/\partial\bar\tau=0$.
After observing that $\Psi(\textsf{g})=e^{-i k \theta} y^{k/2} Y(x+i y)$, we can equivalently express the holomorphy of $Y(\tau)$ through the relation:
\be
e^{+2 i \theta}\left(-i y\frac{\partial}{\partial x}+y\frac{\partial}{\partial y}-\frac{i}{2}\frac{\partial}{\partial\theta}\right)~\Psi(\textsf{g})=0~~~.
\label{r1}
\ee
The differential operator of the above equation can be regarded as an element of the $SL(2,\mathbb{R})$ algebra, whose generators are $(E,F,H)$ satisfying $[E,F]=H$, $[H,E]=+2 E$, $[H,F]=-2 F$. These generators can be realized in terms of operators acting on functions $\Psi(\textsf{g})=\Psi(x,y,\theta)$. We have:
\bea
E&=&e^{-2 i \theta}\left(i y\frac{\partial}{\partial x}+y\frac{\partial}{\partial y}+\frac{i}{2}\frac{\partial}{\partial\theta}\right)\,, \nn\\
F&=&e^{+2 i \theta}\left(-i y\frac{\partial}{\partial x}+y\frac{\partial}{\partial y}-\frac{i}{2}\frac{\partial}{\partial\theta}\right)\,, \nn\\
H&=&i\frac{\partial}{\partial\theta}~~~.
\eea
Thus we can formulate the requirement of holomorphy of $Y(\tau)$ by saying that $\Psi(\textsf{g})$ is annihilated by the lowering operator $F$ of the $SL(2,\mathbb{R})$ algebra. Within this algebra, the Casimir operator $\Delta$, commuting with all the other elements, can be
conventionally chosen as:
\be
\Delta=H^2-2 H+4 EF=4 y\frac{\partial^2}{\partial x \partial\theta}+4 y^2(\frac{\partial^2}{\partial x^2}+\frac{\partial^2}{\partial y^2})~~~.
\ee
To make contact with the general properties of automorphic forms, we observe that $\Psi(\textsf{g})$ is an eigenfunction of $\Delta$:
\be
\Delta\, \Psi(\textsf{g})=k(k-2)\Psi(\textsf{g})~~~.
\label{r2}
\ee
Conversely, by requiring eq.~\eqref{r2} to hold for a function $\Psi(\textsf{g})=e^{-i k \theta} y^{k/2} Y(x,y)$ we find that $Y(x,y)$ should satisfy:
\be
\left[i y k\left(\frac{\partial}{\partial x}+i \frac{\partial}{\partial y}\right)-y^2\left(\frac{\partial^2}{\partial x^2}+\frac{\partial^2}{\partial y^2}\right)\right]Y(x,y)=0~~~,
\label{r3}
\ee
or equivalently $\frac{\partial}{\partial \tau}y^k\frac{\partial}{\partial \overline{\tau}} Y(x, y)=0$. This represents a weaker constraint than holomorphy, eq.~\eqref{r1}. Indeed, while  holomorphic functions $Y(x+iy)$ are solutions of eq.~\eqref{r3}, non-holomorphic solutions are also allowed. The general solution $Y(x, y)$ of eq.~\eqref{r3}, a harmonic weak Maa{\ss} form, admits the unique decomposition $Y=Y^++Y^-$, where
$Y^+$ is the holomorphic part and $Y^-$ represents the non-holomorphic completion. The holomorphic part $Y^+$ lacks modularity and is a mock modular form.
\item[$4.$]
Classical modular forms $Y(\tau)$ are also required to be holomorphic when $\tau\to\infty$,
and, more general, at cusps. Since $\Psi(\textsf{g})=e^{-i k \theta} y^{k/2} Y(x+i y)$, such
property translates into a suitable grow condition on $\Psi(\textsf{g})$.
\end{itemize}
This example shows that there are different ways of dealing with a modular form. We can see $Y(\tau)$ as a function on $G$ invariant under $K$, satisfying the transformation property of eq.~\eqref{af} under $G_d$. Equivalently we can exchange $Y(\tau)$ for an automorphic form $\Psi(\textsf{g})$ which is invariant under $G_d$, possessing suitable transformation property under $K$. The groups $G$ and its maximal compact subgroup $K$ define the moduli space $G/K$, while the discrete subgroup $G_d$ of $G$ and the automorphy factor $j(\textsf{g},\tau)$ specify the transformation properties of the automorphic form.

The definition of automorphic form relies on the choice of a Lie group $G$, a maximal compact subgroup $K$ of $G$, a discrete subgroup $G_d$ of $G$. Moreover, as shown by the previous example, it is useful to introduce an automorphy factor $j(\textsf{g},\tau)$ satisfying a cocycle condition:
\be
j(\textsf{g}_1\textsf{g}_2,\tau)=j(\textsf{g}_1,\textsf{g}_2\tau)j(\textsf{g}_2,\tau)~~,~~~~~~~~~~g_{1,2}\in G,~\tau\in G/K~~~.
\label{autom}
\ee
An automorphic form for $G_d$ is a smooth complex function $\Psi(\textsf{g})$ that
\begin{itemize}
\item[]{1.} is invariant under the action of
the discrete group $G_d$:
\be
\Psi(\gamma \textsf{g})=\Psi(\textsf{g}),~~~~~~\gamma\in G_d~~~,
\ee
\item[]{2.} is $K$-finite: $\Psi(\textsf{g k})$, with $\textsf{k}$ varying in $K$, span a finite dimensional vector space~\cite{Borel2}. In all cases of interest discussed in this paper, such a condition
is realized through the relation:
\be
\label{Kfinite}
~~~~~\Psi(\textsf{g}~\textsf{k})=j(\textsf{k},\tau_0)^{-1}~\Psi(\textsf{g})\,,~~~~~~\textsf{k}\in K\,,~~~~~\textsf{k}\tau_0=\tau_0~~~,
\ee
which defines the transformation property of $\Psi(\textsf{g})$ under $K$. In all such cases the space
obtained by $\Psi(\textsf{g k})$, varying $\textsf{k}$ in $K$, is one-dimensional.
\item[]{3.} $\Psi(\textsf{g})$ is required to be an eigenfunction of the algebra ${\cal D}$ of invariant differential operators on $G$, that is an eigenfunction of all the Casimir operators of $G$.
\item[]{4.} The definition is completed by suitable growth conditions~\cite{Borel,Borel2}.
\end{itemize}
By varying $G$, $K$, $G_d$ and $j(\textsf{g},\tau)$ we have access to generalizations of classical modular forms. In particular we are interested in a framework naturally embedding more moduli fields. Such a framework is the common denominator of many string compactifications and provides an interesting setup for the description of fermion masses. Moreover, for a general choice of $G$, $K$, $G_d$ and $j(\textsf{g},\tau)$, we can always move from automorphic forms to their counterparts $Y(\tau)$, defined by:
\be
Y(\tau)=j(\textsf{g},\tau_0)~\Psi(\textsf{g})~~~,
\label{daf}
\ee
and obeying:
\be
Y(\gamma\tau)=j(\gamma,\tau)~Y(\tau)~~~.
\ee
Indeed when discussing the general framework of automorphic supersymmetric theories we prefer to make use
of the forms $Y(\tau)$, which, with a slight abuse of language, we will refer to as automorphic or modular forms.

There is an important point that is worth stressing.
As we have seen in our example,  the differential condition mentioned in point 3. is strictly related to the holomorphy of the forms $Y(\tau)$. Nevertheless, the condition of being an eigenvalue of all Casimir operators of $G$ does not inevitably imply holomorphy of $Y(\tau)$ and the theory of automorphic forms $\Psi(\textsf{g})$ embraces also non-holomorphic forms $Y(\tau)$~\footnote{The holomorphic part of such forms, mock modular forms, lack modularity but play an important
role in recent developments, being related to the degeneracies of quantum black holes in string theories~\cite{Dabholkar:2012nd}. Non-holomorphic forms arise in
the leading-order low-energy correction to the four-graviton scattering amplitude in ten-dimensional type IIB string theory~\cite{Green:1997tv,Kiritsis:1997em,Green:1998by,Pioline:1998mn,Green:2010wi}. For
a brief overview of how this part of Ramanujan's work has influenced physics see ref.~\cite{Harvey:2019htf} and ref.~\cite{Fleig:2015vky}.}.
Thus, at least in principle, the framework of automorphic forms
might not necessarily require a supersymmetric theory where Yukawa couplings are described by an holomorphic superpotential, and could open the way to more general possibilities, in particular non-supersymmetric realizations.
It is not the purpose of this paper to explore this interesting direction and, in the present discussion, we will restrict our attention to holomorphic $Y(\tau)$ and supersymmetric theories.

\section{Automorphic Supersymmetric Theory}
\label{SS}
Here we generalize the well-known framework of modular invariant supersymmetry theories~\cite{Ferrara:1989bc,Ferrara:1989qb}. The construction, naturally involving more moduli $\tau$, is based on the following data:
\begin{itemize}
\item[]{1.} A continuous group $G$
and a maximal compact subgroup $K$ of $G$.
\item[]{2.}
A discrete, in general noncompact, subgroup $\Gamma$ of $G$ and a normal subgroup $G_d$ of $\Gamma$ of finite index in $\Gamma$. The quotient $\Gamma/G_d$ is a finite group, whose unitary representations are
exploited to define the transformations of matter supermultiplets.
\item[]{3.}
A family of cocycles $j(\textsf{g},\tau)^k$ where $j(\textsf{g},\tau)$ satisfies the condition in eq.~\eqref{autom} and $k$ is an integer. We require that the dependence of $j(\textsf{g}_0,\tau)^k$ on $\tau$, $g_0$ being a fixed element of $G$,  is holomorphic.
\end{itemize}
The resulting theory, which we define in the next subsections, is a (global or local) supersymmetric $\sigma$ model, where moduli parametrize the coset $G/K$ and transformations of matter fields are specified by $\Gamma$, $G_d$ and $j(\textsf{g},\tau)$. The theory is required to be invariant under $\Gamma$, the ``duality" or modular group, identified as part of the flavour symmetry group. Among all coset spaces $G/K$, of special interest are those of noncompact type, first appeared in various supergravity theories in the 70s and then in string theory where moduli spaces of toroidal compactification are given by noncompact groups modded out by their maximal compact subgroups and discrete duality groups. Other moduli spaces have analogous descriptions~\cite{Maharana:1992my}. A well-defined class of coset spaces $G/K$ that are
also K\"ahler manifolds, and are suitable to the supersymmetric setup of our interest, are the hermitian symmetric spaces. Hermitian symmetric spaces are manifolds equipped with a Riemannian metric and an integrable, almost complex, structure which preserves the metric. At each point $p$ of an hermitian symmetric space there is a reflection $s_p$ $(s_p^2=1)$ preserving the hermitian structure and having $p$ as unique fixed point, $s_p p=p$.
Every hermitian symmetric space $M$ is a K\"ahler manifold and is a coset space of the type $M=G/K$ for some connected Lie group $G$ and a compact subgroup $K$ of $G$. We provide a brief description of these spaces in the appendix~\ref{HSS}. They have been completely classified. In particular, the noncompact irreducible ones fall in this list~\cite{calabi1960compact}:
\begin{align}
\label{listHSS}
\nonumber
&\frac{U(m,n)}{U(m)\times U(n)}~~,~~~\frac{SO^*(2m)}{U(m)}~~,~~~\frac{Sp(2m)}{U(m)}~~,~~~\frac{SO(m,2)}{SO(m)\times SO(2)}~~,\\
&~~~~~~~~~~~~~~~~~~~~~\frac{E_{6,-14}}{SO(10)\times SO(2)}~~, ~~~\frac{E_{7,-25}}{E_6\times U(1)} \,.
\end{align}
Any other noncompact hermitian symmetric space contains one of the previous spaces as a factor. Our general construction applies, in particular, to all such manifolds, as we discuss in the next subsections.
\subsection{Moduli space and transformation laws of the fields}
\label{SS1}
We define as moduli space the coset ${\cal H}=G/K$~\footnote{In the simplest string theory compactifications, such as for instance a toroidal compactification, the duality group $\Gamma$ is part of the diffeomorphisms of the theory. As a consequence, the moduli space is the quotient between $G/K$ and $\Gamma$ and it classifies inequivalent complex structures of the
compactified space. In the present bottom-up approach we adhere to a more physical - yet equivalent - picture  and we define the moduli space as the whole $G/K$. The duality group $\Gamma$ can be interpreted as a gauge symmetry related to the redundancy of the description. The quotient between $G/K$ and $\Gamma$ is characterized by the fundamental domain for $\Gamma$, describing the set of inequivalent vacua.}. The generic element of ${\cal H}$ is denoted by $\tau$, and it can also be represented by the action of $\textsf{g}\in G$ on an element $\tau_0$ left invariant by $K$:
\be
\tau=\textsf{g}~\tau_0\,,~~~~~~~~~~~~~~~~\textsf{k}~\tau_0=\tau_0\,,~~~~~\textsf{k}\in K\,.
\label{corr}
\ee
We look for an ${\cal N}=1$ supersymmetric theory invariant under transformations of $\Gamma$ and depending on a set $\Phi=(\tau,\varphi)$ of chiral supermultiplets of the theory, including the moduli $\tau$, here taken as dimensionless and gauge-singlet. Canonical dimensions can be recovered by the definition $\sigma=\Lambda\tau$, $\Lambda$ denoting some convenient mass scale. The other chiral supermultiplets $\varphi$ are in general separated into sectors $\varphi^{(I)}$. Neglecting the VEVs of the chiral supermultiplets $\varphi$, the vacua of the theory are parametrized by
the moduli space ${\cal H}$, modded out by the duality group $\Gamma$.
The quotient ${\cal H}/\Gamma$ can be described by a fundamental domain ${\cal F}$ for $\Gamma$, which is a connected region of ${\cal H}$ such that each point of ${\cal H}$ can be mapped into ${\cal F}$ by a $\Gamma$ transformation, but no two points in the interior of ${\cal F}$ are related under $\Gamma$. The space ${\cal H}/\Gamma$ is represented by ${\cal F}$ with certain boundary points identified.

In the case of global supersymmetry, the action for Yukawa interactions reads:
\be
{\cal S}=\int d^4 x d^2\theta d^2\bar \theta~ K(\Phi,\bar \Phi)+\int d^4 x d^2\theta~ w(\Phi)+\text{h.c.}\,,
\label{action}
\ee
where $K(\Phi,\bar\Phi)$, the K\"ahler potential, is a real gauge-invariant function of the chiral superfields $\Phi$ and their conjugates and $w(\Phi)$, the superpotential, is a holomorphic gauge-invariant function of the chiral superfields $\Phi$. To ensure invariance under transformations of the group $\Gamma$
we should specify how $\tau$ and $\varphi$ transform.
\begin{itemize}
\item[]{1.}
The group $\Gamma$ acts on $\tau$ as:
\be
\label{mod1}
\tau\xrightarrow{\gamma}\gamma~\tau~\,,~~~~~~~~~~~~\gamma\in\Gamma~~~.
\ee
\item[]{2.}
To define the action of $\Gamma$ on the matter multiplets $\varphi$ we resort to the quotient $\Gamma/G_d$, a finite group, and to the family of cocycles $j(\textsf{g},\tau)^k$. There is a natural homomorphism between $\Gamma$ and $\Gamma/G_d$ and any element $\gamma\in \Gamma$ corresponds to a well-defined element of
$\Gamma/G_d$. Being finite, the group $\Gamma/G_d$ admits unitary representations $\rho(\gamma)$. The supermultiplets $\varphi^{(I)}$ of each sector are assumed to transform in a representation $\rho^{(I)}$ of $\Gamma/G_d$, with a weight $k_I$:
\be
\label{mod2}
\varphi^{(I)}\xrightarrow{\gamma} j(\gamma,\tau)^{k_I}~ \rho^{(I)}(\gamma)\varphi^{(I)}~~~,~~~~~~~~~~~~\gamma\in\Gamma~~~.
\ee
\end{itemize}
Eqs.~(\ref{mod1}) and (\ref{mod2}) generalize the transformation laws (\ref{mod}) of modular invariant supersymmetric theories.
\subsection{K\"ahler potential}
\label{SS2}
Minimal kinetic terms for moduli and matter fields can be introduced as follows. We define the function
$Z(\tau,\bar\tau)$:
\be
\label{Z}
Z(\tau,\bar\tau)=[j^\dagger(\textsf{g},\tau_0)j(\textsf{g},\tau_0)]^{-1}~~~,
\ee
where the dependence on $\tau$ is through the element $\textsf{g}$ via the correspondence in eq.~\eqref{corr}. We have required that $j(\textsf{g}_0,\tau)$, with $g_0$ fixed, is an holomorphic function of $\tau$, but in general $j(\textsf{g},\tau_0)$ depends through $\textsf{g}$ both on $\tau$ and $\bar\tau$. Under the group $\Gamma$, this function transforms as:
\be
Z(\tau,\bar\tau)\xrightarrow{\gamma}[j^\dagger(\gamma\textsf{g},\tau_0)j(\gamma\textsf{g},\tau_0)]^{-1}=
[j^\dagger(\gamma,\tau)j(\gamma,\tau)]^{-1}Z(\tau,\bar\tau)~~~.
\ee
A candidate K\"ahler potential for moduli is:
\be
\label{candidateK}
K(\tau,\bar\tau)=-h\log Z(\tau,\bar\tau)~~~,
\ee
where $h$ is a real constant whose sign is chosen to guarantee local positivity of the metric for the moduli $\tau$. Up to a K\"ahler transformation $K(\tau,\bar\tau)$ is invariant under $\Gamma$:
\be
K(\tau,\bar\tau)\xrightarrow{\gamma}K(\tau,\bar\tau)+h\log j(\gamma,\tau)+h\log j^\dagger(\gamma,\tau)~~~.
\ee
We observe that the minimal K\"ahler potential $K(\tau,\bar\tau)$ is also invariant under the full continuous group $G$. Notice that, for $G=SL(2,\mathbb{R})$, $K=SO(2)$ and $j(\textsf{g},\tau_0)=c\tau_0+d$, we get
$Z(\tau,\bar\tau)=-i(\tau-\bar\tau)/2$ and we recover the well-known minimal kinetic term of the single modulus formalism. A minimal K\"ahler potential for the matter multiplets, invariant under $\Gamma$, can be defined as:
\be
K(\varphi,\bar\varphi)=\sum_I Z(\tau,\bar\tau)^{k_I} |\varphi^{(I)}|^2~~~.
\ee
A candidate minimal K\"ahler potential is given by:
\be
\label{kmin}
K_{\tt min}(\Phi,\bar\Phi)=-h\log Z(\tau,\bar\tau)+\sum_I Z(\tau,\bar\tau)^{k_I} |\varphi^{(I)}|^2~~~.
\ee
By construction, the above potential is invariant under $\Gamma$ up to a K\"ahler transformation for a general choice of $G$, $K$, $\Gamma$, $G_d$ and $j(\textsf{g},\tau)$, nonetheless the positivity of the metric for the moduli $\tau$ evaluated from $Z(\tau,\bar\tau)$ should be explicitly checked case by case. We also stress that this is not the most general K\"ahler potential invariant under $\Gamma$. Invariance under $\Gamma$ allows to add to $K_{\tt min}(\Phi,\bar\Phi)$ many more terms, that cannot be excluded or constrained in a pure bottom-up approach. In general these terms can modify the flavour properties of the theory such as physical fermion masses and mixing angles. Additional assumptions or inputs from a top-down approach are needed in order to reduce the arbitrariness of the predictions.
\subsection{Superpotential}
\label{SS3}
To analyze the requirements for the invariance of the superpotential under $\Gamma$, we expand $w(\Phi)$ in powers of the supermultiplets $\varphi^{(I)}$:
\be
w(\Phi)=\sum_n Y_{I_1...I_n}(\tau)~ \varphi^{(I_1)}... \varphi^{(I_n)}~~~.
\label{wexp}
\ee
To guarantee invariance of the $n$-th order term, the functions $Y_{I_1...I_n}(\tau)$ should obey:
\be
\label{Ytras}
Y_{I_1...I_n}(\gamma\tau)=j(\gamma,\tau)^{k_Y(n)} \rho^{(Y)}(\gamma)~Y_{I_1...I_n}(\tau)~~~,
\ee
with $k_Y(n)$ and $\rho^{(Y)}$ such that:
\begin{enumerate}
\item[1.]
The weight $k_Y(n)$ compensates the total weight of the product $\varphi^{(I_1)}... \varphi^{(I_n)}$:
\be
k_Y(n)+k_{I_1}+....+k_{I_n}=0~~~.
\label{compensate}
\ee
\item[2.]
The product $\rho^{(Y)}\times \rho^{{(I_1)}}\times ... \times \rho^{{(I_n)}}$ contains an invariant singlet.
\end{enumerate}
When we restrict to transformation $\gamma$ of the group $G_d$ in eq.~\eqref{Ytras}, we obtain:
\be
\label{Ytras1}
Y_{I_1...I_n}(\gamma\tau)=j(\gamma,\tau)^{k_Y(n)}~Y_{I_1...I_n}(\tau)\,,~~~~~~\gamma\in G_d~~~,
\ee
Thus the function
\be
\Psi(\textsf{g})=j(\textsf{g},\tau_0)^{-k_Y(n)}Y_{I_1...I_n}(\textsf{g}\tau_0)
\ee
is an automorphic form for $G_d$ and the Yukawa couplings of the theory are strictly related to such automorphic forms.
\subsection{Local Supersymmetry}
\label{SS4}
This setup can be easily extended to the case of $N=1$ local supersymmetry where K\"ahler potential and superpotential are not independent
functions since the theory depends on the combination
\be
{\cal G}(\Phi,\bar\Phi)=K(\Phi,\bar \Phi)+\log w(\Phi)+\log w(\bar\Phi)~~~.
\ee
The modular invariance of the theory can be realized in two ways. Either $K(\Phi,\bar \Phi)$ and $w(\Phi)$ are separately modular invariant or
the transformation of $K(\Phi,\bar \Phi)$ under the modular group is compensated by that of $w(\Phi)$. An example of this second possibility is given by the Kahler potential of eq.~\eqref{kmin}, with the superpotential $w(\Phi)$ transforming as
\be
w(\Phi)\to e^{i\alpha(\gamma)} j(\gamma,\tau)^{-h} w(\Phi)
\ee
In the expansion (\ref{wexp}) the Yukawa couplings $Y_{I_1...I_n}(\tau)$ should now transform as
\be
Y_{I_1...I_n}(\gamma\tau)= e^{i\alpha(\gamma)} j(\gamma,\tau)^{k_Y(n)} \rho^{(Y)}(\gamma)~Y_{I_1...I_n}(\tau)~~~,
\ee
with $k_Y(n)+k_{I_1}+....+k_{I_n}=-h$ and the representation $\rho^{(Y)}$ subject to the requirement 2. When we have $k_{I_1}+....+k_{I_n}=-h$, we get $k_Y(n)=0$ and the functions $Y_{I_1...I_n}(\tau)$ are $\tau$-independent constants. This occurs for supermultiplets belonging to the untwisted sector in the orbifold compactification of the heterotic string.
\section{Siegel Modular Forms}
\label{S3}
Having defined the general framework of automorphic supersymmetric theories, it is useful to illustrate it by discussing a specific realization.
Varying the choice of $G$, $K$, $\Gamma$, $G_d$ and $j(\textsf{g},\tau)$, many possibilities can be explored. In the remaining part of this work we will exemplify the general formalism outlined above by choosing $G=Sp(2g,\mathbb{R})$, $K=U(g)=Sp(2g,\mathbb{R})\cap O(2g,\mathbb{R})$ and $\Gamma=Sp(2g,\mathbb{Z})$, where $g$ is a generic positive integer. This is the third case of the list (\ref{listHSS}), denoted as $\mathbf{III}_{g}$ in table~\ref{tab:coset_space} of appendix~\ref{HSS}. The related automorphic forms are provided by Siegel modular forms.
Choosing $g=2$, the moduli space $G/K$ modded out by the discrete group $Sp(4,\mathbb{Z})$, parametrizes
Riemann surfaces of genus 2~\cite{AlvarezGaume:1986pe}, offering a simple generalization of the choice $G=SL(2,\mathbb{R})$ and $K=SO(2)$,
where $G/K$, modded out by the discrete group $SL(2,\mathbb{Z})$, parametrizes complex structures on a torus.
\subsection{Moduli space}
\label{S31}
The moduli space of Siegel modular forms, that will be denoted as $\mathcal{H}_g$, arises by choosing $G=Sp(2g,\mathbb{R})$ and $K=Sp(2g,\mathbb{R})\cap O(2g,\mathbb{R})$, ($g=1,2,...$).
It corresponds to the case $\mathbf{III}_{g}$ in table~\ref{tab:coset_space} of appendix~\ref{HSS}. The group $Sp(2g,\mathbb{R})$ is the symplectic group. Its elements are $2g\times 2g$ real matrices
of the type
\be
\textsf{g}=
\left(
\begin{array}{cc}
A&B\\
C&D
\end{array}
\right)~~~,
\ee
leaving invariant the symplectic form $J$:
\be
J=
\left(
\begin{array}{cc}
0&\mathbb{1}_g\\
-\mathbb{1}_g&0
\end{array}
\right)~~~.
\ee
Asking $\textsf{g}^t~J~\textsf{g}=J$,
where the superscript $t$ denotes the transpose,
we have the following conditions on the $g\times g$ matrices $A$, $B$, $C$ and $D$:
\be
A^t C=C^t A~~~,~~~~~~B^t D=D^t B~~~,~~~~~~~A^t D-C^t B=\mathbb{1}_g~~~.
\label{p2}\ee
The matrix $J$ has determinant $+1$ and has an inverse given by $J^{-1}=J^t=-J$. Both $J$ and identity matrix are symplectic matrices. The inverse of the element $\textsf{g}$ is also a symplectic matrix with
\begin{equation}
\textsf{g}^{-1}= \begin{pmatrix} D^t & -B^t \\ -C^t & A^t \end{pmatrix}\,.
\end{equation}
The symplectic group $Sp(2g,\mathbb{R})$ has a maximal compact subgroup, $K=Sp(2g,\mathbb{R})\cap O(2g,\mathbb{R})$, the group of orthogonal symplectic matrices $\textsf{k}\in K$ of the type~\footnote{$Sp(2g,\mathbb{R})\cap O(2g,\mathbb{R})$ is isomorphic to $U(g)$, as follows from $(A+iB)$ being a unitary $g\times g$ matrix.}:
\be
\textsf{k}=
\left(
\begin{array}{cc}
A&B\\
-B&A
\end{array}
\right)~~~,~~~~~~~A^t A+B^t B=\mathbb{1}_g~~~,~~A^tB=B^tA~~~.
\ee
\label{Sympl}
An element $\textsf{g}$ of $Sp(2g,\mathbb{R})$ can be uniquely decomposed as:
\be
\textsf{g}=\left(
\begin{array}{cc}
\sqrt{Y}& X \sqrt{Y^{-1}}\\
0&\sqrt{Y^{-1}}
\end{array}
\right)~\textsf{k}~~~,
\label{gS}
\ee
where $X$ and $Y$ are real symmetric $g\times g$ matrices, $Y$ is positive definite $(Y>0)$ and $\textsf{k}$ is an element of $Sp(2g,\mathbb{R})\cap O(2g,\mathbb{R})$. The matrices $X$ and $Y$ offer a parametrization of the moduli space $\mathcal{H}_g=G/K$. An element $\tau$ of the moduli space $\mathcal{H}_g$ is described by a symmetric complex $g\times g$ matrix $\tau$ with positive definite imaginary part:
\begin{equation}
\mathcal{H}_g=\Big\{\tau \in GL(g,\mathbb{C})~\Big|~  \tau^t = \tau ,\quad \texttt{Im} (\tau) > 0 \Big\}\,.
\end{equation}
The space $\mathcal{H}_g$ is called Siegel upper half plane, a natural generalization of the well-known complex upper half plane $\mathcal{H}$~\footnote{$\mathcal{H}_g$ is analytically equivalent to the bounded symmetric domain $\mathcal{D}$ of $Sp(2g,\mathbb{R})/U(g)$, shown in table~\ref{tab:coset_space}, appendix \ref{HSS}. They are explicitly related by the generalized Cayley transformation: $\tau \mapsto z=(\tau-i\mathbb{1}_g)(\tau+i\mathbb{1}_g)^{-1}$, $z \mapsto \tau=i(\mathbb{1}_g + z)(\mathbb{1}_g -z)^{-1}$.}. The integer $g$ is called genus. Similarly to the case of classical modular forms we define the action of $Sp(2g,\mathbb{R})$ on $\tau$ as:
\be
\tau\to \textsf{g} \tau=(A \tau+B)(C\tau +D)^{-1}~~~.
\ee
We recover the one-to-one correspondence between the elements of $\mathcal{H}_g$ and those of the coset $G/K$ by acting with the generic element $\textsf{g}\in Sp(2g,\mathbb{R})$ in eq.~\eqref{gS} on the fixed element $i \mathbb{1}_g\in \mathcal{H}_g$:
\be
\textsf{g}~ i \mathbb{1}_g=\left(
\begin{array}{cc}
\sqrt{Y}& X \sqrt{Y^{-1}}\\
0&\sqrt{Y^{-1}}
\end{array}
\right)~\textsf{k}\cdot i \mathbb{1}_g=X+i Y=\tau~~~,
\ee
where we have exploited the fact that $\textsf{k}\cdot i \mathbb{1}_g=i \mathbb{1}_g$, for any $\textsf{k}\in K$.
The complex dimension of $\mathcal{H}_g$ is $g(g+1)/2$. Choosing $g=1$ we go back to $\mathcal{H}$.
\subsection{Automorphy factor}
\label{S32}
In analogy with classical modular forms we can define a class of automorphy factors for the $G=Sp(2g,\mathbb{R})$ and $\mathcal{H}_g=Sp(2g,\mathbb{R})/Sp(2g,\mathbb{R})\cap O(2g,\mathbb{R})$:
\be
j_k(\textsf{g},\tau)=[{\tt det}(C\tau+D)]^k~~~,
\label{saf}
\ee
where $k$, an integer number, is called the weight. It can be readily checked that $j_k(\textsf{g},\tau)$ in eq.~\eqref{saf} satisfies the cocycle condition of eq.~\eqref{autom}.
\subsection{Siegel modular group}
\label{S33}
As modular group $\Gamma$ we can choose a discrete subgroup of $G=Sp(2g,\mathbb{R})$. A reference choice is
the Siegel modular group $\Gamma_g=Sp(2g,\mathbb{Z})$. A set of generators is provided by:
\be
\label{genSieg}
S=\left(
\begin{array}{cc}
0&\mathbb{1}_g\\
-\mathbb{1}_g&0
\end{array}
\right)~~~,~~~~T_i=\left(
\begin{array}{cc}
\mathbb{1}_g&B_i\\
0&\mathbb{1}_g
\end{array}
\right)~~~,
\ee
where $\{B_i\}$ is a basis for the $g\times g$ integer symmetric matrices and $S$ coincides with the invariant symplectic form $J$ satisfying $S^2=-\mathbb{1}_{2g}$. Other discrete subgroups of $G=Sp(2g,\mathbb{R})$, relevant to our purposes are the principal congruence subgroups $\Gamma_g(n)$ of level $n$,
defined as:
\begin{equation}
\label{eq:def_principal congruence subgroup}
\Gamma_g(n)=\Big\{\gamma \in \Gamma_g ~\Big|~  \gamma \equiv \mathbb{1}_{2g} \,\texttt{mod}\, n\Big\}\,,
\end{equation}
where $n$ is a generic positive integer, and $\Gamma_g(1)=\Gamma_g$.
The group $\Gamma_g(n)$ is a normal subgroup of $\Gamma_g$, and the quotient group $\Gamma_{g, n}=\Gamma_g/ \Gamma_g(n)$, which is known as finite Siegel modular group, has finite order~\cite{Koecher1954Zur,thesis_Fiorentino}:
\begin{equation}
\label{eq:subgroup_index}
|\Gamma_{g,n}|= n^{g(2g+1)}\prod_{p|n}\prod_{1\leq k\leq g} (1 - \dfrac{1}{p^{2k}})\,,
\end{equation}
where the product is over the prime divisors $p$ of $n$.
For the simplest case, $g=1$, we have
\begin{eqnarray}
|\Gamma_{1,n}|=n^3\prod_{p|n}(1 - \dfrac{1}{p^{2}})\,.
\end{eqnarray}
This is consistent with the dimension formula of $SL(2, \mathbb{Z}_n)$~\cite{Gunning1962,Schoeneberg}.
\subsection{Fundamental domain}
\label{S34}
Two symplectic matrices $\textsf{g}_1$, $\textsf{g}_2\in Sp(2g, \mathbb{R})$ have the same action on $\mathcal{H}_g$ if and only if $\textsf{g}_1=\pm\textsf{g}_2$~\cite{Freitag1991,Freitag1983}, therefore only the group $Sp(2g, \mathbb{R})/\{\pm\mathbb{1}\}$ acts effectively on $\mathcal{H}_g$. A fundamental domain $\mathcal{F}_g$ in ${\cal H}_g$ for the Siegel modular group $\Gamma_g$ is a connected region of $\mathcal{H}_g$ such that each point of  $\mathcal{H}_g$ can be mapped into $\mathcal{F}_g$ by a $\Gamma_g$ transformation, but no two points in the interior of $\mathcal{F}_g$ are related under $\Gamma_g$. It is considerably more complicated than the $g=1$ case. Siegel provided an explicit description of $\mathcal{F}_g$, which we report in appendix~\ref{A1}. In section~\ref{S5} we show the explicit form of a fundamental domain ${\cal F}_2$, at genus 2.
\subsection{Siegel Modular Forms}
\label{S35}
Siegel modular forms $f(\tau)$ of integral weight $k$ and level $n$ at genus $g$ are holomorphic functions on the Siegel half upper half plane  $\mathcal{H}_g$ transforming under $\Gamma_g(n)$ as
\begin{equation}
\label{SiegelForms}
f(\gamma \tau)=\det(C\tau+D)^k f(\tau) \,, \quad\quad \gamma=\begin{pmatrix}
A & B \\
C & D
\end{pmatrix} \in \Gamma_g(n) \,.
\end{equation}
When $n=1,2$, we have $-\mathbb{1}_{2g} \in \Gamma_g(n)$ and the above definition gives:
\begin{equation}
f(-\mathbb{1}_{2g} \tau)=f(\tau)=(-1)^{kg} f(\tau)\,.
\end{equation}
Therefore Siegel modular forms at genus $g$ of weight $k$ and level $n=1,2$ vanish if $kg$ is odd.
The elements $T_i^n$, where $\{T_i\}$ are the generators of eq.~\eqref{genSieg}, belong to $\Gamma_g(n)$. From eq.~\eqref{SiegelForms}, we see that
\begin{equation}
f(T_i^n \tau)=f(\tau+n B_i)=f(\tau) \quad~~~,
\end{equation}
and $f(\tau)$ admits an expansion as a Fourier series:
\begin{equation}
f(\tau)=\sum_{N\in Sym_g^s(\mathbb{Q}),N\geq 0} a(N)e^{\frac{2\pi i}{n}\text{Tr}(N\tau)}\,,
\end{equation}
where the sum extends to $Sym^s_g(\mathbb{Q})$, the set of half-integral matrices~\footnote{A symmetric $g\times g$ matrix $N$ is called half-integral if $2N$ is an integral matrix with even diagonal entries.}, and $N \geq 0$ means that $N$ is positive semi-definite.
The complex linear space $\mathcal{M}_k(\Gamma_g(n))$ of Siegel modular forms of given weight $k$, level $n$ and genus $g$ is finite dimensional~\cite{Bruinier2008The} and there are no non-vanishing
forms of negative weight~\cite{Bruinier2008The}.

Similarly to the case $g=1$~\cite{Feruglio:2017spp}, it is possible to choose a basis $\{f_i(\tau)\}$ in the space $\mathcal{M}_k(\Gamma_g(n))$ such that the action of $\Gamma_g$ on the elements of the basis is described by a unitary representation $\rho_{\mathbf{r}}$ of the finite Siegel modular group $\Gamma_{g, n}=\Gamma_g/\Gamma_g(n)$:
\begin{equation}
\label{SMF_decomposition}
f_i(\gamma \tau) = \det(C\tau+D)^k \rho_\mathbf{r}(\gamma)_{ij} f_j(\tau), \quad\quad \gamma=\begin{pmatrix}
A & B \\
C & D
\end{pmatrix} \in \Gamma_g\,.
\end{equation}
At variance with eq.~\eqref{SiegelForms}, where only transformations of $\Gamma_g(n)$ were considered, in the previous equation the full Siegel modular group $\Gamma_g$ is acting. Eq.~\eqref{SMF_decomposition} shows that the forms $\{f_i(\tau)\}$ of given weight, level and genus have good transformation properties also with respect to $\Gamma_g$.
We prove this relation in the appendix~\ref{app:decomposition-SMF}.
The full set of Siegel modular forms with respect to $\Gamma_g(n)$ form a positive graded ring $\mathcal{M}(\Gamma_g(n))= \bigoplus_{k \geq 0} \mathcal{M}_k(\Gamma_g(n))$.
\section{Siegel modular invariant supersymmetric theory }
\label{S4}
We apply the general formalism of section~\ref{SS} to the case
$G=Sp(2g,\mathbb{R})$, $K=Sp(2g,\mathbb{R})\cap O(2g,\mathbb{R})$,
$\Gamma=Sp(2g,\mathbb{Z})$, $G_d=\Gamma_g(n)$ and $j(\textsf{g},\tau)={\tt det}(C\tau+D)$. We first consider the case of rigid supersymmetry where we focus on Yukawa interactions. The action:
\be
{\cal S}=\int d^4 x d^2\theta d^2\bar \theta~ K(\Phi,\bar \Phi)+\int d^4 x d^2\theta~ w(\Phi)+\text{h.c.}\,,
\label{action}
\ee
is required to be invariant under transformations of the Siegel modular group $\Gamma_g$. To define the action of $\Gamma_g$ on the matter multiplets $\varphi$ we choose a particular level $n$. Thus, throughout the whole construction, both the genus $g$ and the level $n$ are kept fixed.
The supermultiplets $\varphi^{(I)}$ of each sector $I$ are assumed to transform in a representation $\rho^{(I)}$ of the finite Siegel modular group $\Gamma_{g,n}$, with a weight $k_I$:
\be
\left\{
\begin{array}{l}
\tau\to \gamma \tau=(A \tau+B)(C\tau +D)^{-1}\,,
\\[0.2 cm]
\varphi^{(I)}\to [\det(C\tau+D)]^{k_I} \rho^{(I)}(\gamma) \varphi^{(I)}~\,,
\end{array}
\right.~~~~~~~~~\gamma=\begin{pmatrix}
A & B \\
C & D
\end{pmatrix}\in\Gamma_g\,.
\label{eq:tmg-1}
\ee
Due to the cocycle condition in eq.~\eqref{autom} and the properties of $\rho^{(I)}$, the above definition satisfies the group law. The supermultiplets $\varphi^{(I)}$ are not modular forms and real values of $k_I$ are a priori allowed.
The invariance of the action ${\cal S}$ under eq.~\eqref{eq:tmg-1} requires the invariance of the superpotential $w(\Phi)$ and the invariance of the Kahler potential up to a Kahler transformation:
\be
\left\{
\begin{array}{l}
w(\Phi)\to w(\Phi)\\[0.2 cm]
K(\Phi,\bar \Phi)\to K(\Phi,\bar \Phi)+f(\Phi)+f(\bar{\Phi})
\end{array}
\right.~~~.
\ee
The requirement of invariance of the K\"ahler potential can be easily satisfied. We find that the combination $Z(\tau,\bar\tau)$ of eq.~\eqref{Z} is equal to $\det[(-i\tau+i\tau^\dagger)/2]$. Thus a minimal K\"ahler potential for the moduli $\tau$ is:
\be
K_\tau=-h~ \Lambda^2 \log\det(-i\tau+i\tau^\dagger),~~~~~~~~~h>0~~~,
\ee
where $h$ is a dimensionless constant and $\Lambda$ is some reference mass scale. Notice that the above minimal K\"ahler potential $K_{\tau}$ exactly matches the K\"ahler potential $K(z,z^*)$ of eq.~\eqref{KahlerPotential} after performing the Cayley transformation. This generalizes the K\"ahler potential of the case $g=1$. Under $\textsf{g}\in Sp(2g,\mathbb{R})$ the combination $(\tau-\tau^\dagger)$ transforms as:
\be
(\tau-\tau^\dagger)\to (C\tau^\dagger+D)^{-1T}~(\tau-\tau^\dagger)~(C\tau+D)^{-1},~~~~~~~~~~~\textsf{g}=\begin{pmatrix}
A & B \\
C & D
\end{pmatrix}~~~,
\ee
and we find:
\be
K_{\tau}\to K_{\tau}+h~\Lambda^2\log\det (C\tau^\dagger+D)^T+h~\Lambda^2\log\det (C\tau+D)~~~,
\ee
which shows the invariance of $K$ under the full symplectic group $Sp(2g,\mathbb{R})$ up to a K\"ahler transformation.
A minimal K\"ahler potential for matter multiplets $\varphi^{(I)}$ transforming under $\Gamma_g$
as in eq.~\eqref{eq:tmg-1} is given by:
\be
K_\varphi=\sum_I [\det(-i\tau+i\tau^\dagger)]^{k_I} | \varphi^{(I)}|^2~~~.
\ee
It is invariant under transformations of $\Gamma_g$.
The overall K\"ahler potential is minimally described by:
\be
K=K_\tau+K_\varphi~~~.
\label{minK}
\ee
To study the invariance of the superpotential $w(\Phi)$ under the Siegel modular group, we closely follow the steps outlined in section~\ref{SS3}. We consider the expansion of $w(\Phi)$ in power series of the supermultiplets $\varphi^{(I)}$ given in eq.~\eqref{wexp}. For the $p$-th order term to be modular invariant the functions $Y_{I_1...I_p}(\tau)$ should transform as Siegel modular forms with weight $k_Y(p)$ in the representation $\rho^{(Y)}$ of $\Gamma_{g,n}$:
\be
Y_{I_1...I_p}(\gamma\tau)=[\det(C\tau+D)]^{k_Y(p)}\rho^{(Y)}(\gamma)~Y_{I_1...I_p}(\tau)~~~,
\ee
with $k_Y(p)$ and $\rho^{(Y)}$ such that:
\begin{enumerate}
\item[1.]
The weight $k_Y(p)$ should compensate the overall weight of the product $\varphi^{(I_1)}... \varphi^{(I_p)}$:
\be
k_Y(p)+k_{I_1}+....+k_{I_p}=0~~~.
\label{compensate}
\ee
\item[2.]
The product $\rho^{(Y)}\times \rho^{{I_1}}\times ... \times \rho^{{I_p}}$ contains an invariant singlet.
\end{enumerate}
The case of local supersymmetry follows straightforwardly from the discussion in section~\ref{SS4}.
\subsection{Invariant loci in moduli space}
\label{S41}
In a generic point $\tau$ of the moduli space ${\cal H}_g$ the discrete symmetry $\Gamma_g$ is completely broken (i.e. $\gamma\tau=\gamma$ has no solution for $\gamma\in\Gamma_g$), but there can be regions where a part of $\Gamma_g$ is preserved. When we consider genus 1, the modular group $SL(2,\mathbb{Z})$ is always broken in the upper half of the complex plane, except for points possessing some residual symmetry. In the standard fundamental domain of $SL(2,\mathbb{Z})$, the inequivalent fixed points are $\tau=i, -1/2+i\sqrt{3}/2,i \infty$, which are left invariant by the subgroups generated by $S$, $ST$ and $T$, respectively~\footnote{~$S=\left(\begin{smallmatrix}
0~&1\\
-1~&0
\end{smallmatrix}\right)$~~,~~$T=\left(\begin{smallmatrix}
1~&1\\
0~&1
\end{smallmatrix}\right)$~~.}. In ${\cal H}_g$ we have a richer variety of possibilities, with the qualitatively new
feature that the regions possessing residual symmetries can be points, lines, surfaces or even spaces of higher dimensions.
We define a region $\Omega$ whose points $\tau$ are individually left invariant by some element $h$ of $\Gamma_g$:
\be
h~\tau=\tau~~~.
\label{fp}
\ee
Since $h_1\,\tau=\tau$ and $h_2\,\tau=\tau$ imply $h_1h_2\,\tau=\tau$, the elements $h$ satisfying eq.~\eqref{fp} form a subgroup $H$ of $\Gamma_g$ and we can write $H\tau=\tau$. Moreover, a trivial element of
$\Gamma_g$ leaving $\tau$ invariant is $-\mathbb{1}_{2g}$ and we also consider the group
$\bar{H}=H/\{\pm \mathbb{1}_{2g}\}$. Both $H$ and $\bar{H}$ are called stabilizers.
In our theory we can consistently restrict the domain of moduli to this region $\Omega$.
The group $N(H)$ that, as a whole, leaves the region $\Omega$ invariant includes the elements $\gamma$ of $\Gamma_g$ such that:
\be
\gamma \tau=\tau'~~~,
\label{N1}
\ee
where $\tau$ and $\tau'$ are both in $\Omega$:
\be
H\tau=\tau~~~,~~~~~~~H\tau'=\tau'~~~.
\label{N2}
\ee
By combining eqs.~\eqref{N1} and \eqref{N2}, we get
\be
\label{eq:normalizer}\gamma^{-1}H\gamma=H~~~,
\ee
which shows that the searched-for group $N(H)$ is the normalizer of $H$.
The condition defining the normalizer is weaker than the one concerning the stabilizer and, in general, $H$ is a proper subgroup of $N(H)$.
When the region $\Omega$ consists of a single isolated point $\tau_0$, as is the case for the fixed points of $SL(2,\mathbb{Z})$ in ${\cal H}$, $H$ and $N(H)$ coincide, since both $\tau$ and $\tau'$ are equal to $\tau_0$ in eq.~\eqref{N1}. Thus the distinction between stabilizer and normalizer is a new feature of the multidimensional moduli space
${\cal H}_g$. As a consequence, in our supersymmetric action
we can restrict the moduli $\tau$ to the region $\Omega$, which supersedes the full moduli space ${\cal H}_g$,
and replace the group $\Gamma_g$ with $N(H)$. An element $\gamma$ of $N(H)$ induces the transformation laws
\be
\left\{
\begin{array}{l}
\tau\to \gamma \tau=(A \tau+B)(C\tau +D)^{-1}
\\[0.2 cm]
\varphi^{(I)}\to [\det(C\tau+D)]^{k_I} \rho^{(I)}(\gamma) \varphi^{(I)}~~~.
\end{array}
\right.~~~~~~~~~\gamma=\begin{pmatrix}
A & B \\
C & D
\end{pmatrix}\in N(H)\,,
\label{tmg}
\ee
where $\rho^{(I)}(\gamma)$ is now a unitary representation of a finite group $N_n(H)$ obtained through the same steps leading to the Siegel finite modular groups $\Gamma_{g,n}$ in section~\ref{S33}. We can define the principal congruence subgroup of $N(H)$, denoted as $N(H, n)$:
\begin{equation}
N(H, n)=\left\{\hat{\gamma}\in N(H)~\Big|~\hat{\gamma}=\mathbb{1}_{2g} \,\texttt{mod}\, n \right\}~~~.
\end{equation}
Obviously $N(H, n)$ is a subgroup of $\Gamma_{g}(n)$, and it is also a normal subgroup of $N(H)$. The finite modular subgroup $N_n(H)$ corresponding to the modular subgroup $N(H)$ is the quotient group $N_n(H)= N(H)/N(H, n)$, and it is a subgroup of finite Siegel modular group $\Gamma_{g,n}$.
In summary, we can consistently truncate the moduli space to the subspace $\Omega$, and substitute
$\Gamma_g$ and $\Gamma_{g,n}$ with $N(H)$ and $N_n(H)$, respectively.
\section{Genus 2 Siegel modular invariant theories}
\label{S5}
In this section we analyze in detail the case $g=2$, which offers the simplest non trivial generalization of modular invariant supersymmetric theories studied in~\cite{Ferrara:1989bc,Ferrara:1989qb}. The moduli space ${\cal H}_2$ has complex dimension 3 and describes 3 moduli:
\be
\tau=\left(
\begin{array}{cc}
\tau_1&\tau_3\\
\tau_3&\tau_2
\end{array}
\right)~~,~~~~		\det(\texttt{Im}(\tau))>0 \,,~~~~\,\text{Tr}(\texttt{Im}(\tau))>0~~~.
\ee
A set of generators of the Siegel modular group $\Gamma_2$ is given by
\begin{align}
\label{eq:generators_SP4}
& T_1 =\begin{pmatrix}
\mathbb{1}_2 & B_1 \\
0 & \mathbb{1}_2
\end{pmatrix},\quad
T_2 =\begin{pmatrix}
\mathbb{1}_2 & B_2 \\
0 & \mathbb{1}_2
\end{pmatrix}, \quad
T_3 =\begin{pmatrix}
\mathbb{1}_2 & B_3 \\
0 & \mathbb{1}_2 \\
\end{pmatrix}, \quad S =\begin{pmatrix}
0 & \mathbb{1}_2 \\
-\mathbb{1}_2 & 0 \\
\end{pmatrix}\,,
\end{align}
with
\begin{align}
&B_1 =\begin{pmatrix}
1 & 0 \\
0 & 0 \\
\end{pmatrix},\quad B_2 =\begin{pmatrix}
0 & 0 \\
0 & 1 \\
\end{pmatrix}, ~\quad B_3 =\begin{pmatrix}
0 & 1 \\
1 & 0 \\
\end{pmatrix}\,.
\end{align}
The fundamental domain ${\cal F}_2={\cal H}_2/\Gamma_2$
can be defined by the following inequalities~\cite{Gottschling1959Explizite,J2017Efficient}:
\begin{equation}
\mathcal{F}_2 = \left\{ \tau \in \mathcal{H}_2  ~~\Bigg|~~
\begin{cases}
|\texttt{Re}(\tau_1)| \leq 1/2, \quad |\texttt{Re}(\tau_3)| \leq 1/2. \quad |\texttt{Re}(\tau_2)| \leq 1/2, \\
\texttt{Im}(\tau_2) \geq \texttt{Im}(\tau_1) \geq 2 \texttt{Im}(\tau_3) \geq 0 \\
|\tau_1| \geq 1, \quad |\tau_2| \geq 1, \quad |\tau_1+\tau_2-2\tau_3 \pm 1| \geq 1 \\
|\det(\tau + \mathcal{E}_i)| \geq 1
\end{cases} \right\}\,,
\end{equation}
where the set $\{\mathcal{E}_i\}$ includes the following 15 matrices:
\begin{align}
\nonumber
&\begin{pmatrix}
0 & 0 \\
0 & 0
\end{pmatrix}, \quad \begin{pmatrix}
\pm 1 & 0 \\
0 & 0
\end{pmatrix}, \quad \begin{pmatrix}
0 & 0 \\
0 & \pm 1
\end{pmatrix}, \quad \begin{pmatrix}
\pm 1 & 0 \\
0 & \pm 1
\end{pmatrix}, \\
&\begin{pmatrix}
\pm 1 & 0 \\
0 & \mp 1
\end{pmatrix}, \quad \begin{pmatrix}
0 & \pm 1 \\
\pm 1 & 0
\end{pmatrix}, \quad \begin{pmatrix}
\pm 1 & \pm 1 \\
\pm 1 & 0
\end{pmatrix}, \quad \begin{pmatrix}
0 & \pm 1 \\
\pm 1 & \pm 1
\end{pmatrix} .
\end{align}
Gottschling found that $\mathcal{F}_2$ has 28 boundary pieces~\cite{Gottschling1959Explizite}.
\subsection{Finite modular groups of genus 2}
\label{S52}
The dimension of the finite Siegel modular groups $\Gamma_{2,n}$ is
\begin{eqnarray}
\nonumber&&|\Gamma_{2,n}|=n^{10}\prod_{p|n}(1 - \dfrac{1}{p^{2}})(1 - \dfrac{1}{p^{4}})\,,
\end{eqnarray}
where the product is over the prime divisors $p$ of $n$. $|\Gamma_{2,n}|$ rapidly grows with $n$: for example $|\Gamma_{2,2}|=720$, $|\Gamma_{2,3}|=51840$. The group $\Gamma_{2,2}$ is isomorphic to $S_6$, and $\Gamma_{2, 3}$ is $Sp(4, F_3)$ or the double covering of Burkhardt group. In the rest of our paper we will focus on level $n=2$.
The finite Siegel modular group $\Gamma_{2,2}=S_6$ can be regarded as the permutation group of six objects. It can be generated by two permutations $F_1 \equiv (123456)$ and $F_2 \equiv (12)$ which obey
\begin{equation}
F_2^2=F_1^6=(F_2F_1)^5=(F_2F_1^3)^4=(F_2F_1^4F_2F_1^2)^2=1\,.
\end{equation}
The finite Siegel modular group $\Gamma_{2,2}=S_6$ can also be obtained from the $Sp(4,\mathbb{Z})$ generators $S$, $T_1$, $T_2$ and $T_3$ in eq.~\eqref{eq:generators_SP4} by imposing the following conditions,
\begin{align}
\nonumber&S^2=T_1^2=T_2^2=T_3^2=1\,,~~T_1T_2=T_2T_1\,,~~T_1T_3=T_3T_1\,,~~T_2T_3=T_3T_2\,,\\
&(S T_1)^6=(ST_2T_3)^5=((S T_1)^2T_3)^3T_2=((ST_2)^2T_3)^3T_1=1\,.
\end{align}
The two generators $F_1$ and $F_2$ can be expressed in terms of $T_{1,2,3}$ and $S$ as:
\begin{equation}
F_1=ST_3\,~~~~~ F_2=T_1\,.
\end{equation}
The relation between the two set of generators are summarized in table~\ref{tab:Relation}.
\begin{table}[!h]
\centering
\resizebox{1.0\textwidth}{!}{
\begin{tabular}{|c|c|c|c|}\hline\hline
 $T_1$ & $T_2$ & $T_3$ & $S$  \\ \hline
 $(12)$ & $(45)$ & $(12)(36)(45)$ & $(26)(35)$  \\ \hline
 $F_2$ & $F_1^3F_2F_1^3$ & $F_2F_1F_2(F_1^{-1}F_2F_1^{-1})^2(F_2F_1)^2F_1^2$ & $F_2F_1F_2(F_1^{-1}F_2F_1^{-1})^2(F_2F_1)^2F_1$  \\ \hline\hline
\end{tabular} }
\caption{\label{tab:Relation} The relationship between the two sets of generators $T_{1,2,3}$, $S$ and $F_{1,2}$ of the finite Seigel modular group $S_6$. }
\end{table}

The $S_6$ group has two one-dimensional, four five-dimensional, two nine-dimensional, two ten-dimensional and one sixteen-dimensional irreducible representations~\cite{S6}. It would be desirable to assign left-handed lepton fields to a three-dimensional irreducible representation of the finite modular group. This assignment has been shown to lead to the strongest and most predictive constraints
on neutrino masses and mixing parameters. From this point of view, the simplest choice $n=2$ would seem inadequate, since $S_6$ does not possess three-dimensional irreducible representations.
However, as discussed in section~\ref{S41}, we can consistently restrict our theory to a subregion $\Omega$ of the moduli space ${\cal H}_2$ left invariant by a subgroup $N(H)$ of $\Gamma_2$, $H$ being the stabilizer of $\Omega$. In the restricted theory $\Gamma_2$ and $\Gamma_{2,2}$ are replaced by $N(H)$ and the finite group $N_2(H)$, respectively. If $N_2(H)$ has three-dimensional irreducible representations, we have the necessary
ingredient to construct a predictive model. We proceed by inspecting the classification of the fixed points of $\Gamma_2$ in $\mathcal{H}_2$ and their residual symmetries.
\subsection{Invariant loci in ${\cal H}_2$}
\label{S51}
The fixed points of $\Gamma_2$ in Siegel upper half plane $\mathcal{H}_2$ have been classified by Gottschling~\cite{gottschling1961fixpunkte,gottschling1961fixpunktuntergruppen,gottschling1967uniformisierbarkeit}.
\begin{table}[!t]
\centering
\resizebox{0.92\textwidth}{!}{
\begin{tabular}{|c|c|c|c|c|} \hline\hline
 \multirow{2}{*}{Fixed points $\tau$} &  \multicolumn{2}{c|}{Stabilizer} & \multirow{2}{*}{Normalizer $N(H)$ } & \multirow{2}{*}{$N_2(H)$} \\ \cline{2-3}
  ~&~ $\bar{H}$ & $H$ ~&~ &~ \\ \hline\hline
 $\begin{pmatrix} \tau_1 & 0 \\ 0 & \tau_2 \end{pmatrix}$ & $Z_2$
 &$Z_2\times Z_2$& Eq.~\eqref{NorTwo1} & $(S_3\times S_3)\rtimes Z_2$ \\ \hline
 $\begin{pmatrix} \tau_1 & \tau_3 \\ \tau_3 & \tau_1 \end{pmatrix}$ &$Z_2$ & $Z_2\times Z_2$&Eq.~\eqref{NorTwo2} & $S_4 \times Z_2$  \\ \hline\hline
 $\begin{pmatrix} i & 0 \\ 0 & \tau_2 \end{pmatrix}$&$Z_4$ &$Z_4\times Z_2$& Eq.~\eqref{NorOne1} & $D_{12}$  \\  \hline
 $\begin{pmatrix} \omega & 0 \\ 0 & \tau_2 \end{pmatrix}$&$Z_6$ &$Z_6\times Z_2$& Eq.~\eqref{NorOne2} & $S_3 \times Z_3$  \\ \hline
 $\begin{pmatrix} \tau_1 & 0 \\ 0 & \tau_1 \end{pmatrix}$&$Z_2\times Z_2$ &$D_8$& Eq.~\eqref{NorOne3} & $D_{12}$  \\ \hline
 $\begin{pmatrix} \tau_1 & 1/2 \\ 1/2 & \tau_1 \end{pmatrix}$&$Z_2\times Z_2$ &$D_8$& Eq.~\eqref{NorOne4} & $D_8\times Z_2$ \\ \hline
 $\begin{pmatrix} \tau_1 & \tau_1 /2 \\ \tau_1 /2 & \tau_1 \end{pmatrix}$&$S_3$ &$D_{12}$& Eq.~\eqref{NorOne5} & $S_3 \times S_3$  \\ \hline\hline
$\begin{pmatrix} \zeta & \zeta+\zeta^{-2} \\ \zeta+\zeta^{-2} & -\zeta^{-1} \end{pmatrix} $ &$Z_5$ &$Z_{10}$& $Z_{10}$ & $Z_{5}$    \\ \hline
$\begin{pmatrix} \eta & \frac{1}{2}(\eta -1) \\ \frac{1}{2}(\eta -1) & \eta \end{pmatrix} $& $S_4$& $GL(2,3)$ & $GL(2,3)$   & $S_4$  \\ \hline
$\begin{pmatrix} i & 0 \\ 0 & i \end{pmatrix}$& $(Z_4\times Z_2)\rtimes Z_2$ & $(Z_4\times Z_4)\rtimes Z_2$& $(Z_4\times Z_4)\rtimes Z_2$  & $D_8$ \\ \hline
$\begin{pmatrix} \omega & 0 \\ 0 & \omega \end{pmatrix}$& $S_3\times Z_6
$ & $[72,30]$ & $[72,30]$ & $Z_3\times S_3$ \\ \hline
$\dfrac{i\sqrt{3}}{3}\begin{pmatrix} 2 & 1 \\ 1 & 2 \end{pmatrix}$& $D_{12}$ & $(Z_6\times Z_2)\rtimes Z_2$ & $(Z_6\times Z_2)\rtimes Z_2$ & $D_{12}$  \\ \hline
$\begin{pmatrix} \omega & 0 \\ 0 & i \end{pmatrix}$& $Z_{12}$ & $Z_{12}\times Z_2$ & $Z_{12}\times Z_2$ & $Z_6$ \\ \hline\hline
\end{tabular} }
\caption{\label{FixedPoints} All inequivalent fixed points of $Sp(4,\mathbb{Z})$ in the Siegel upper half plane $\mathcal{H}_2$. They are divided into three classes according to the dimension of the region they live in. The complex moduli are denoted by $\tau_1\,,\tau_2\,, \tau_3$, and $\zeta= e^{2\pi i /5},~\eta=\frac{1}{3}(1+i2\sqrt{2}),~\omega= e^{2\pi i/3}$. The generators of the normalizers $N(H)$ of the two-dimensional, one-dimensional and zero-dimensional cases are listed in the corresponding equations of appendix~\ref{SN}. Note that the group with \texttt{GAP} id [72,30] is isomorphic to
$Z_3\times((Z_2\times Z_6)\rtimes Z_2)$. }
\end{table}
In table~\ref{FixedPoints} we show the inequivalent fixed points of the fundamental domain $\mathcal{F}_2$ and their stabilizers $\bar{H}$ and normalizers $N(H)$. In appendix~\ref{SN} we list the generators of
$\bar{H}$ and $N(H)$. We see that the fixed points come in three classes, depending on the complex dimension of the region $\Omega$ contained in $\mathcal{H}_2$. There are two inequivalent sets of fixed points filling two-dimensional regions. The restriction of the theory to either of these regions results in Yukawa couplings depending on two moduli. There are five inequivalent sets of fixed points filling one-dimensional regions. The corresponding theory depends on a single modulus. Finally, there are six independent isolated fixed points, thus all moduli are frozen in the related theory. Equivalent invariant loci in the Siegel upper plane $\mathcal{H}_2$ can be found by applying modular transformations to those of the fundamental domain $\mathcal{F}_2$. For the case of $g=1$, a complete analysis of the fixed points and the phenomenological implications in neutrino mass models can be found in~\cite{Novichkov:2018yse,Gui-JunDing:2019wap}. Since $\mathcal{F}_2$ parametrizes Riemann surfaces of genus 2, the above invariant loci correspond to genus 2 Riemann surfaces of special type. For instance, the modular subspace with $\tau_3=0$ describes surfaces that factorize into two tori with moduli $\tau_1$ and $\tau_2$. Such a limiting case helps us to understand how our general framework includes and extends factorizable moduli spaces. For the interested reader, we describe the locus $\tau_3=0$ in appendix~\ref{subsec:tau3=0}.

\subsubsection{\label{subsec:tau1=tau2} The modular subspace with $\tau_1=\tau_2$  }
\noindent
From table~\ref{FixedPoints}, we see that among the regions with complex dimension one and two,
there is a single choice allowing for three-dimensional irreducible representations. It is the
one defined by the condition $\tau_1=\tau_2$:
\be
\label{Omega}
\Omega=\left\{\tau=
\begin{pmatrix}
\tau_1 & \tau_3 \\ \tau_3 & \tau_1
\end{pmatrix}\Bigg| \tau\in{\cal H}_2
\right\}~~~.
\ee
The requirement that the imaginary part of $\tau$ is positive definite implies $\texttt{Im}(\tau_1)>0$ and $\texttt{Im}(\tau_1)>|\texttt{Im}(\tau_3)|$. The stabilizer $H$ of the generic point $\tau$ of $\Omega$ is $Z_2\times Z_2=
 \{\pm\mathbb{1}_4,\pm h\}$:
\begin{equation}
\label{h}
h=\begin{pmatrix} 0&1&0&0 \\ 1&0&0&0 \\ 0&0&0&1 \\ 0&0&1&0\end{pmatrix}\,.
\end{equation}
The elements of the normalizer $N(H)$ of $H$ in $Sp(4,\mathbb{Z})$ fulfill:
\begin{equation}
\hat{\gamma}_{+} h = h \hat{\gamma}_{+}~~~\text{or}~~~\hat{\gamma}_{-} h =-h \hat{\gamma}_{-}~~~,
\end{equation}
resulting in
\begin{equation}
\label{NorTwo2}
\hat{\gamma}_{+}=\begin{pmatrix} a_1&a_2&b_1&b_2 \\ a_2&a_1&b_2&b_1 \\ c_1&c_2&d_1&d_2 \\ c_2&c_1&d_2&d_1\end{pmatrix},~~~\hat{\gamma}_{-}=\begin{pmatrix} a_1&a_2&b_1&b_2 \\ -a_2&-a_1&-b_2&-b_1 \\ c_1&c_2&d_1&d_2 \\ -c_2&-c_1&-d_2&-d_1\end{pmatrix}\,,
\end{equation}
with
\begin{equation}
a_1d_1+a_2d_2-b_1c_1-b_2c_2=1,\quad a_1d_2+a_2d_1-b_1c_2-b_2c_1=0\,.
\end{equation}
We can choose as generators of $N(H)$ the elements:
\begin{align}
\label{generators_Sp2}
\nonumber&
G_1= T_1T_2=\begin{pmatrix} 1&0&1&0 \\ 0&1&0&1 \\ 0&0&1&0 \\ 0&0&0&1\end{pmatrix},~~~
G_2= T_3=\begin{pmatrix} 1&0&0&1\\ 0&1&1&0 \\ 0&0&1&0 \\ 0&0&0&1 \end{pmatrix},\\
&G_3= S=\begin{pmatrix} 0&0&1&0\\ 0&0&0&1 \\ -1&0&0&0 \\ 0&-1&0&0 \end{pmatrix},~~~ G_4=\begin{pmatrix} 1&0&0&0 \\ 0&-1&0&0 \\ 0&0&1&0 \\ 0&0&0&-1\end{pmatrix}\,.
\end{align}
By the procedure of canonical projection and group quotient, we find the finite modular subgroup $N_2(H)$ is isomorphic to $S_4\times Z_2$ with \texttt{GAP} Id [48,48]~\cite{gap}. When referred to $N_2(H)$, the generators $G_{1,2,3}$ obey the relations:
\begin{equation}G_1^2=G_2^2=G_3^2=(G_1G_2)^2=(G_1G_3)^3=(G_1G_2G_3)^4=1\,.
\end{equation}
The element $G_4$ is indistinguishable from the identity of $N_2(H)$, since $G_4=\mathbb{1}_4~(\text{mod}~2)$. It is convenient to work with another set of generators: $\mathcal{S}=G_1$, $\mathcal{T}=(G_3G_2)^4$ and $\mathcal{V}=(G_3G_2)^3$, which satisfy the following multiplication rules
\begin{equation}
\mathcal{S}^2=\mathcal{T}^3=(\mathcal{S}\mathcal{T})^4=1,~~\mathcal{V}^2=1,~~~\mathcal{S}\mathcal{V}=\mathcal{V}\mathcal{S},~~~
\mathcal{T}\mathcal{V}=\mathcal{V}\mathcal{T}\,.
\end{equation}
The group has four singlet representations $\mathbf{1}$, $\mathbf{1}'$, $\mathbf{\hat{1}}$, $\mathbf{\hat{1}'}$, two doublet representations $\mathbf{2}$, $\mathbf{\hat{2}}$, and four triplet representations $\mathbf{3}$, $\mathbf{3'}$, $\mathbf{\hat{3}}$ and $\mathbf{\hat{3}'}$. The elements $\mathcal{S}$ and $\mathcal{T}$ generate the $S_4$ subgroup, and $\mathcal{V}$ generates $Z_2$. More details about the group $S_4\times Z_2$ can be found in the appendix~\ref{app:S4xZ2-group}.

In the remaining part of our work we focus on a theory whose moduli space is the region $\Omega$ of eq.~\eqref{Omega}, the modular group is $N(H)$ and the finite modular group is $N_2(H)$. As we can see from table~\ref{FixedPoints}, while the generic point in $\Omega$ is invariant under the element $h$
in eq.~\eqref{h} (and its opposite), there are complex lines and points of enhanced symmetry. For example, the points of the line $\tau_3=0$ in $\Omega$ are also invariant under $G_4$, those of the line $\tau_3=1/2$ are invariant under $G_2G_4$ and so on.
\subsection{Modular forms of genus 2 and level 2}
\label{S53}
For $g=n=2$ and even weight $k$, the dimension of the complex linear space $\mathcal{M}_k(\Gamma_2(2))$ is~\cite{dimformula}
\begin{equation}
\text{dim}~\mathcal{M}_k(\Gamma_2(2))=\frac{(k+1)(k^2+2k+12)}{12}=1,5,15,35,...~~~~~~~~~~~~(k=0,2,4,6,...)~~~.
\end{equation}
Modular forms of vanishing weight are constant. There are five linearly independent modular forms of weight 2. They can be expressed in terms of polynomials of the second Theta constant $\Theta[\sigma](\tau)$
and are explicitly given in appendix~\ref{A2}, where we show that they transform according to one of the irreducible five-dimensional representation of $\Gamma_{2,2}=S_6$. Higher weight modular forms can be obtained from
polynomials of modular forms of lower weight. Consider any set $f^{(k)}_i(\tau)$ forming a basis of ${\cal M}_k(\Gamma_2(2))$. We have
\be
f^{(k)}_i(\gamma\tau)=[\det(C\tau+D)]^k~f^{(k)}_i(\tau)~~~,
\ee
for any $\tau$ in ${\cal H}_2$ and $\gamma$ of $\Gamma_2(2)$. When we restrict the moduli space to $\Omega$ of eq.~\eqref{Omega} and the transformations to $N(H,2)\subset N(H)$, the previous equality still holds.
Indeed $N(H,2)$ is a subgroup of $\Gamma_2(2)$ and $\Omega$ is closed under the action of the whole $N(H)\supset N(H,2)$. This means that the modular forms of genus 2, level 2, weight $k$ with support in $\Omega$ can be obtained by restricting $f^{(k)}(\tau)$ to $\Omega$ and replacing $\Gamma_2(2)$ with $N(H,2)$. Moreover, exactly as discussed in appendix~\ref{app:decomposition-SMF}, when considering a transformation
$\gamma\in N(H)$ we have
\be
f^{(k)}_i(\gamma\tau)=[\det(C\tau+D)]^k~\rho(\gamma)_{ij}~f^{(k)}_j(\tau)~~~,
\ee
where, up to a change of basis, $\rho(\gamma)$ is a unitary representation of the finite modular group $N_2(H)=N(H)/N(H,2)$. When we restrict $\tau$ to $\Omega$, the five linearly independent modular forms of weight 2 given in appendix~\ref{A2} collapse to four. They can be organized into an invariant singlet and an irreducible triplet of the finite Siegel modular subgroup $N_2(H)= S_4\times Z_2$:
\begin{align}
\label{moularformsC2xS4}
\nonumber&\mathbf{1}: ~~~Y_{\mathbf{1}}(\tau)=p_0(\tau)+3p_3(\tau)\equiv Y_4(\tau)\,,\\
& \mathbf{3}': ~~~Y_{\mathbf{3}'}(\tau)
= \begin{pmatrix}
p_0(\tau)+4p_1(\tau)-p_3(\tau) \\
p_0(\tau)-2p_1(\tau)-p_3(\tau)-2i\sqrt{3}p_4(\tau) \\
p_0(\tau)-2p_1(\tau)-p_3(\tau)+2i\sqrt{3}p_4(\tau)
\end{pmatrix}\equiv\begin{pmatrix}
Y_1(\tau) \\ Y_2(\tau) \\ Y_3(\tau)
\end{pmatrix} \,.
\end{align}
The unitary matrix $\rho_\mathbf{3'}(\gamma)$ is given in appendix~\ref{app:S4xZ2-group}, while
the forms $\{p_i(\tau)\}$ $(i=0,1,3,4)$ are given by:
\begin{align}
\label{newp}
\nonumber&p_0=\Theta[00]^4(\tau)+2\Theta[01]^4(\tau)+\Theta[11]^4(\tau) \,,\\
\nonumber&p_1=2~\Theta[01]^2(\tau)\left(\Theta[00]^2(\tau)+\Theta[11]^2(\tau)\right)\,,\\
\nonumber&p_3=2\left(\Theta[00]^2(\tau)\Theta[11]^2(\tau)+\Theta[01]^4(\tau)\right)\,, \\
&p_4=4 \Theta[00](\tau)\Theta[01]^2(\tau)\Theta[11](\tau)\, ,
\end{align}
in terms of the relevant second Theta constant $\Theta[\sigma](\tau)$ that can be found in appendix~\ref{A2}. In the same appendix we present the $q$-expansion of $Y_i(\tau)$ $(i=1,2,3,4)$ and higher-weight modular forms.

As can be seen from table~\ref{FixedPoints}, there are only four independent fixed points of zero dimension in the region $\Omega$ of eq.~\eqref{Omega}. If the moduli $\tau_1$ and $\tau_3$ are stabilized by some mechanism to these fixed points, the normalizer is enhanced and the Siegel modular forms at these points are completely fixed. Concretely, the alignments of the weight 2 modular multiplets $Y_\mathbf{3'}(\tau)$ are of the following type:
\begin{align}
\label{eq:vev}
&\tau=\begin{pmatrix} \eta & \frac{1}{2}(\eta -1) \\ \frac{1}{2}(\eta -1) & \eta \end{pmatrix}\,:
~~ Y_\mathbf{3'}(\tau) \propto \begin{pmatrix}
1 \\ (-\frac{3}{17}+i\frac{6}{17})\omega^2 \\ (-\frac{3}{17}+i\frac{6}{17})\omega
\end{pmatrix}\,, \\
&\tau=\begin{pmatrix} i & 0 \\ 0 & i \end{pmatrix}\,:
~~~~~~~~~~~~~~~~~~~~~ Y_\mathbf{3'}(\tau) \propto \begin{pmatrix}
1 \\-\frac{1}{5}\omega \\ -\frac{1}{5}\omega^2
\end{pmatrix}\,, \\
&\tau=\begin{pmatrix} \omega & 0 \\ 0 & \omega \end{pmatrix}\,:
~~~~~~~~~~~~~~~~~~~~~ Y_\mathbf{3'}(\tau) \propto \begin{pmatrix}
1 \\ \omega \\ -\frac{1}{2}\omega^2
\end{pmatrix}\,, \\
&\tau= \frac{i\sqrt{3}}{3}\begin{pmatrix} 2 & 1 \\ 1 & 2 \end{pmatrix}\,:
~~~~~~~~~~~~~~~~ Y_\mathbf{3'}(\tau) \propto \begin{pmatrix}
1 \\ -\frac{1}{3}\omega \\ -\frac{1}{3}\omega^2
\end{pmatrix}\,.
\end{align}
Moreover, we see from table~\ref{FixedPoints} that there are three independent sets of fixed points spanning a region of dimension one in $\Omega$ of eq.~\eqref{Omega}. From the relations in eq.~\eqref{eq:relations_MF_d=1}, in each fixed point the four modular forms $Y_{1,2,3,4}(\tau)$ are found to satisfy the following constraints:
\begin{align}
\label{eq:alignments_3}
\nonumber
&\tau=\begin{pmatrix} \tau_1 & 0\\ 0 & \tau_1 \end{pmatrix}:
~~~~ Y_1(\tau) = 2\omega^2 Y_2(\tau)+ 2\omega Y_3(\tau) + 3Y_4(\tau)\,, \\
\nonumber
&\tau=\begin{pmatrix} \tau_1 & 1/2\\ 1/2 & \tau_1 \end{pmatrix}:
~~~~ Y_2(\tau)=Y_3(\tau)  \,, \\
&\tau=\begin{pmatrix} \tau_1 & \tau_1/2 \\ \tau_1/2 & \tau_1 \end{pmatrix}:
~~~~ Y_1(\tau)=Y_4(\tau)  \,.
\end{align}
Notice that a $Z_2\times Z_2$ subgroup generated by $\mathcal{V}$ and $(\mathcal{T}\mathcal{S}\mathcal{T})^2\mathcal{S}$ is preserved at the  second fixed point above.

\subsection{K\"ahler potential for moduli}
\label{S54}
For the case of $g=2$, the minimal K\"ahler potential of eq.~\eqref{minK} reads:
\be
K=-h~\Lambda^2\log\left((\tau_3-\bar \tau_3)^2-(\tau_1-\bar \tau_1)(\tau_2-\bar \tau_2)\right)+
\sum_I \left((\tau_3-\bar \tau_3)^2-(\tau_1-\bar \tau_1)(\tau_2-\bar \tau_2)\right)^{k_I} | \varphi^{(I)}|^2~.
\ee
Observe that when $\tau_3=0$, the dependence on ${\tt Im}(\tau_{1,2})$ factorizes, reproducing the K\"ahler potential expected in compactification on two independent tori.
From $K$ we can derive the kinetic terms. For instance, in the moduli sector we obtain:
\be
{\cal L}_{\tt kin}=K^i_j~\partial_\mu \tau_i \partial^\mu \bar \tau^j~~~,
\ee
where the matrix $K^i_j$ is given by:
\be
K^i_j=\xi
\left(
\begin{array}{ccc}
{\tt Im}(\tau_2)^2&{\tt Im}(\tau_3)^2&-2{\tt Im}(\tau_2){\tt Im}(\tau_3)\\
{\tt Im}(\tau_3)^2&{\tt Im}(\tau_1)^2&-2{\tt Im}(\tau_1){\tt Im}(\tau_3)\\
-2{\tt Im}(\tau_2){\tt Im}(\tau_3)&-2{\tt Im}(\tau_1){\tt Im}(\tau_3) & 2\left({\tt Im}(\tau_3)^2+{\tt Im}(\tau_1){\tt Im}(\tau_2)\right)
\end{array}
\right)~~~,
\ee
with
\be
\xi=\frac{h~\Lambda^2}{4\left(-{\tt Im}(\tau_3)^2+{\tt Im}(\tau_1){\tt Im}(\tau_2)\right)^2}~~~.
\ee
When ${\tt Im}(\tau_3)\ne 0$, the K\"ahler matrix $K^i_j$ is not diagonal and the moduli are coupled, a feature not present in factorized tori.

\section{Explicit models}
\label{model}
In this last section we present some concrete examples of models for fermion masses, separately in the leptonic and in the quark sectors. These are models invariant under rigid supersymmetry, where we restrict to $\Omega$ of eq.~\eqref{Omega} as moduli space and where the role of flavour symmetry is played by the discrete group $N(H)$ generated by the elements given in eq.~\eqref{generators_Sp2}. The Yukawa couplings depend on Siegel modular forms of genus 2 and level 2 restricted to $\Omega$. We will adopt minimal K\"ahler potentials, aware of the fact that this choice implicitly set to zero
additional input parameters related to non-minimal choices. We neglect corrections coming from the breaking of supersymmetry and from the renormalization group flow. In models with modular invariance, both have been shown to be negligible in ample portions of the parameter space~\cite{Criado:2018thu}. In our models the flavons are the four fields $\texttt{Re}(\tau_1)$, $\texttt{Im}(\tau_1)$, $\texttt{Re}(\tau_3)$, $\texttt{Im}(\tau_3)$, which will be treated as free parameters, varied to maximize the agreement between data and theory, by a $\chi$-square-based minimization procedure. Additional parameters are related to
the number of independent invariants allowed by the symmetry. In this respect, we have selected our matter multiplets in such a way that only weight 2 and weight 4 Siegel modular forms enter the Yukawa couplings. If weight 6 or higher modular forms were used, more independent couplings would be involved because of the presence of additional modular forms in the representations $\mathbf{1}$ and $\mathbf{3}'$.
So far we have not tried to optimize the predictability of our models, which are meant to provide only a general test of our construction. In particular, the model dealing with quark masses and mixing angles involves
twelve free parameters, thus exceeding the number of the physical quantities to be reproduced. In more predictive models the number of independent parameters could be reduced by imposing $CP$ invariance,
spontaneously broken by the choice of $\tau$~\cite{Novichkov:2019sqv}.
\subsection{Lepton model I}
\label{lem1}
In the first model, we assume that neutrino masses arises from the type-I seesaw mechanism. The left-handed lepton $L$, right-handed neutrinos $N^c$ and right-handed charged leptons $E^c$ all transform as triplet $\mathbf{3'}$ under the finite Siegel modular group $S_4\times Z_2$. This assignment is of particular interest in view of its compatibility with the embedding in a grand unified theory. The modular transformation properties and weights for the fields are:
\begin{align}
\nonumber&\rho_{E^c}=\rho_{N^c}= \rho_{L} = \mathbf{3'},~~~~ \rho_{H_u}=\rho_{H_d}=\mathbf{1}\,,\\
&k_{H_u}=k_{H_d}=0,~~~ k_{E^c}=k_{N^c}=0,~ ~k_{L}=-2\,.
\end{align}
Hence the superpotential for charged leptons and neutrinos can be written as
\begin{align}
\nonumber w_e &=  \alpha (E^c L Y_\mathbf{3'})_\mathbf{1}H_d + \beta (E^c L Y_\mathbf{1})_\mathbf{1}H_d\,, \\
 w_\nu &= g_1 (N^c L Y_\mathbf{3'})_\mathbf{1} H_u + g_2 (N^c L Y_{\mathbf{1}})_\mathbf{1} H_u + \Lambda(N^c N^c)_\mathbf{1}\,.
\end{align}
Using the Clebsch-Gordon coefficients of the $S_4\times Z_2$ symmetry group listed in appendix~\ref{app:S4xZ2-group}, we get the charged lepton and neutrino mass matrices:
\begin{align}
\label{lept1}
\nonumber&~~~~~~~~~~~~~~~~M_e=  \begin{pmatrix}
2 \alpha Y_1 + \beta Y_4 & -\alpha Y_3 & -\alpha Y_2 \\
-\alpha Y_3 & 2\alpha Y_2 & -\alpha Y_1+ \beta Y_4 \\
-\alpha Y_2 & -\alpha Y_1 + \beta Y_4 & 2\alpha Y_3
\end{pmatrix}v_d  \,, \\
&M_D=\begin{pmatrix}
2g_1Y_1 + g_2 Y_4  & -g_1 Y_3 & -g_1 Y_2 \\
-g_1 Y_3 & 2g_1 Y_2 & -g_1 Y_1 + g_2 Y_4 \\
-g_1 Y_2 & -g_1 Y_1 + g_2 Y_4 & 2g_1 Y_3
\end{pmatrix}v_u\,,\quad
M_N= \Lambda \begin{pmatrix}
1 & 0 & 0 \\
0 & 0 & 1 \\
0 & 1 & 0
\end{pmatrix}\,.
\end{align}
The light neutrino mass matrix $m_{\nu}$ is given by the seesaw formula
\begin{equation}
M_\nu= -M_D^T M_N^{-1} M_D\,.
\end{equation}
Besides the complex moduli $\tau_1$ and $\tau_3$, the charged lepton mass matrix $M_e$ depends only on a dimensionless parameter $\beta/\alpha$ and the overall scale $\alpha v_d$. Analogously the neutrino mass matrix $M_{\nu}$ depends on the parameter $g_2/g_1$ and the overall scale $g_1^2v^2_u/\Lambda$. We can get rid of the phases of $\alpha$ and $g_1$ through field redefinitions. At low energies we are left with 6 real Lagrangian parameters plus the 2 complex VEV of $\tau_1$ and $\tau_3$.
The best fit values of the input parameters are determined to be:
\begin{align}
\label{bf1}
\nonumber &\tau_1 =0.04017+0.89185i\,,\quad \tau_3=0.49053+0.00792i\,,\quad\beta/\alpha= 2.10415 - 0.14380i\,,\\
 &g_2/g_1 = -0.96942 - 3.32507i\,,\quad \alpha v_d = 136.26910 ~\text{MeV}\,,\quad g_1^2v^2_u/\Lambda= 2.71970 ~\text{meV}\,,
\end{align}
where $\tau_1$ and $\tau_3$ are treated as random complex numbers varying in the fundamental domain. The lepton mixing parameters and neutrino masses are predicted to be
\begin{align}
\nonumber &\sin^2\theta_{12}=0.3068\,,\quad \sin^2\theta_{13}=0.02219\,,\quad \sin^2\theta_{23}=0.5753\,,\quad \delta_{CP}=1.09\pi\,,\\
\nonumber &\alpha_{21}= 0.05\pi \,,\quad \alpha_{31} = 0.03\pi\,,\quad m_e/m_\mu=0.00476\,,\quad m_\mu/m_\tau = 0.06071\,,\\
\label{eq:nu-pred-modelI} & m_1=120.75 ~\text{meV}\,,\quad m_2 =121.06 ~\text{meV}\,,\quad m_3 =130.69 ~\text{meV}\,,
\end{align}
which are compatible with the experimental data at $1\sigma$ level~\cite{Esteban:2020cvm}. The light neutrino masses are quasi-degenerate, and the sum of neutrino masses is $m_1+m_2+m_3=372.5$ meV, this is marginally compatible with the latest bound $\sum_i m_i<(120\sim600)$ meV given by Planck~\cite{Aghanim:2018eyx}. From the values of neutrino masses and mixing parameters given in eq.~\eqref{eq:nu-pred-modelI}, one can extract the predictions for the effective neutrino masses $m_{\beta}$ in beta decay and $m_{\beta\beta}$ neutrinoless double beta decay:
\begin{equation}
m_{\beta}=121.07~\text{meV}\,,\quad m_{\beta\beta}=120.43 ~\text{meV}\,,
\end{equation}
which are below the present most stringent upper limits $m_{\beta}<1.1$ eV from KATRIN~\cite{Aker:2019uuj} and $m_{\beta\beta}<(61\sim165)$ meV from KamLAND-Zen~\cite{KamLAND-Zen:2016pfg}. The prediction for $m_{\beta\beta}$ is within the reach of future neutrinoless double beta decay experiments.

We see that the value of $\tau_3$ preferred by data, eq.~\eqref{bf1}, is very close to $1/2$, and the moduli $\tau$ are near the fixed point $\begin{psmallmatrix} \tau_1 & 1/2\\ 1/2 & \tau_1 \end{psmallmatrix}$ of table~\ref{FixedPoints}, where a $Z_2\times Z_2$ subgroup generated by the elements $\mathcal{V}$ and $(\mathcal{T}\mathcal{S}\mathcal{T})^2\mathcal{S}$ is preserved. Indeed, at this fixed point we have $Y_2(\tau)=Y_3(\tau)$ and both $M_e$ and $m_\nu$ become $\mu-\tau$ symmetric~\footnote{For a review of the $\mu-\tau$ symmetry, see for instance ref.~\cite{Altarelli:2010gt} and references therein.}. This specific arrangement does not allow to faithfully reproduce the data: for instance, we get $\theta_{13}=0$. Both the values of $\theta_{13}$ and $\delta_{CP}$  matching the experimental data arise from a small deviation of $\tau$ from the fixed point, an intriguing feature already observed for other parameters in the context of modular flavour symmetries. An interesting property of this model is that the observed hierarchy among charged lepton masses is achieved with Lagrangian parameters $\alpha$ and $\beta$ of the same order. All the dimensionless parameters of the model are
of order one, a significant result in the context of the flavour puzzle.
\subsection{Lepton model II}
The neutrino masses are described by the Weinberg operator in this model. The first two generations of the right-handed charged leptons $E^c_{D}=(E^c_1,~E^c_2)^{T}$ are assumed to transform as a doublet $\mathbf{2}$ under $S_4\times Z_2$, and the third generation $E^c_3$ transforms as $\mathbf{1}$. The representation and weight assignments of the fields are:
\begin{align}
\nonumber&\rho_{E^c}=\mathbf{2}\oplus\mathbf{1},~~~ \rho_{L} = \mathbf{3'},~~~\rho_{H_u}=\rho_{H_d}=\mathbf{1}\,,\\
&k_{H_u}=k_{H_d}=0,~~~ k_{E_D^c}=-3,~~k_{E_3^c}=k_{L}=-1\,.
\end{align}
The superpotential of the lepton sector includes:
\begin{align}
\nonumber w_e &=  \alpha (E_D^c L Y^{(4)}_{\mathbf{3'}a})_\mathbf{1}H_d + \beta (E_D^c L Y^{(4)}_{\mathbf{3'}b})_\mathbf{1}H_d + \gamma (E_3^c L Y_\mathbf{3'})_\mathbf{1}H_d\,, \\
w_\nu &= \frac{g_1}{\Lambda} (L L Y_{\mathbf{3}'})_\mathbf{1}H_u H_u + \frac{g_2}{\Lambda}(L L Y_\mathbf{1})_\mathbf{1}H_u H_u\,.
\end{align}
We can assume real $\alpha$, $\gamma$ and $g_1$ parameters, since their phases can be absorbed by the lepton fields, while the phases of $\beta$ and $g_2$ cannot be removed by exploiting field redefinitions. The charged lepton and neutrino mass matrices read:
\begin{align}
\nonumber&M_e=  \begin{pmatrix}
 \alpha Y^{(4)}_{\mathbf{3'}a,2} + \beta  Y^{(4)}_{\mathbf{3'}b,2} & \alpha Y^{(4)}_{\mathbf{3'}a,1} + \beta  Y^{(4)}_{\mathbf{3'}b,1} & \alpha Y^{(4)}_{\mathbf{3'}a,3} + \beta  Y^{(4)}_{\mathbf{3'}b,3} \\
\alpha Y^{(4)}_{\mathbf{3'}a,3} + \beta  Y^{(4)}_{\mathbf{3'}b,3} & \alpha Y^{(4)}_{\mathbf{3'}a,2} + \beta  Y^{(4)}_{\mathbf{3'}b,2} & \alpha Y^{(4)}_{\mathbf{3'}a,1} + \beta  Y^{(4)}_{\mathbf{3'}b,1} \\
\gamma Y_1 & \gamma Y_3 & \gamma Y_2
\end{pmatrix}v_d   \,, \\
& M_\nu=\begin{pmatrix}
2g_1Y_1 + g_2 Y_4  & -g_1 Y_3 & -g_1 Y_2 \\
-g_1 Y_3 & 2g_1 Y_2 & -g_1 Y_1 + g_2 Y_4 \\
-g_1 Y_2 & -g_1 Y_1 + g_2 Y_4 & 2g_1 Y_3
\end{pmatrix} \frac{v_u^2}{\Lambda}\,.
\end{align}
The predictions depend on seven Lagrangian parameters plus the complex values of $\tau_1$ and $\tau_3$. A good agreement between the model predictions and the experimental data can be achieved for the following choice of parameter values:
\begin{align}
\nonumber &\tau_1 =0.25861+1.04092i\,,\quad \tau_3=-0.49800+0.49265i\,,\\
\nonumber &\beta/\alpha= 0.13339+0.48532i\,,\quad \gamma/\alpha=0.00431\,,\quad g_2/g_1 = 0.78219+2.61802i\,,\\
& \alpha v_d = 171.32533~ \text{MeV}\,,\quad g_1^2v^2_u/\Lambda= 6.95550~\text{meV}\,.
\end{align}
Accordingly the lepton mixing parameters and neutrino masses are determined to be:
\begin{align}
\nonumber &\sin^2\theta_{12}=0.3040\,,\quad \sin^2\theta_{13}=0.02219\,,\quad \sin^2\theta_{23}=0.5699\,,\quad \delta_{CP}=1.50\pi\,,\\
\nonumber & \alpha_{21}= 0.16\pi \,,~~ \alpha_{31} = 0.70\pi\,,~~ m_e/m_\mu=0.00480 ,~~m_\mu/m_\tau=0.05646\,,  \\
\nonumber&m_1=14.51~\text{meV}\,,\quad m_2 =16.87~\text{meV}\,,\quad m_3 =52.21~\text{meV}\,,\\
&m_\beta=16.98~\text{meV} \,,\quad m_{\beta\beta}=15.03~\text{meV}\,.
\end{align}
It is notable that the Dirac CP phase $\delta_{CP}$ is approximately  $3\pi/2$. The neutrino masses are of normal hierarchy type and they are quite tiny. All the experimental bounds in neutrino oscillation, tritium beta decays, neutrinoless double decay and cosmology are nicely fulfilled.

\begin{figure}[t!]
\centering
\includegraphics[width=0.98\linewidth]{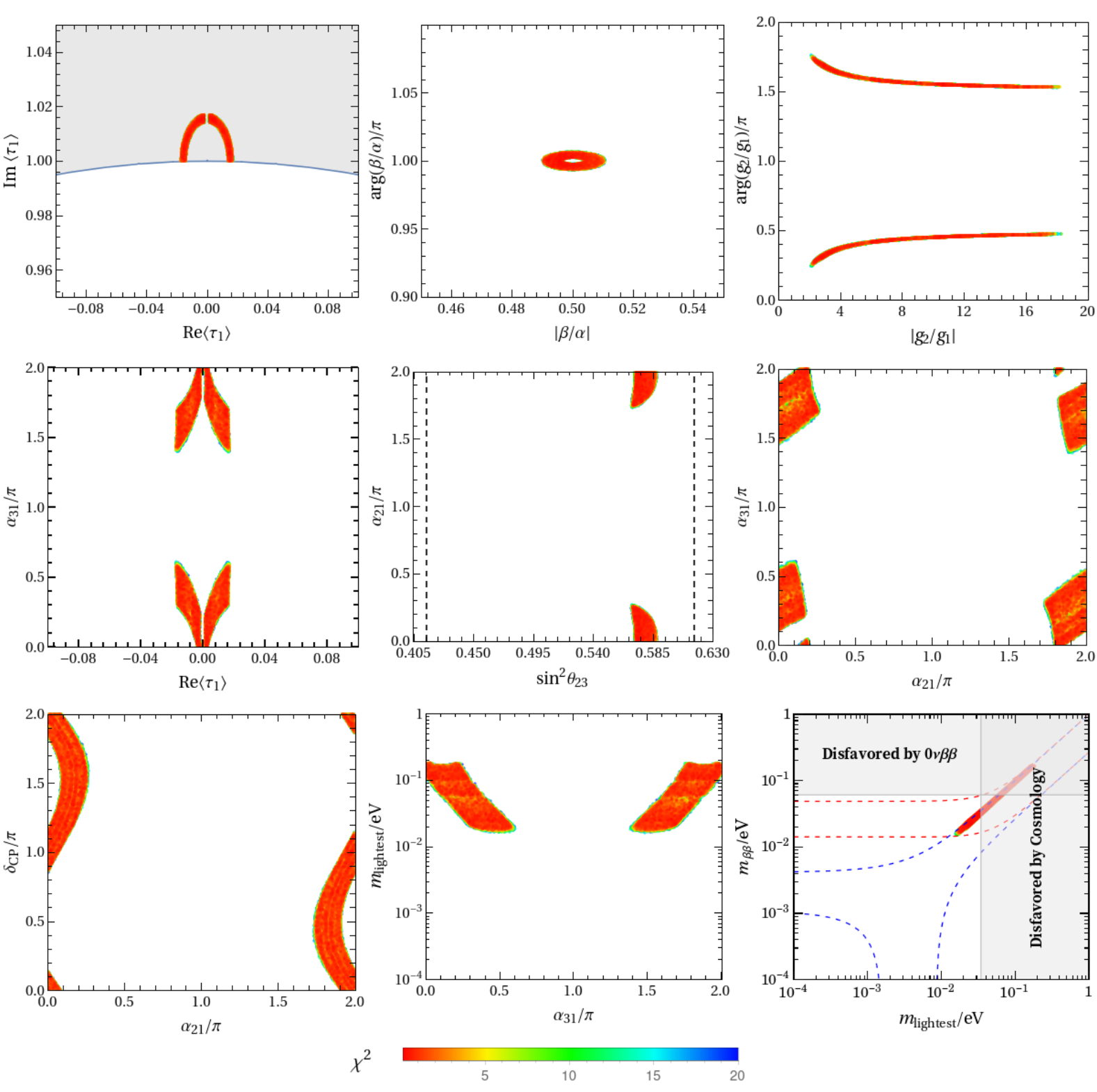}
\caption{The predicted correlations among the input free parameters, neutrino mixing angles and CP violating phases in the \textbf{Model II}, where the moduli are restricted to the subspace with $\tau_1=\tau_2$ and $\tau_3=0$. The last panel shows the prediction for the effective Majorana neutrino mass $m_{\beta\beta}$ in neutrinoless double decay. The vertical black dashed lines denote the $3\sigma$ regions of $\sin^2\theta_{23}$~\cite{Esteban:2020cvm}. The most general allowed regions expected for normal ordering and inverted ordering neutrino masses are indicated by red and blue dashed lines respectively. The horizontal grey band denote the current experimental bound $m_{\beta\beta}<(61\sim165)$ meV from KamLAND-Zen~\cite{KamLAND-Zen:2016pfg}, and the vertical grey exclusion band denotes the current bound $\sum_im_i<120$ meV at 95\% confidence level obtained by the Planck collaboration~\cite{Aghanim:2018eyx}.}
\label{fig:model2}
\end{figure}

Unfortunately, treating the moduli as free parameters considerably reduces
the predictive power of the model. It would be of great help to supplement the approach with some mechanism able to limit the freedom in moduli space. We explore such possibility by restricting the complex moduli to the fixed points of table~\ref{FixedPoints}, compatible with the condition $\tau_1=\tau_2$. We found that by setting $\tau_3=0$ the observed lepton masses and mixing angles can still be accommodated. The best fit values of the remaining nine input parameters are given by:
\begin{equation}
\begin{gathered}
\tau_1 = 0.01541 + 1.00011i \,,~~~~\beta/\alpha= -0.49134+0.00224i\,,~~~\gamma/\alpha=0.00227\,,\\
g_2/g_1 = 1.58430-2.16369i\,,~~~~ \alpha v_d = 259.61745~ \text{MeV}\,,\quad g_1^2v^2_u/\Lambda= 6.97565~\text{meV}\,,
\end{gathered}
\end{equation}
giving rise to the following predictions for lepton masses and mixing parameters:
\begin{align}
\nonumber&\sin^2\theta_{12}=0.3037\,,\quad \sin^2\theta_{13}=0.02219\,,\quad \sin^2\theta_{23}=0.5708\,,\quad \delta_{CP}=1.53 \pi\,,\\
\nonumber&\alpha_{21}= 0.25 \pi\,,\quad \alpha_{31} = 1.70 \pi\,,\quad  m_e/m_\mu=0.00480 \,,\quad m_\mu/m_\tau=0.05747\,, \\
\nonumber&m_1 =22.23~\text{meV}\,,\quad m_2=23.84~\text{meV} \,,\quad m_3=54.88~\text{meV}\,,\\
& m_{\beta}=23.92~\text{meV}\,, \quad m_{\beta\beta}=20.57~\text{meV} \,.
\end{align}
All the mixing angles as well as the neutrino squared mass differences are in the experimentally preferred $1\sigma$ range~\cite{Esteban:2020cvm}, and the Dirac CP phase $\delta_{CP}$ is around $3\pi/2$. Furthermore, we have comprehensively explored the parameter space of this model. Requiring the three lepton mixing angles and neutrino squared mass splittings $\Delta m^2_{21}$, $\Delta m^2_{31}$ to lie in the experimentally allowed $3\sigma$ regions~\cite{Esteban:2020cvm}, we get the correlations between the free parameters and observable quantities shown in figure~\ref{fig:model2}. We observe that the atmospheric mixing angle is predicted to be in the narrow range $0.569\leq\sin^2\theta_{23}\leq0.588$.

Furthermore, the phenomenologically viable region in the $\tau_1$ plane is around $\tau_1=i$ so that the realistic values of the moduli $\tau$ are close to the zero-dimensional fixed point  $\begin{psmallmatrix} i & 0 \\ 0 & i \end{psmallmatrix}$.

\subsection{A quark model}
In this section, we apply our framework to the quark sector. The left-handed quarks $Q$ are assumed to transform as a triplet $\mathbf{3'}$ under the finite modular group $S_4\times Z_2$, both right-handed up quarks and right-handed down quarks are assigned to the direct sum of a doublet $\mathbf{2}$ and a singlet $\mathbf{1}$:
\begin{align}
\nonumber&\rho_{u^c_D}=\rho_{d^c_D}=\mathbf{2},~~~\rho_{u^c_3}=\rho_{d^c_3}=\mathbf{2}=\mathbf{1} ,\quad \rho_{Q} = \mathbf{3'},\quad \rho_{H_u}=\rho_{H_d}=\mathbf{1}\,, \\
&k_{H_u}=k_{H_d}=0,\quad k_Q=4-k_{u_D^c}=4-k_{d_D^c}=2-k_{u_3^c}=2-k_{d_3^c}\,,
\end{align}
where $u^c_{D}=(u^c_1,~u^c_2)^{T}$, $d^c_{D}=(d^c_1,~d^c_2)^{T}$, and $k_Q$ is a generic real number. With this assignment the superpotentials $w_u$ and $w_d$ have the same structure:
\begin{align}
\nonumber w_u &=  \alpha_u (u_D^c Q Y^{(4)}_{\mathbf{3'}a})_\mathbf{1}H_u + \beta_u (u_D^c L Y^{(4)}_{\mathbf{3'}b})_\mathbf{1}H_u + \gamma_u (u_3^c Q Y_\mathbf{3'})_\mathbf{1}H_u\,, \\
\label{eq:quark}w_d &=  \alpha_d (d_D^c Q Y^{(4)}_{\mathbf{3'}a})_\mathbf{1}H_d + \beta_d (d_D^c L Y^{(4)}_{\mathbf{3'}b})_\mathbf{1}H_d + \gamma_d (d_3^c Q Y_\mathbf{3'})_\mathbf{1}H_d\,,
\end{align}
where the parameters $\alpha_u$, $\gamma_u$, $\alpha_d$ and $\gamma_d$ can be assumed real since their phases are unphysical. On the contrary, the phases of $\beta_u$ and $\beta_d$  cannot be removed by a field redefinition. From eq.~\eqref{eq:quark}, we obtain the up and down quark mass matrices:
\begin{align}
\nonumber&M_u=  \begin{pmatrix}
\alpha_u Y^{(4)}_{\mathbf{3'}a,2} + \beta_u Y^{(4)}_{\mathbf{3'}b,2} & \alpha_u Y^{(4)}_{\mathbf{3'}a,1} + \beta_u Y^{(4)}_{\mathbf{3'}b,1} & \alpha_u Y^{(4)}_{\mathbf{3'}a,3} + \beta_u Y^{(4)}_{\mathbf{3'}b,3} \\
\alpha_u Y^{(4)}_{\mathbf{3'}a,3} + \beta_u Y^{(4)}_{\mathbf{3'}b,3} & \alpha_u Y^{(4)}_{\mathbf{3'}a,2} + \beta_u  Y^{(4)}_{\mathbf{3'}b,2} & \alpha_u Y^{(4)}_{\mathbf{3'}a,1} + \beta_u  Y^{(4)}_{\mathbf{3'}b,1} \\
\gamma_u Y_1 & \gamma_u Y_3 & \gamma_u Y_2
\end{pmatrix}v_u  \,, \\
& M_d=  \begin{pmatrix}
 \alpha_d Y^{(4)}_{\mathbf{3'}a,2} + \beta_d  Y^{(4)}_{\mathbf{3'}b,2} & \alpha_d Y^{(4)}_{\mathbf{3'}a,1} + \beta_d  Y^{(4)}_{\mathbf{3'}b,1} & \alpha_d Y^{(4)}_{\mathbf{3'}a,3} + \beta_d Y^{(4)}_{\mathbf{3'}b,3} \\
\alpha_d Y^{(4)}_{\mathbf{3'}a,3} + \beta_d  Y^{(4)}_{\mathbf{3'}b,3} & \alpha_d Y^{(4)}_{\mathbf{3'}a,2} + \beta_d  Y^{(4)}_{\mathbf{3'}b,2} & \alpha_d Y^{(4)}_{\mathbf{3'}a,1} + \beta_d  Y^{(4)}_{\mathbf{3'}b,1} \\
\gamma_d Y_1 & \gamma_d Y_3 & \gamma_d Y_2
\end{pmatrix}v_d  \,.
\end{align}
We find that the agreement between model predictions and experimental data is optimized by the following values of the free parameters
\begin{align}
\nonumber &\tau_1 =-0.41097+2.03917i\,,~ \tau_3=0.01711+1.19070i\,,~\beta_u/\alpha_u= -0.18760-0.03857i\,,\\
\nonumber& ~\gamma_u/\alpha_u=20.75350\,, ~ \beta_d/\alpha_d=-0.15847-0.02620i\,,~\gamma_d/\alpha_d=0.15466\,,\\
&~\alpha_u v_u = 2.39972~ \text{GeV}\,,~ \alpha_d v_d= 0.40724~\text{GeV}\,.
\end{align}
The predictions for the quark masses and quark mixing parameters are given by
\begin{align}
\nonumber &~~~~~~~~\theta^q_{12}=0.2275\,,\quad \theta^q_{13}=0.0031\,,\quad \theta^q_{23}=0.0388\,,\quad \delta^q_{CP}=68.4^\circ \,,\\
 & m_u/m_c= 0.00197 \,,\quad m_c/m_t = 0.00272\,,\quad m_d/m_s=0.05042,~~m_s/m_b=0.02030\,.
\end{align}
All these observable quantities fall in the $3\sigma$ range of the values obtained at the GUT scale in the minimal SUSY breaking scenario with SUSY breaking scale $M_{\text{SUSY}}= 1$ TeV and $\tan\beta=7.5$, $\bar{\eta}_b=0.09375$~\cite{Antusch:2013jca}. Since we are treating $\tau_{1,3}$ as free input variables, in this example the number of free parameters exceeds the number of observable quantities and there is no predictive power. This shows the relevance of any dynamical mechanism able to determine the moduli and reduce the arbitrariness of the predictions.
\subsection{Towards a quark-lepton unified description}
In the lepton model described in section~\ref{lem1}, all matter fields transform in the $\mathbf{3'}$ representation under the finite Siegel modular group $S_4\times Z_2$. Moreover the charged lepton mass matrix possesses hierarchical eigenvalues for parameters $\alpha$ and $\beta$ of the same order of magnitude. This might be considered as a promising starting point to describe the whole fermion spectrum in terms of a unique choice for $\tau_1$ and $\tau_3$. Nevertheless, by exploiting for quarks superfields the same assignment used for leptons in section~\ref{lem1}, namely $\rho_{u^c}=\rho_{d^c}=\rho_{Q} = \mathbf{3'},\quad k_{u^c}=k_{d^c}=-2-k_Q$, we could not find a value of the model parameters providing a good fit of the data. However, when restricting the fit to the quark sector alone, values of $\tau_1$ and $\tau_3$ very close to those in eq.~\eqref{bf1} for the lepton sector give rise to a picture in a qualitative agreement with the experimental values. In particular, setting to approximately zero $V_{ub}$ and $V_{cb}$, the Cabibbo angle can be correctly reproduced. At the same time, all quark mass ratios, but $m_c/m_t$, fall in the experimentally allowed range. The predicted value of $m_c/m_t$ is close to 0.02, out of the experimental range, but exhibiting an hierarchy in the correct direction. This leaves open the possibility that additional small effects, like for instance a departure of the K\"ahler potential from the minimal choice adopted here, could fix the failures of this simple scheme.

\section{Discussion}
\label{S7}

If relevant to the solution of the flavour puzzle, flavour symmetries are well hidden in the data and realized in a broken phase~\cite{Feruglio:2019ktm}. So far, the complete arbitrariness of the symmetry breaking sector has represented a formidable obstacle in the realization of a predictive framework. An interesting possibility to reduce the ambiguity of a pure bottom-up approach comes from string theory where the background over which the string propagates provides a natural setup for the symmetry breaking sector. For instance, typical compactifications of extra dimensions require the presence of moduli, four-dimensional scalar fields transforming under discrete duality symmetries that can be part of the flavour symmetry group. The simplest example of two extra dimensions compactified on a torus or an orbifold gives rise to target space modular invariance, inherited by the low-energy effective theory.

Modular invariant supersymmetric theories have been shown successful in constraining Yukawa couplings and in removing the arbitrariness related to the traditional flavon sector. Nevertheless there are several motivations to go beyond the case of a single modulus. In a bottom-up perspective, it is difficult to realize viable models where both the neutrino and the charged lepton sectors are correctly reproduced by a unique modulus. In the existing examples of this type, ad-hoc free parameters are needed to fit the charged lepton masses. Also the simultaneous description of quarks and leptons seem to require more than one modulus. From a top-down viewpoint, low energy theories arising from the compactification of six extra dimensions
typically display dependence on a variety of moduli, many of them of geometrical type.

It is therefore of great interest to search for an extension of modular invariant supersymmetric theories where several moduli can occur. The most straightforward extension involves a product of separate one-dimensional moduli spaces, thus parametrizing a surface factorized into independent tori~\cite{Ferrara:1989qb,deMedeirosVarzielas:2019cyj}.
In the present work we have shown that a much more general extension exists, which has its roots in the theory of automorphic forms of several variables. The moduli space is a coset $G/K$ originating from a Lie group $G$ and a subgroup $K$. Automorphic forms are periodic functions under the action of a discrete subgroup $G_d$  of $G$. Classical modular forms are recovered for $G=SL(2,\mathbb{R})$, $K=SO(2)$ and $G_d$ the modular group or one of its
subgroups. We have shown how, for a generic choice of $G$, $K$ and $G_d$, we can define an automorphic supersymmetric theory where moduli span $G/K$ and $G_d$ is closely related to the flavour symmetry group. This picture nicely fits the framework of various supergravity theories and string theory where moduli spaces of the most common compactifications are given by noncompact groups modded out by their maximal compact subgroups and discrete duality groups. Our approach is purely bottom-up, much as the one of modular invariant supersymmetric theories discovered in the late eighties. In particular, in the general case, we have built a minimal K\"ahler potential and the most general superpotential.

We have shown how the construction specializes when $G=Sp(2g,\mathbb{R})$, $K=U(g)$ and $G_d$ a subgroup of $Sp(2g,\mathbb{Z})$. The automorphic forms are Siegel modular forms. When $g=2$ the moduli space has three, non-factorizable, complex dimensions. Its fundamental region under $Sp(2g,\mathbb{Z})$ has many invariant loci, regions with a non-trivial
residual symmetry of complex dimension two, one or zero, a feature not present when the modulus is unique. We have exploited this fact to show how we can consistently restrict the theory to one of these loci and build phenomenologically viable models. As a proof of principle we have presented several models for lepton and quark masses and mixing angles.

There are several aspects that have been disregarded or only briefly mentioned in our work. So far, moduli have mostly been treated as part of the independent free parameters to be varied in order to match the experimental data. Moving from the single modulus to the multi moduli case, such viewpoint is no more tenable in a setup aiming at a good degree of predictability. Our discussion makes explicit the importance of finding some criterium to select the correct moduli VEV. Inspired by the observation that minima of modular invariant functionals typically occur in regions of
enhanced residual symmetries, we have explored such possibility in our explicit examples.
At least in the lepton sector, the models we have built can accomodate the data when moduli either coincide with fixed points or are very close to them, an intriguing feature that has already appeared in the context of single moduli models. Features like the smallness of $\theta_{13}$ or the non-vanishing of $\delta_{CP}$ seem to be related to small departure of the moduli from fixed points. We also came close to a unified model for both quark and lepton masses described in terms of a single point of enhanced symmetry in moduli space, though additional corrections are required to achieve a realistic description. These results represent just a first step and clearly there is a large room for a more systematic and comprehensive analysis. Another well-known aspect which threats predictability is the choice of a more general K\"ahler potential~\cite{Feruglio:2017spp,Chen:2019ewa}. As in the single modulus case, it is reasonable to expect deviations of the most general K\"ahler potential from the minimal choice postulated here. In a bottom-up approach there is no compelling reason to give preference to the minimal option and the predictions can depend on an additional set of parameters. In the single modulus case, finite modular invariance can be just part of a more general symmetry, dubbed "eclectic flavour group", which also includes an
ordinary flavour group leaving moduli invariant~\cite{Baur:2019kwi,Baur:2019iai,Nilles:2020nnc}. Restrictions imposed by such bigger symmetry might result in a more constrained K\"aler potential~\cite{Nilles:2020kgo}. Examples of this type have been shown
to occur in low-energy realizations of string theory~\cite{Nilles:2020tdp,Baur:2020jwc}. It would be interesting to see which eclectic flavour groups arise in the more general context studied here. Multi-moduli can also offer new possibilities to realize hierarchical mass spectra and to describe more efficiently  the observed charged lepton and quark masses. In a specific model, we achieved an excellent agreement in the lepton sector with all dimensionless input parameters of order one. A much more ambitious but very interesting step would be to build a consistent modular invariant non-supersymmetric quantum field theory. In principle the framework of automorphic forms, not necessarily implying holomorphy of the
functions with good modular transformation properties, might provide the correct tools to carry out such a construction.

``Are all the dimensionless parameters that characterize the physical universe calculable in principle or are some merely determined by historical or quantum mechanical accident and uncalculable ?''~\cite{Duff:2001zk} If fitting the general framework of automorphic forms described above, Yukawa couplings might display at the same time both the characters of
a purely accidental event, consequence of an unknown mechanism of vacuum selection, and those of a highly constrained and mathematical beautiful construction.

\begin{appendices}
\setlength{\extrarowheight}{0.0 cm}

\section{Hermitian Symmetric Spaces}
\label{HSS}
Well known examples of moduli spaces relevant to our construction are hermitian symmetric spaces \cite{book}. Hermitian spaces are manifolds equipped with a Riemannian metric and an integrable, almost complex, structure which preserves the metric. At each point $p$ of an hermitian symmetric space there is a reflection $s_p$ $(s_p^2=1)$ preserving the hermitian structure and having $p$ as unique fixed point, $s_p p=p$.

Every hermitian symmetric space $M$ is a K\"ahler manifold and is a coset space of the type $M=G/K$ for some connected
Lie group $G$ and a compact subgroup $K$ of $G$. The Lie algebra ${\cal G}$ of $G$ decomposes as ${\cal G}={\cal V}\oplus{\cal A}$,
where ${\cal V}$ is the Lie algebra of $K$, and ${\cal V}$ and ${\cal A}$ are orthogonal with respect to the killing form
\be
B(X,Y)={\tt tr}(Ad(X) Ad(Y))~~~.
\ee
We have $B(V,A)=0$, for any $V(A)\in{\cal V}({\cal A})$.
The reflection $s_p$ induces an automorphism of ${\cal G}$ such that $V+A\to V-A$. As a consequence the Lie algebra satisfies $[{\cal V},{\cal V}]\subset {\cal V}$, $[{\cal V},{\cal A}]\subset {\cal A}$ and $[{\cal A},{\cal A}]\subset {\cal V}$. Since $K$ is compact, with an appropriate sign convention, we get $B(V,V)<0$ for any $V\in{\cal V}$.

Depending on the sign of $B(A,A)$ $(A\in{\cal A})$, an hermitian symmetric space $M$ can be
of compact type $(B(A,A)<0)$, of noncompact type $(B(A,A)>0)$ or of euclidean type $(B(A,A)=0)$.
In general none of these cases applies and $M$ decomposes as a product $M=M_c\times M_{nc}\times M_e$, where the three factors
are hermitian symmetric spaces of compact, noncompact and euclidean type, respectively.
A hermitian symmetric space is irreducible if it is not the product of two hermitian symmetric spaces of lower dimension.

An irreducible hermitian symmetric space is a coset space of the type
$G/K$ where $G$ is a simple, connected Lie group and $K$ is a connected maximal compact subgroup of $G$.
Irreducible hermitian symmetric spaces has been classified \cite{wolf,calabi1960compact}.
$G/K$ is a finite dimensional complex manifold. Irreducible hermitian symmetric spaces of noncompact type are listed in table \ref{tab:coset_space}. Spaces $\mathbf{I} - \mathbf{IV}$ are called classical, while $\mathbf{V}$ and  $\mathbf{VI}$ are called exceptional.
\begin{table}[th!]
\centering
\resizebox{1.0\textwidth}{!}{
\begin{tabular}{|c|c|c|c|c|}\hline\hline
Type & Group $G$ & Compact subgroup $K$ & $\dim_{\mathbb{C}}G/K$ & Symmetric domain $\mathcal{D}$  \\ \hline
$\mathbf{I}_{m,n}$ & $U(m,n)$ & $U(m)\times U(n)$ & $mn$ & $\Big\{z\in M_{m,n}(\mathbb{C})~|~ 1_n - z^\dagger z > 0 \Big\}$  \\ \hline
$\mathbf{II}_{m}$ & $SO^*(2m)$ & $U(m)$ & $\frac{1}{2}m(m-1)$ & $\Big\{z\in M^{\text{skew}}_{m,m}(\mathbb{C})~|~ 1_m - z^\dagger z > 0\Big\}$   \\ \hline
 $\mathbf{III}_{m}$ & $Sp(2m)$ & $U(m)$ & $\frac{1}{2}m(m+1)$ & $\Big\{z\in M^{\text{sym}}_{m,m}(\mathbb{C})~|~ 1_m - z^\dagger z > 0\Big\}$   \\ \hline
 $\mathbf{IV}_{m}$ & $SO(m,2)$ & $SO(m)\times SO(2)$ & $m$ & $\Big\{z\in \mathbb{C}^m~|~ 1 + |z^t z|^2 -2 z^\dagger z > 0,~z^\dagger z <1 \Big\}$   \\ \hline
$\mathbf{V}$ &  $E_{6,-14}$ & $SO(10)\times SO(2)$ & $16$ & $\mathcal{D} \subset \mathbb{O}^2_{\mathbb{C}}$   \\ \hline
$\mathbf{VI}$ &  $E_{7,-25}$ & $E_6\times U(1)$ & $27$ & $\mathcal{D} \subset H_3(\mathbb{O}_{\mathbb{C}})$   \\ \hline \hline
\end{tabular} }
\caption{\label{tab:coset_space} Irreducible hermitian symmetric manifolds of noncompact type, their complex dimension, and the bounded symmetric domain. With $M_{m,n}(\mathbb{C})$ we denote the set of $m$-by-$n$ complex matrices, and $M^{\text{skew(sym)}}_{m,m}(\mathbb{C})$ refers to the subset  of $m$-by-$m$ antisymmetric(symmetric) complex matrices; $\mathbb{O}_{\mathbb{C}}$ refers to the complexification of octonions (also called the complex Cayley algebra), and $H_3(\mathbb{O}_{\mathbb{C}})$ is the Hermitian $3\times 3$ matrices with entries in $\mathbb{O}_{\mathbb{C}}$. The symmetric domains of the last two exceptional types described in Ref.~\cite{Viviani2014A}.}
\end{table}
Irreducible hermitian symmetric spaces of compact type can be obtained from the noncompact ones, by means of a transformation
on the generators of the Lie algebra ${\cal G}$: $(V,A)\to (V,i A)$. All the new generators $X$ satisfy $B(X,X)<0$ and the resulting
space is of compact type.

Irreducible hermitian symmetric spaces of noncompact type are complex analytically equivalent to a bounded symmetric domain (Cartan domain) $\mathcal{D}$, listed in table~\ref{tab:coset_space}. This observation is very useful, since
all bounded symmetric domains admit an hermitian metric of K\"aler type (Bergman metric), which is explicitly known.
For classical spaces, such a metric can be derived by the following K\"ahler potentials~\cite{calabi1960compact}:
\begin{align}
\label{KahlerPotential}
\nonumber
& \mathbf{I}_{m ,n} :~~~ K(z,z^*)=-\log \det (1_n - z^\dagger z)\,, \\
\nonumber
& \mathbf{II}_{m} :~~~ K(z,z^*)=- \frac{1}{2}\log \det (1_m - z^\dagger z)\,, \\
\nonumber
& \mathbf{III}_{m} :~~~ K(z,z^*)=- \frac{1}{2}\log \det (1_m - z^\dagger z) \,,\\
& \mathbf{IV}_{m} :~~~ K(z,z^*)=- \log(1 +  |z^t z|^2- 2z^\dagger z) \,.
\end{align}
We have explicitly shown that our general ansatz for the K\"ahler potential of $\tau$, eq.~\eqref{candidateK}, coincides with the above
expression in case $\mathbf{III}_{m}$, after relating the variables $\tau$ and $z$ through a Cayley transformation. There are different parameterizations of the above coset spaces. In addition to the bounded symmetric domains, there are more commonly used unbounded models:
\begin{align}
\nonumber
& \mathbf{I}_{m \leq n} : \mathcal{H}_U=\Big\{(z,u) ~|~ i(z-z^\dagger) + u^\dagger u <0 , \quad z\in M_{n}(\mathbb{C}),~u\in M_{n-m,n}(\mathbb{C}) \Big\}  \,,\\
\nonumber
& \mathbf{II}_{m} : \mathcal{H}_{O^*}=\Big\{z ~|~ i(z-z^\dagger) > 0 , \quad z\in M_{m}(\mathbb{H})\Big\} \,, \\
\nonumber
& \mathbf{III}_{m} : \mathcal{H}_{g}=\Big\{z ~|~ -i(z-z^*) > 0 , \quad z\in M^{sym}_{m}(\mathbb{C})\Big\}  \,,\\
& \mathbf{IV}_{m} : \mathcal{H}_{O}=\Big\{(z,u) ~|~ i(z-z^*)+|u|^2 < 0 , \quad z\in \mathbb{C},~ u \in \mathbb{C}^m \Big\} \,.
\end{align}
Note that for type $\mathbf{I}_{m,m}$, $\mathcal{H}_{U}$ is the so-called hermitian upper half-space, $\mathcal{H}_{O^*}$ is the so-called quaternion upper half-space, and $\mathcal{H}_g$ is the so-called Siegel upper half plane. The bounded symmetric domain  can be mapped to the unbounded model by a Cayley transformation.

For the classical cases $\mathbf{I}_{m,n}$, $\mathbf{II}_m$ and $\mathbf{III}_m$, the noncompact group $G$ can be realized by $2\times 2$ block matrices:
\begin{align}
\textsf{g}=\begin{pmatrix}
A & B \\
C & D
\end{pmatrix} \,,
\end{align}
where for the case $\mathbf{I}_{m,n}$, $A$ is $m$-by-$m$, $B$ is $m$-by-$n$, $C$ is $n$-by-$m$ and $D$ is $n$-by-$n$ matrices. For cases $\mathbf{II}_m$ and $\mathbf{III}_m$, $A, B, C, D$ are all $m$-by-$m$ matrices.
They act on the corresponding bounded or unbounded domain through the generalized linear fractional transformations
\begin{equation}
\begin{pmatrix}
A & B \\
C & D
\end{pmatrix} z = (A z + B)(C z + D)^{-1} \,.
\end{equation}
The matrices $A, B, C, D$ satisfy appropriate conditions, that depend on the parametrization chosen for $G/K$. The metric defined by the K\"ahler potentials of eq.~\eqref{KahlerPotential} is invariant under the action of the group $G$. The action of $G$ on $\mathcal{H}$ induces a non-trivial  factor of automorphy $(\textsf{g},  z) \mapsto j(\textsf{g},z)$, which satisfy eq.~\eqref{autom}. Then we can define the automorphic forms corresponding to the group $G$ with respect to $G_d$, similar to eq.~\eqref{af}. Some concrete examples of automorphic forms on these noncompact groups $G$ with respect to certain arithmetic subgroup $G_d$ have been discussed by mathematicians~\cite{shimura1978arithmetic,hofmann2011automorphic,klocker2005modular}.

\section{Siegel fundamental domain for $\Gamma_g$}
\label{A1}
A fundamental domain $\mathcal{F}_g$ for the action of $\Gamma_g$ on $\mathcal{H}_g$ ca be defined
as follows \cite{siegel1943symplectic}:
\begin{equation}
\mathcal{F}_g = \left\{\tau \in \mathcal{H}_g  \,\Bigg|
\begin{cases}
\texttt{Im}(\tau) \,\text{is reduced in the sense of Minkowski}, \\
|\det(C\tau + D)| \geq 1 \quad \forall \gamma \in \Gamma_g,\\
|\texttt{Re}(\tau_{ij})| \leq 1/2;
\end{cases} \right\}\,.
\end{equation}
Here Minkowski reduced means that $\texttt{Im}(\tau)$ satisfies the two properties: \\ 1) $h^t \texttt{Im}(\tau) h \geq \texttt{Im}(\tau)_{kk} \quad (\forall h =(h_1,\dots,h_g) \in \mathbb{Z}^g) $  for $1 \leq k \leq g$ whenever $h_1,\dots,h_g $ are coprime ; \\
2)$ \texttt{Im}(\tau)_{k,k+1} \geq 0 \,$ for $0 \leq  k \leq g-1$.

It can be shown that in the case $g=1$, $\mathcal{F}_1$ is the $SL(2,\mathbb{Z})$ fundamental domain $\mathcal{F}_1 = \left\{\tau \in \mathcal{H}_1 ~|~ |\texttt{Re}(\tau)| \leq 1/2, |\tau| \geq 1  \right\}$.

\section{Action of $\Gamma_g$ on $\mathcal{M}_k(\Gamma_g(n))$}
\label{app:decomposition-SMF}
In the following, we shall use $\gamma$ and $h$ to denote a generic element of $\Gamma_g$ and $\Gamma_g(n)$ respectively, that is $\gamma \in \Gamma_g$, $h \in \Gamma_g(n)$.
In the finite-dimensional complex vector space $\mathcal{M}_k(\Gamma_g(n))$ we
select a multiplet of linearly independent Siegel modular forms $f(\tau)\equiv(f_1(\tau),\,f_2(\tau),\,\dots,\,f_m(\tau))^T $ with $m=\texttt{dim}~\mathcal{M}_{k}(\Gamma_g(n))$.
We define the functions $F_{\gamma}(\tau)= j^{-1}_k(\gamma,\,\tau) f(\gamma\tau)$, where
$j_k(\gamma,\tau)$ stands for the automorphy factor $\det(C\tau+D)^k$, satisfying the cocycle relation~\cite{Freitag1991},
\begin{equation}
\label{cocycle}
j_k(\gamma\gamma', \tau)=j_k(\gamma,\gamma'\tau)j_k(\gamma',\tau)\,.
\end{equation}
Under $h\in\Gamma_g(n)$, $F_{\gamma}(\tau)$ transforms as
\begin{equation}
\label{eq:F-SMF}F_{\gamma}(h\tau)=j_k(h,\,\tau)F_{\gamma}(\tau)\,,
\end{equation}
which implies that the holomorphic functions $F_{\gamma}(\tau)$ are actually Siegel modular forms of $\Gamma_g(n)$ with weight $k$. Therefore $F_{\gamma}(\tau)$ can be written as linear combination of $f_i(\tau)$, i.e.
\begin{equation}
\label{eq:rho_comb}F_{\gamma}(\tau)=\rho(\gamma)f(\tau)\,,
\end{equation}
where the linear combination matrix $\rho(\gamma)$ only depends on the modular transformation $\gamma$.
Recalling the definition of $F_\gamma(\tau)$ we have
\begin{equation}
\label{eq:decomp_reducible}f(\gamma\tau)=j_k(\gamma,\,\tau)\rho(\gamma)f(\tau)=\det(C\tau+D)^k\rho(\gamma)f(\tau)\,.
\end{equation}
Using eq.~\eqref{eq:decomp_reducible}, we get
\begin{equation}
\label{eq:fg1g2_fir}f(\gamma_1\gamma_2\tau)=j_k(\gamma_1\gamma_2, \tau)\rho(\gamma_1\gamma_2)f(\tau)\,,
\end{equation}
and
\begin{equation} \label{eq:fg1g2_sec}f(\gamma_1\gamma_2\tau)=j_k(\gamma_1,\,\gamma_2\tau)\rho(\gamma_1)f(\gamma_2\tau)
=j_k(\gamma_1\gamma_2, \tau)\rho(\gamma_1)\rho(\gamma_2)f(\tau)\,.
\end{equation}
Comparing eq.~\eqref{eq:fg1g2_fir} with eq.~\eqref{eq:fg1g2_sec}, we arrive at the following result,
\begin{equation}
\label{eq:homomorphism}\rho(\gamma_1\gamma_2)=\rho(\gamma_1)\rho(\gamma_2)\,,
\end{equation}
showing that $\rho$ is a linear representation of the Siegel modular group $\Gamma_g$. From eq.~\eqref{eq:decomp_reducible} and the definition of Siegel modular form in eq.~\eqref{SiegelForms}, we see that $\rho(h)=\mathbb{1}$, when $h$ belongs to $\Gamma_g(n)$.
Therefore $\rho$ is actually a linear representation of the quotient group $\Gamma_{g, n}=\Gamma_g/\Gamma_g(n)$.  By the Maschke's theorem~\cite{rao2006linear}, the representation $\rho$ is completely reducible and it can be decomposed into a direct sum of irreducible unitary representations. Furthermore, applying eq.~\eqref{eq:decomp_reducible} to the element $\gamma=S^2$, we have
\begin{equation}
\rho(S^2)=(-1)^{kg}\,,
\end{equation}
which is always an identity matrix for even $g$. This implies that $S^2$ is identified with the unit element in the representation $\rho$ for even $g$ or even $k$. As a consequence, the Siegel modular forms would be arranged into multiplets of the the inhomogeneous finite modular group $\bar{\Gamma}_{g, n}\equiv\Gamma_{g, n}/\{1,S^2\}$ for $n>2$. Note $\bar{\Gamma}_{g, n}=\Gamma_{g, n}$ for $n=2$. In the case that both $g$ and $k$ are odd numbers, we have $\rho(S^2)=-1$ and thus Siegel form and the matter fields are allowed to transform under $S^2$, as can be seen from eqs.~(\ref{SMF_decomposition}, \ref{eq:tmg-1}), although the moduli $\tau$ don't change under the action of $S^2$. Hence the odd weight Siegel modular forms for odd $g$ could be organized into multiplets of the the homogeneous finite modular group $\Gamma_{g, n}$. The general analysis of odd weight weight modular forms for $g=1$ has been given in~\cite{Liu:2019khw}. If the weight $k$ is a non-integer, the $g=1$ case indicates that at least the Siegel modular group $\Gamma_g$ should be extended to certain covering group and a multiplier should be included in the definition of Siegel modular form~\cite{Liu:2020msy}. In a similar fashion, the Siegel modular forms of $N(H,n)$ can be decomposed into irreducible representations of the finite modular subgroup $N_n(H)$.

\section{Generators of $\bar{H}$ and $N(H)$ for fixed points in ${\cal F}_2$}
\label{SN}
\subsection{Dimension 2}
There are two non-equivalent manifolds of complex dimension 2, i.e. parametrized by two moduli, whose points are left fixed by
subgroups of $\Gamma_2$.
\begin{itemize}
\item[1.] $\tau= \begin{pmatrix} \tau_1 & 0 \\ 0 & \tau_2 \end{pmatrix}$ \,. \\
In this case the stabilizer $\bar{H}=Z_2$ is generated by
\be
h=\begin{pmatrix}
1 & 0 & 0 & 0 \\ 0 & -1 & 0 & 0 \\ 0 & 0 & 1 & 0 \\ 0 & 0 & 0 & -1
\end{pmatrix}\,,
\ee
while the normalizer $N(H)$ has generators:
\begin{align}
\nonumber
&G_1=\begin{pmatrix} 1&0&0&0 \\ 0&0&0&1 \\ 0&0&1&0 \\ 0&-1&0&0\end{pmatrix} ,~~
G_2=\begin{pmatrix}
1&0&0&0\\ 0&1&0&1 \\ 0&0&1&0 \\ 0&0&0&1 \end{pmatrix},~~
G'_1=\begin{pmatrix}
0&0&1&0\\ 0&1&0&0 \\ -1&0&0&0 \\ 0&0&0&1 \end{pmatrix},\\
&\label{NorTwo1} G'_2=\begin{pmatrix}
1&0&1&0\\ 0&1&0&0 \\ 0&0&1&0 \\ 0&0&0&1 \end{pmatrix},~~~ G_3=\begin{pmatrix}
0&1&0&0\\ 1&0&0&0 \\ 0&0&0&1 \\ 0&0&1&0 \end{pmatrix}\,.
\end{align}
The finite modular subgroup $N_2(H)$ can be generated by the generators $G_{1,2}$, $G'_{1,2}$ and $G_3$ which obey the relations
\begin{eqnarray}
\nonumber&& G^2_1=(G_1G_2)^3=G^2_2=1,~~G'^2_1=(G'_1G'_2)^3=G'^2_2=1,~~G^2_3=1\,,\\
\nonumber&& G_1G'_1=G'_1G_1,~ G_1G'_2=G'_2G_1,~ G_2G'_1=G'_1G_2,~ G_2G'_2=G'_2G_2\,,\\
\label{eq:S3xS3-Semi-Z2}&&G_3G_1G^{-1}_3=G'_1,~~G_3G_2G^{-1}_3=G'_2
\end{eqnarray}
The group $N_2(H)$ is isomorphic to $(S_3 \times S_3) \rtimes Z_2$ which is the group Id [72, 40] in the \texttt{GAP} notation~\cite{gap}, and the two $S_3$ subgroups are generated by $G_{1, 2}$ and $G'_{1,2}$ respectively. It has four singlet irreducible representations, one doublet representation and four quartet representations but without triplet representation.
\item[2.] $\tau= \begin{pmatrix} \tau_1 & \tau_3 \\ \tau_3 & \tau_1 \end{pmatrix}$ \,. \\
In the original ref.~\cite{gottschling1967uniformisierbarkeit} the equivalent fixed point $\begin{pmatrix} \tau_1 & 1/2 \\ 1/2 & \tau_2 \end{pmatrix}$ is discussed. The two are related by:
\begin{equation}
\begin{pmatrix}
1 & 0 & 0 & 0 \\ 1 & 0 & 0 & -1 \\ 0 & -1 & 1 & 0 \\ 0 & 1 & 0 & 0
\end{pmatrix} \begin{pmatrix} \tau_1 & 1/2 \\ 1/2 & \tau_2 \end{pmatrix}
=\begin{pmatrix} \tau_1-\dfrac{1}{4\tau_2} & \tau_1+\dfrac{1}{4\tau_2} \\[+0.15in]  \tau_1+\dfrac{1}{4\tau_2} & \tau_1-\dfrac{1}{4\tau_2} \end{pmatrix}~~~.
\end{equation}
The stabilizer $\bar{H}=Z_2$ is generated by
\begin{equation}
h= \begin{pmatrix}
0 & 1 & 0 & 0 \\ 1 & 0 & 0 & 0 \\ 0 & 0 & 0 & 1 \\ 0 & 0 & 1 & 0
\end{pmatrix}\,,
\end{equation}
The generators of the normalizer $N(H)$ can be chosen to be
\begin{align}
\nonumber&
G_1=T_1T_2=\begin{pmatrix} 1&0&1&0 \\ 0&1&0&1 \\ 0&0&1&0 \\ 0&0&0&1\end{pmatrix},~~~
G_2= T_3=\begin{pmatrix} 1&0&0&1\\ 0&1&1&0 \\ 0&0&1&0 \\ 0&0&0&1 \end{pmatrix},\\
\label{NorTwo2_gen}&G_3= S=\begin{pmatrix} 0&0&1&0\\ 0&0&0&1 \\ -1&0&0&0 \\ 0&-1&0&0 \end{pmatrix},~~~ G_4=\begin{pmatrix} 1&0&0&0 \\ 0&-1&0&0 \\ 0&0&1&0 \\ 0&0&0&-1\end{pmatrix}\,.
\end{align}
The finite modular subgroup $N_2(H)$ is isomorphic to $S_4\times Z_2$, Id [48,48] in \texttt{GAP}, and the generators $G_{1,2,3,4}$ obey the relations:
\begin{equation}G_1^2=G_2^2=G_3^2=(G_1G_2)^2=(G_1G_3)^3=(G_1G_2G_3)^4=1\,.
\end{equation}
We also have $G_4=\mathbb{1}_4~(\text{mod}~2)$ and $G_4$ represents the unit element of $N_2(H)$.
It is convenient to express $S_4\times Z_2$ in terms of the three generators $\mathcal{S}=G_1$, $\mathcal{T}=(G_3G_2)^4$ and $\mathcal{V}=(G_3G_2)^3$ which satisfy the multiplication rules:
\begin{equation}
\mathcal{S}^2=\mathcal{T}^3=(\mathcal{S}\mathcal{T})^4=1,~~\mathcal{V}^2=1,~~~\mathcal{S}\mathcal{V}=\mathcal{V}\mathcal{S},~~~
\mathcal{T}\mathcal{V}=\mathcal{V}\mathcal{T}\,.
\end{equation}
The elements $\mathcal{S}$ and $\mathcal{T}$ generate the $S_4$ subgroup, and $\mathcal{V}$ generates $Z_2$.
\end{itemize}
\subsection{Dimension 1}
There are five non-equivalent fixed point manifolds of complex dimension 1.
\begin{itemize}
\item[1.] ${\tau}= \begin{pmatrix} i & 0 \\ 0 & \tau_2 \end{pmatrix}$ \,. \\
In this case the stabilizer $\bar{H}=Z_4$ is generated by:
\begin{equation}
h=\begin{pmatrix}
0 & 0 & 1 & 0 \\ 0 & 1 & 0 & 0 \\ -1 & 0 & 0 & 0 \\ 0 & 0 & 0 & 1
\end{pmatrix} \,,
\end{equation}
and the normalizer $N(H)$ by:
\begin{equation}
\label{NorOne1}
G_1=\begin{pmatrix}
1 & 0 & 0 & 0 \\ 0 & 0 & 0 & -1 \\ 0 & 0 & 1 & 0 \\ 0 & 1 & 0 & 0
\end{pmatrix} ,\quad G_2= \begin{pmatrix}
1 & 0 & 0 & 0 \\ 0 & 1 & 0 & 1 \\ 0 & 0 & 1 & 0 \\ 0 & 0 & 0 & 1
\end{pmatrix} ,\quad R=\begin{pmatrix}
0 & 0 & -1 & 0 \\ 0 & 1 & 0 & 0 \\ 1 & 0 & 0 & 0 \\ 0 & 0 & 0 & 1
\end{pmatrix}\,.
\end{equation}
The finite normalizer $N_2(H)$  is $S_3 \times Z_2= D_{12}$, its group Id is $[12,4]$ in \texttt{GAP} and has four singlet irreducible representations, two doublet representation. The generators $G_{1,2}$ and $R$ of the finite normalizer $N_2(H)$ obey the relations
\begin{equation}
G_1^2=G_2^2=(G_1G_2)^3=R^2=1, ~G_1 R =R G_1,~ G_2 R= R G_2\,.
\end{equation}
\item[2.] $\tau= \begin{pmatrix} \omega & 0 \\ 0 & \tau_2 \end{pmatrix} $ with $\omega=e^{2\pi i/3}$ \,. \\
The stabilizer $\bar{H}=Z_6$ is generated by:
\begin{equation}
h=\begin{pmatrix}
1 & 0 & 1 & 0 \\ 0 & 1 & 0 & 0 \\ -1 & 0 & 0 & 0 \\ 0 & 0 & 0 & 1
\end{pmatrix} \,,
\end{equation}
and the normalizer $N(H)$ by:
\begin{equation}
G_1=\begin{pmatrix}
\label{NorOne2}
1 & 0 & 0 & 0 \\ 0 & 0 & 0 & -1 \\ 0 & 0 & 1 & 0 \\ 0 & 1 & 0 & 0
\end{pmatrix},\quad G_2=\begin{pmatrix}
1 & 0 & 0 & 0 \\ 0 & 1 & 0 & 1 \\ 0 & 0 & 1 & 0 \\ 0 & 0 & 0 & 1
\end{pmatrix},\quad R= \begin{pmatrix}
0 & 0 & -1 & 0 \\ 0 & 1 & 0 & 0 \\ 1 & 0 & 1 & 0 \\ 0 & 0 & 0 & 1
\end{pmatrix}
\end{equation}
The finite normalizer $N_2(H)$ is $S_3 \times Z_3$ , its group Id is $[18,3]$ in \texttt{GAP} and has six singlet irreducible representations, three doublet representation.
The generators $G_{1,2}$ and $R$ of the finite normalizer $N_2(H)$ obey the relations:
\begin{equation}
G_1^2=G_2^2=(G_1G_2)^3=R^3=1, ~G_1 R =R G_1,~ G_2 R= R G_2\,.
\end{equation}
\item[3.] $\tau= \begin{pmatrix} \tau_1 & 0 \\ 0 & \tau_1 \end{pmatrix}$ \,. \\
In this case the stabilizer $H=D_8$ is generated by:
\begin{equation}
h_1=\begin{pmatrix}
0 & 1 & 0 & 0 \\ 1 & 0 & 0 & 0 \\ 0 & 0 & 0 & 1 \\ 0 & 0 & 1 & 0
\end{pmatrix}~~~,~~~~~~~
h_2 = \begin{pmatrix}
0 & -1 & 0 & 0 \\ 1 & 0 & 0 & 0 \\ 0 & 0 & 0 & -1 \\ 0 & 0 & 1 & 0
\end{pmatrix},
\end{equation}
satisfying $h_1^2=h_2^4=(h_2 h_1)^2=\mathbb{1}_4$. We have $\bar{H}=D_8/\{\pm \mathbb{1}_4\}\cong D_4$.
The normalizer $N(H)$ has generators:
\begin{align}
\nonumber&G_1=\begin{pmatrix}
0 & 0 & -1 & 0 \\ 0 & 0 & 0 & -1 \\ 1 & 0 & 0 & 0 \\ 0 & 1 & 0 & 0
\end{pmatrix},\quad G_2=\begin{pmatrix}
1 & 0 & 1 & 0 \\ 0 & 1 & 0 & 1 \\ 0 & 0 & 1 & 0 \\ 0 & 0 & 0 & 1
\end{pmatrix},\\
\label{NorOne3}&R_1= \begin{pmatrix}
0 & 1 & 0 & 0 \\ 1 & 0 & 0 & 0 \\ 0 & 0 & 0 & 1 \\ 0 & 0 & 1 & 0
\end{pmatrix},\quad R_2= \begin{pmatrix}
1 & 0 & 0 & 0 \\ 0 & -1 & 0 & 0 \\ 0 & 0 & 1 & 0 \\ 0 & 0 & 0 & -1
\end{pmatrix}~~.
\end{align}
The finite normalizer $N_2(H) = S_3 \times Z_2$. The generators $G_{1,2}$ and $R_{1,2}$ of the finite normalizer $N_2(H)$  obey the relations:
\begin{equation}
G_1^2=G_2^2=(G_1G_2)^3=R_1^2=1,~G_1 R_1 =R_1 G_1,~ G_2 R_1= R_1 G_2\,.
\end{equation}
Note that the generator $R_2$ is essentially the identity element in $N_2(H)$.

\item[4.] $\tau= \begin{pmatrix} \tau_1 & 1/2 \\ 1/2 & \tau_1 \end{pmatrix}$ \,. \\
In this case the stabilizer $H=D_8$ is generated by:
\begin{equation}
h_1=\begin{pmatrix}
0 & 1 & 0 & 0 \\ 1 & 0 & 0 & 0 \\ 0 & 0 & 0 & 1 \\ 0 & 0 & 1 & 0
\end{pmatrix}~~~,~~~~~~~
h_2 = \begin{pmatrix}
0 & 1 & -1 & 0 \\ -1 & 0 & 0 & 1 \\ 0 & 0 & 0 & 1 \\ 0 & 0 & -1 & 0
\end{pmatrix},
\end{equation}
satisfying $h_1^2=h_2^4=(h_2 h_1)^2=\mathbb{1}_4$.
We have $\bar{H}=D_8/\{\pm \mathbb{1}_4\}\cong D_4$.
The normalizer $N(H)$ has generators:
\begin{equation}
\label{NorOne4}
G_1=\begin{pmatrix}
1 & 0 & 0 & -1 \\ 0 & -1 & 1 & 0 \\ 0 & 0 & 1 & 0 \\ 0 & 0 & 0 & -1
\end{pmatrix},~~ G_2=\begin{pmatrix}
1 & 0 & 1 & 0 \\ -1 & 0 & -1 & 1 \\ -1 & -1 & 0 & 0 \\ 1 & -1 & 1 & -1
\end{pmatrix},~~ R= \begin{pmatrix}
1 & 0 & -1 & -1 \\ 0 & -1 & 1 & 1 \\ 0 & 0 & 1 & 0 \\ 0 & 0 & 0 & -1
\end{pmatrix}\,.
\end{equation}
Thus the finite normalizer $N_2(H)$  is $D_{8} \times Z_2$, its group Id is $[16,11]$ in \texttt{GAP} and it has eight singlet irreducible representations, two doublet representations.
The generators $G_{1,2}$ and $R$ in the finite normalizer $N_2(H)$ obey the relations:
\begin{equation}
G_1^2=G_2^4=(G_1G_2)^2=R^2=1, ~G_1 R =R G_1,~ G_2 R= R G_2\,.
\end{equation}
\item[5.] $\tau= \begin{pmatrix} \tau_1 & \tau_1 /2 \\ \tau_1 /2 & \tau_1 \end{pmatrix} $ \,. \\
In this case the stabilizer $\bar{H}=S_3$ is generated by:
\begin{equation}
h_1=\begin{pmatrix}
-1 & 1 & 0 & 0 \\ 0 & 1 & 0 & 0 \\ 0 & 0 & -1 & 0 \\ 0 & 0 & 1 & 1
\end{pmatrix}~~~,~~~~~~~
h_2 = \begin{pmatrix}
1 & 0 & 0 & 0 \\ 1 & -1 & 0 & 0 \\ 0 & 0 & 1 & 1 \\ 0 & 0 & 0 & -1
\end{pmatrix},
\end{equation}
satisfying $h_1^2=h_2^2=(h_1 h_2)^3=\mathbb{1}_4$.
The generators of group $N(H)$ can be chosen to be:
\begin{align}
\nonumber
&G_1=\begin{pmatrix}
0 ~& 0 ~& 0 ~& -1 \\ 0 ~& 0 ~& 1 ~& 0 \\ 0 ~& 1 ~& 0 ~& 0 \\ -1 ~& 0 ~& 0 ~& 0
\end{pmatrix},\quad G_2=\begin{pmatrix}
1 ~& 0 ~& 0 ~& 0 \\ 0 ~& 1 ~& 0 ~& 0 \\ 2 ~& -1 ~& 1 ~& 0 \\ -1 ~& 2 ~& 0 ~& 1
\end{pmatrix},\\
\label{NorOne5}&G_1'=\begin{pmatrix}
0 ~& 1 ~& 0 ~& 0 \\ 1 ~& 0 ~& 0 ~& 0 \\ 0 ~& 0 ~& 0 ~& 1 \\ 0 ~& 0 ~& 1 ~& 0
\end{pmatrix},~~~~~\quad G_2'=\begin{pmatrix}
1 ~& 0 ~& 0 ~& 0 \\ 1 ~& -1 ~& 0 ~& 0 \\ 0 ~& 0 ~& 1 ~& 1 \\ 0 ~& 0 ~& 0 ~& -1
\end{pmatrix}\,.
\end{align}
The finite normalizer $N_2(H)$ is isomorphic to  $S_3 \times S_3$. The generators $G_{1,2}$ and $G_{1,2}'$ of the finite normalizer $N_2(H)$ satisfy the following relations
\begin{align}
\nonumber
&G_1^2=G_2^2=(G_1G_2)^3=1,~~G_1'^2=G_2'^2=(G_1'G_2')^3=1\,,\\
&G_1G_1'=G_1'G_1,~ G_2G_2'=G_2'G_2,~ G_1G_2'=G_2'G_1, ~G_2G_1'=G_1'G_2\,.
\end{align}
\end{itemize}
\subsection{Dimension zero}
There are six non-equivalent isolated fixed points. We recall that in this case $N(H)=H$, while $\bar{H}=H/\{\pm \mathbb{1}_4\}$.
\begin{itemize}
\item[1.]$\tau= \begin{pmatrix} \zeta & \zeta+\zeta^{-2} \\ \zeta+\zeta^{-2} & -\zeta^{-1} \end{pmatrix} $ with $\zeta= e^{2\pi i /5}$ \,. \\
In this case the stabilizer $\bar{H}$ is the cyclic group $Z_5$, generated by:
\begin{equation}
h=\begin{pmatrix}
0 & -1 & -1 & -1 \\ 0 & 0 & -1 & 0 \\ 0 & 0 & 0 & -1 \\ 1 & 0 & 0 & 1
\end{pmatrix} \,.
\end{equation}
\item[2.]$\tau= \begin{pmatrix} \eta & \frac{1}{2}(\eta -1) \\ \frac{1}{2}(\eta -1) & \eta \end{pmatrix} $ with $\eta=\dfrac{1}{3}(1+i2\sqrt{2})$ \,. \\
In this case the stabilizer $\bar{H}$ is group $S_4$, generated by:
\begin{equation}
h_1=\begin{pmatrix}
0 & 1 & 0 & 0 \\ 1 & 0 & 0 & 0 \\ 0 & 0 & 0 & 1 \\ 0 & 0 & 1 & 0
\end{pmatrix}\,,~~h_2=\begin{pmatrix}
-1 & 1 & 1 & 0 \\ 1 & 0 & 0 & 1 \\ -1 & 0 & 0 & 0 \\ 1 & -1 & 0 & 1
\end{pmatrix}\,,
\end{equation}
obeying the relations: $h_1^2=h_2^4=(h_1 h_2)^3=1$.
\item[3.]$\tau= \begin{pmatrix} i & 0 \\ 0 & i \end{pmatrix} $ \,. \\
In this case the stabilizer $\bar{H}$ is group $(Z_4\times Z_2)\rtimes Z_2$ generated by:
\begin{equation}
h_1=\begin{pmatrix}
0 & 0 & 1 & 0 \\ 0 & -1 & 0 & 0 \\ -1 & 0 & 0 & 0 \\ 0 & 0 & 0 & -1
\end{pmatrix}\,,~~h_2=\begin{pmatrix}
0 & 0 & -1 & 0 \\ 0 & 0 & 0 & 1 \\ 1 & 0 & 0 & 0 \\ 0 & -1 & 0 & 0
\end{pmatrix}\,,~~h_3=\begin{pmatrix}
0 & 1 & 0 & 0 \\ 1 & 0 & 0 & 0 \\ 0 & 0 & 0 & 1 \\ 0 & 0 & 1 & 0
\end{pmatrix}\,.
\end{equation}
The generators obey the relations:
\begin{equation}
h_1^4=h_2^2=h_3^2=1,~~~h_1 h_2 =h_2 h_1,~~~h_2 h_3 =h_3 h_2,~~~h_3 h_1 h_3^{-1}=h_1 h_2\,.
\end{equation}
\item[4.]$\tau= \begin{pmatrix} \omega & 0 \\ 0 & \omega \end{pmatrix} $ with $\omega=e^{2\pi i/3}$ \,. \\
In this case the stabilizer $\bar{H}$ is group $S_3\times Z_6$ generated by:
\begin{equation}
h_1=\begin{pmatrix}
0 & 0 & 0 & -1 \\ 1 & 0 & 1 & 0 \\ 0 & 1 & 0 & 1 \\ -1 & 0 & 0 & 0
\end{pmatrix}\,,~~h_2=\begin{pmatrix}
0 & 1 & 0 & 0 \\ 1 & 0 & 0 & 0 \\ 0 & 0 & 0 & 1 \\ 0 & 0 & 1 & 0
\end{pmatrix}\,,~~h_3=\begin{pmatrix}
0 & 0 & 1 & 0 \\ 0 & 0 & 0 & -1 \\ -1 & 0 & -1 & 0 \\ 0 & 1 & 0 & 1
\end{pmatrix}\,,
\end{equation}
satisfying the relations:
\begin{equation}
h_3^6=h_1^2=h_2^2=(h_1 h_2)^3=1~~,~~~ h_1 h_3=h_3 h_1~~,~~~ h_2 h_3 =h_3 h_2 \,.
\end{equation}
\item[5.]$\tau= \dfrac{i\sqrt{3}}{3}\begin{pmatrix} 2 & 1 \\ 1 & 2 \end{pmatrix} $ \,. \\
In this case the stabilizer is group $S_3\times Z_2 $, and are generated by:
\begin{equation}
h_1=\begin{pmatrix}
0 & 0 & 0 & 1 \\ 0 & 0 & 1 & 1 \\ 1 & -1 & 0 & 0 \\ -1 & 0 & 0 & 0
\end{pmatrix}\,,~~h_2=\begin{pmatrix}
0 & 0 & 1 & 1 \\ 0 & 0 & 1 & 0 \\ 0 & -1 & 0 & 0 \\ -1 & 1 & 0 & 0
\end{pmatrix}\,,~~h_3=\begin{pmatrix}
0 & 0 & 0 & 1 \\ 0 & 0 & -1 & 0 \\ 0 & -1 & 0 & 0 \\ 1 & 0 & 0 & 0
\end{pmatrix}\,.
\end{equation}
that obey the relations:
\begin{equation}
h_3^2=h_1^2=h_2^2=(h_1 h_2)^3=1~~,~~~ h_1 h_3=h_3 h_1~~, ~~~h_2 h_3 =h_3 h_2 \,.
\end{equation}
\item[6.]$\tau= \begin{pmatrix} \omega & 0 \\ 0 & i \end{pmatrix} $ with  $\omega= e^{2\pi i/3}$ \,. \\
In this case the stabilizer $\bar{H}$ is cyclic group $Z_{12}$, generated by
\begin{equation}
h=\begin{pmatrix}
0 & 0 & 1 & 0\\ 0 & 0 & 0 & 1 \\ -1 & 0 & -1 & 0 \\ 0 & -1 & 0 & 0
\end{pmatrix} \,.
\end{equation}
\end{itemize}
\section{The modular subspace with $\tau_3=0$  }
\label{subsec:tau3=0}
In this case, the modular subspace is of the form $\Omega=\begin{pmatrix} \tau_1 & 0 \\ 0 & \tau_2 \end{pmatrix}$ with $\texttt{Im}(\tau_1)>0$ and $\texttt{Im}(\tau_2)>0$. The stabilizer of this modular subspace space is $H=\{\pm \mathbb{1}_4, \pm h\}$ with
\begin{equation}
 h=\begin{pmatrix} 1&0&0&0 \\ 0&-1&0&0 \\ 0&0&1&0 \\ 0&0&0&-1\end{pmatrix}\,.
\end{equation}
It is straightforward to check that $H=\{\pm \mathbb{1}_4, \pm h\}$ is the most general modular transformations which leaves $\tau=\begin{pmatrix} \tau_1 & 0 \\ 0 & \tau_2 \end{pmatrix}$ invariant. The modular subgroup $N(H)$ acting on $\Omega$ is determined by eq.~\eqref{eq:normalizer}, then we find the element $\hat{\gamma}\in N(H)$
fulfills the following identities
\begin{equation}
\hat{\gamma}_{+} h = h \hat{\gamma}_{+}~~~\text{or}~~~\hat{\gamma}_{-} h =-h \hat{\gamma}_{-}\,,
\end{equation}
which gives rise to the most general form of $\hat{\gamma}$ as
\begin{equation}
\label{NorTwo1}
\hat{\gamma}_{+}=\begin{pmatrix} a_1&0&b_1&0 \\ 0&a_4&0&b_4 \\ c_1&0&d_1&0 \\ 0&c_4&0&d_4\end{pmatrix}~~~\text{or}~~~
\hat{\gamma}_{-}=
\begin{pmatrix}
 0 & a_4 & 0 & b_4 \\
a_1 & 0 & b_1 & 0 \\
 0 & c_4 & 0 & d_4 \\
c_1 & 0 & d_1 & 0
\end{pmatrix}\,,
\end{equation}
with
\begin{equation}
a_1d_1-b_1c_1=1,\quad a_4d_4-b_4d_4=1,~~a_{1,4}, b_{1,4}, c_{1,4}, d_{1,4}\in \mathbb{Z}\,.
\end{equation}
Notice that $\hat{\gamma}_{+}$ and $\hat{\gamma}_{-}$ are related with each other,
\begin{equation}
\hat{\gamma}_{+}=\begin{pmatrix} 0&1&0&0 \\ 1&0&0&0 \\ 0&0&0&1 \\ 0&0&1&0\end{pmatrix}\hat{\gamma}_{-}\,.
\end{equation}
It is easy to check that action of $\hat{\gamma}_{+}$ and $\hat{\gamma}_{-}$ on $\Omega$ is
\begin{eqnarray}
\nonumber&&
\hat{\gamma}_{+}\begin{pmatrix} \tau_1 & 0 \\ 0 & \tau_2 \end{pmatrix} = \begin{pmatrix}
\dfrac{a_{1}\tau_1 +b_{1}}{c_{1}\tau_1 +d_{1}} & 0 \\
0 & \dfrac{a_{4}\tau_2 + b_{4}}{c_{4}\tau_2 + d_{4}}
\end{pmatrix}\,,\\
&&
\hat{\gamma}_{-}\begin{pmatrix} \tau_1 & 0 \\ 0 & \tau_2 \end{pmatrix} = \begin{pmatrix}
\dfrac{a_{4}\tau_2 + b_{4}}{c_{4}\tau_2 + d_{4}} & 0 \\
0 & \dfrac{a_{1}\tau_1 +b_{1}}{c_{1}\tau_1 +d_{1}}
\end{pmatrix}\,,
\end{eqnarray}
Therefore the action of $\hat{\gamma}_{+}$ is equivalent to two independent $SL(2, \mathbb{Z})$ transformations $\tau_1\rightarrow\dfrac{a_{1}\tau_1 +b_{1}}{c_{1}\tau_1 +d_{1}}$ and $\tau_2\rightarrow\dfrac{a_{4}\tau_2 + b_{4}}{c_{4}\tau_2 + d_{4}}$, and consequently the group formed by $\hat{\gamma}_{+}$ is isomorphic to $SL(2, \mathbb{Z})\times SL(2, \mathbb{Z})$. Another kind of element $\hat{\gamma}_{-}$ maps $\tau_1\rightarrow\dfrac{a_{4}\tau_2 + b_{4}}{c_{4}\tau_2 + d_{4}}$ and $\tau_2\rightarrow\dfrac{a_{1}\tau_1 +b_{1}}{c_{1}\tau_1 +d_{1}}$, it relates the complex modulus $\tau_1$ with $\tau_2$.
From the well-known generators of $SL(2,\mathbb{Z})$: $S=\begin{pmatrix}0&1\\ -1&0\end{pmatrix},\,T=\begin{pmatrix}1&1\\0&1\end{pmatrix}$, we can construct the generators of $N(H)$ as:
\begin{align}
\nonumber
&G_1=\begin{pmatrix} 1&0&0&0 \\ 0&0&0&1 \\ 0&0&1&0 \\ 0&-1&0&0\end{pmatrix} ,~~
G_2=\begin{pmatrix}
1&0&0&0\\ 0&1&0&1 \\ 0&0&1&0 \\ 0&0&0&1 \end{pmatrix},~~
G'_1=\begin{pmatrix}
0&0&1&0\\ 0&1&0&0 \\ -1&0&0&0 \\ 0&0&0&1 \end{pmatrix},\\
&\label{generators_Sp1} G'_2=\begin{pmatrix}
1&0&1&0\\ 0&1&0&0 \\ 0&0&1&0 \\ 0&0&0&1 \end{pmatrix},~~~ G_3=\begin{pmatrix}
0&1&0&0\\ 1&0&0&0 \\ 0&0&0&1 \\ 0&0&1&0 \end{pmatrix}\,.
\end{align}
The finite modular subgroup $N_2(H)$ can be generated by the generators $G_{1,2}$, $G'_{1,2}$ and $G_3$ which obey the relations
\begin{eqnarray}
\nonumber&& G^2_1=(G_1G_2)^3=G^2_2=1,~~G'^2_1=(G'_1G'_2)^3=G'^2_2=1,~~G^2_3=1\,,\\
\nonumber&& G_1G'_1=G'_1G_1,~ G_1G'_2=G'_2G_1,~ G_2G'_1=G'_1G_2,~ G_2G'_2=G'_2G_2\,,\\
\label{eq:S3xS3-Semi-Z2-fir}&&G_3G_1G^{-1}_3=G'_1,~~G_3G_2G^{-1}_3=G'_2
\end{eqnarray}
The elements $G_{1, 2}$ and $G'_{1,2}$ are generators of two $S_3$ groups which are commutable with each other, consequently $G_{1, 2}$ and $G'_{1,2}$ generate $S_3\times S_3$. The element $G_3$ is of order 2, the last line of eq.~\eqref{eq:S3xS3-Semi-Z2-fir} define a homomorphism of $G_3$ on $S_3\times S_3$. As a consequence, the finite Siegel modular subgroup generated by $G_{1,2}$, $G'_{1,2}$ and $G_3$ is isomorphic to $(S_3 \times S_3) \rtimes Z_2$ with group Id [72, 40] in \texttt{GAP}~\cite{gap}. It has four singlet irreducible representations, one doublet representation and four quartet representations but without triplet representation. If we only focus on the group of $\hat{\gamma}_{+}$, the finite modular subgroup $N_2(H)$ would be generated by $G_{1,2}$ and $G'_{1,2}$, it is isomorphic to $S_3\times S_3$. Accordingly the two moduli $\tau_1$ and $\tau_2$ are independent from each other, and the geometry of the region $\Omega_{+}$ is a factorizable torus $T^2\times T^2$.

\section{The finite Siegel modular group $S_4\times Z_2$}
\label{app:S4xZ2-group}
The finite modular group $S_4\times Z_2$ with \texttt{GAP} Id  [48,48] can be generated by three elements $\mathcal{S}$, $\mathcal{T}$ and $\mathcal{V}$ with the presentation:
\begin{equation}
\mathcal{S}^2=\mathcal{T}^3=(\mathcal{S}\mathcal{T})^4=1,~~\mathcal{V}^2=1,~~~\mathcal{S}\mathcal{V}=\mathcal{V}\mathcal{S},~~~
\mathcal{T}\mathcal{V}=\mathcal{V}\mathcal{T}\,.
\end{equation}
The elements $\mathcal{S}$, $\mathcal{T}$ generate the $S_4$ subgrop, and $Z_2$ is generated by $\mathcal{V}$. This group is a subgroup of $\Gamma_{2,2}=S_6$, and the three generators $\mathcal{S}$, $\mathcal{T}$ and $\mathcal{V}$ can be expressed in terms of permutations $\mathcal{S}=(12)(45)$, $\mathcal{T}=(153)(264)$  and $\mathcal{V}=(14)(25)(36)$.
This group has 10 conjugacy classes which can be expressed in terms of $\mathcal{S}$, $\mathcal{T}$ and $\mathcal{V}$ as follows:
\begin{equation}
\begin{array}{l}
1C_1 = \{ 1 \}\,, \\
1C_2 =\{\mathcal{V}\}\,, \\
3C_2= \left\{(\mathcal{S}\mathcal{T})^2,(\mathcal{T}\mathcal{S})^2,(\mathcal{T}\mathcal{S}\mathcal{T})^2 \right\}\,, \\
3C'_2= \left\{(\mathcal{S}\mathcal{T})^2\mathcal{V},(\mathcal{T}\mathcal{S})^2\mathcal{V},(\mathcal{T}\mathcal{S}\mathcal{T})^2\mathcal{V} \right\} \,, \\
6C_2= \left\{\mathcal{S}, \mathcal{T}^2\mathcal{S}\mathcal{T}, \mathcal{T}\mathcal{S}\mathcal{T}^2, \mathcal{S}\mathcal{T}(\mathcal{T}\mathcal{S})^2,(\mathcal{S}\mathcal{T})^2\mathcal{T}\mathcal{S}, (\mathcal{T}\mathcal{S}\mathcal{T})^2\mathcal{S}\right\}\,, \\
6C'_2= \left\{\mathcal{S}\mathcal{V}, \mathcal{T}^2\mathcal{S}\mathcal{T}\mathcal{V}, \mathcal{T}\mathcal{S}\mathcal{T}^2\mathcal{V}, \mathcal{S}\mathcal{T}(\mathcal{T}\mathcal{S})^2\mathcal{V},(\mathcal{S}\mathcal{T})^2\mathcal{T}\mathcal{S}\mathcal{V}, (\mathcal{T}\mathcal{S}\mathcal{T})^2\mathcal{S}\mathcal{V}\right\} \,, \\
6C_4= \left\{\mathcal{S}\mathcal{T}^2,\mathcal{S}\mathcal{T},\mathcal{T}^2\mathcal{S},\mathcal{T}\mathcal{S},\mathcal{T}^2\mathcal{S}\mathcal{T}^2,\mathcal{T}\mathcal{S}\mathcal{T}\right\} \,, \\
6C'_4= \left\{\mathcal{S}\mathcal{T}^2\mathcal{V},\mathcal{S}\mathcal{T}\mathcal{V},\mathcal{T}^2\mathcal{S}\mathcal{V},\mathcal{T}\mathcal{S}\mathcal{V},\mathcal{T}^2\mathcal{S}\mathcal{T}^2\mathcal{V},\mathcal{T}\mathcal{S}\mathcal{T}\mathcal{V}\right\} \,,\\
8C_3= \left\{\mathcal{T},\mathcal{T}^2,\mathcal{S}\mathcal{T}^2\mathcal{S},\mathcal{S}\mathcal{T}\mathcal{S},(\mathcal{T}\mathcal{S})^2\mathcal{T}^2, (\mathcal{S}\mathcal{T})^2\mathcal{T},\mathcal{T}(\mathcal{T}\mathcal{S})^2,\mathcal{T}^2(\mathcal{S}\mathcal{T})^2 \right\}\,, \\
8C_6= \left\{\mathcal{T}\mathcal{V},\mathcal{T}^2\mathcal{V},\mathcal{S}\mathcal{T}^2\mathcal{S}\mathcal{V},\mathcal{S}\mathcal{T}\mathcal{S}\mathcal{V},(\mathcal{T}\mathcal{S})^2\mathcal{T}^2\mathcal{V},(\mathcal{S}\mathcal{T})^2\mathcal{T}\mathcal{V},\mathcal{T}(\mathcal{T}\mathcal{S})^2\mathcal{V},\mathcal{T}^2(\mathcal{S}\mathcal{T})^2 \mathcal{V}\right\}\,,
\end{array}
\end{equation}
where $nC_k$ denotes a conjugacy class with $n$ elements and the subscript $k$ refers to the order of the elements. The group has four singlet representations $\mathbf{1}$, $\mathbf{1}'$, $\mathbf{\hat{1}}$, $\mathbf{\hat{1}'}$, two doublet representations $\mathbf{2}$, $\mathbf{\hat{2}}$, and four triplet representations $\mathbf{3}$, $\mathbf{3'}$, $\mathbf{\hat{3}}$ and $\mathbf{\hat{3}'}$. The explicit form of the generators $\mathcal{S}$, $\mathcal{T}$ and $\mathcal{V}$ in each irreducible representations ate given by
\begin{equation*}
\begin{array}{cccc}
 &~~ \rho(\mathcal{S}) & \rho(\mathcal{T}) & \rho(\mathcal{V})\\ [+0.1in]
 \mathbf{1} &~~ 1 & 1 & 1 \\[+0.1in]
 \mathbf{1}' &~~ -1 & 1 & 1\\[+0.1in]
 \mathbf{\hat{1}} &~~ 1 & 1 & -1 \\[+0.1in]
 \mathbf{\hat{1}'} &~~ -1 & 1 & -1 \\[+0.1in]
 \mathbf{2} &~~ \left(
\begin{array}{cc}
 0 & 1 \\
 1 & 0 \\
\end{array}
\right) ~~&~~ \left(
\begin{array}{cc}
 \omega & 0 \\
 0 & \omega^2 \\
\end{array}
\right) ~~&~~ \mathbb{1}_2\\[+0.1in]
 \mathbf{\hat{2}} &~~ \left(
\begin{array}{cc}
 0 & 1 \\
 1 & 0 \\
\end{array}
\right) ~~&~~ \left(
\begin{array}{cc}
 \omega & 0 \\
 0 & \omega^2 \\
\end{array}
\right) ~~&~~ -\mathbb{1}_2 \\[+0.1in]
 \mathbf{3} &~~ -\dfrac{1}{3}\left(
\begin{array}{ccc}
 -1 & 2 & 2 \\
 2 & 2 & -1 \\
 2 & -1 & 2 \\
\end{array}
\right)  ~~&~~ \left(
\begin{array}{ccc}
 1 & 0 & 0 \\
 0 & \omega^2 & 0 \\
 0 & 0 & \omega \\
\end{array}
\right) ~~&~~ \mathbb{1}_3 \\[+0.1in]
 \mathbf{3'} &~~  \dfrac{1}{3}\left(
\begin{array}{ccc}
 -1 & 2 & 2 \\
 2 & 2 & -1 \\
 2 & -1 & 2 \\
\end{array}
\right)  ~~&~~ \left(
\begin{array}{ccc}
 1 & 0 & 0 \\
 0 & \omega^2 & 0 \\
 0 & 0 & \omega \\
\end{array}
\right) ~~&~~ \mathbb{1}_3\\[+0.1in]
 \mathbf{\hat{3}} &~~ -\dfrac{1}{3}\left(
\begin{array}{ccc}
 -1 & 2 & 2 \\
 2 & 2 & -1 \\
 2 & -1 & 2 \\
\end{array}
\right)  ~~&~~ \left(
\begin{array}{ccc}
 1 & 0 & 0 \\
 0 & \omega^2 & 0 \\
 0 & 0 & \omega \\
\end{array}
\right) ~~&~~ -\mathbb{1}_3 \\ [+0.1in]
 \mathbf{\hat{3}'} &~~ \dfrac{1}{3}\left(
\begin{array}{ccc}
 -1 & 2 & 2 \\
 2 & 2 & -1 \\
 2 & -1 & 2 \\
\end{array}
\right)  ~~&~~ \left(
\begin{array}{ccc}
 1 & 0 & 0 \\
 0 & \omega^2 & 0 \\
 0 & 0 & \omega \\
\end{array}
\right) ~~&~~ -\mathbb{1}_3
\end{array}
\label{eq:repmat_C2S4}
\end{equation*}
with $\omega = e^{2\pi i/3}$. We report below the multiplication rules between different irreducible representation of $S_4\times Z_2$,
\begin{align}
\nonumber&\mathbf{1'} \otimes \mathbf{1'} =\mathbf{\hat{1}} \otimes \mathbf{\hat{1}} =\mathbf{\hat{1}'} \otimes \mathbf{\hat{1}'} =\mathbf{1},~\mathbf{1'} \otimes \mathbf{\hat{1}} =\mathbf{\hat{1}'},~\mathbf{1'} \otimes \mathbf{\hat{1}'} =\mathbf{\hat{1}},~\mathbf{\hat{1}} \otimes \mathbf{\hat{1}'} =\mathbf{1'}\,,\\
\nonumber&\mathbf{1'} \otimes \mathbf{2} =\mathbf{\hat{1}'} \otimes \mathbf{\hat{2}}=\mathbf{\hat{1}} \otimes \mathbf{\hat{2}}=\mathbf{2},~
\mathbf{1'} \otimes \mathbf{\hat{2}} =\mathbf{\hat{1}'} \otimes \mathbf{2}=\mathbf{\hat{1}} \otimes \mathbf{2}=\mathbf{\hat{2}}\,,\\
\nonumber&\mathbf{1'} \otimes \mathbf{3'} = \mathbf{\hat{1}} \otimes \mathbf{\hat{3}}=\mathbf{\hat{1}'} \otimes \mathbf{\hat{3}'}=\mathbf{3}, ~\mathbf{1'} \otimes \mathbf{3} = \mathbf{\hat{1}} \otimes \mathbf{\hat{3}'}=\mathbf{\hat{1}'} \otimes \mathbf{\hat{3}}=\mathbf{3'}, \\
\nonumber&\mathbf{1'} \otimes \mathbf{\hat{3'}} = \mathbf{\hat{1}} \otimes \mathbf{3}=\mathbf{\hat{1}'} \otimes \mathbf{3'}=\mathbf{\hat{3}}, ~\mathbf{1'} \otimes \mathbf{\hat{3}} = \mathbf{\hat{1}} \otimes \mathbf{3'}=\mathbf{\hat{1}'} \otimes \mathbf{3}=\mathbf{\hat{3}'}\,,\\
\nonumber&\mathbf{2} \otimes \mathbf{2} =\mathbf{\hat{2}} \otimes \mathbf{\hat{2}} =\mathbf{1}\oplus\mathbf{1'}\oplus\mathbf{2},~\mathbf{2} \otimes \mathbf{\hat{2}} =\mathbf{\hat{1}}\oplus\mathbf{\hat{1}'}\oplus\mathbf{\hat{2}}\,,\\
\nonumber&\mathbf{2} \otimes \mathbf{3} =\mathbf{2} \otimes \mathbf{3'} =\mathbf{3}\oplus\mathbf{3'},~\mathbf{2} \otimes \mathbf{\hat{3}} =\mathbf{2} \otimes \mathbf{\hat{3}'} =\mathbf{\hat{3}}\oplus\mathbf{\hat{3}'},\\
\nonumber&\mathbf{\hat{2}} \otimes \mathbf{3}=\mathbf{\hat{2}} \otimes \mathbf{3'}=\mathbf{\hat{3}}\oplus\mathbf{\hat{3}'},~\mathbf{\hat{2}} \otimes \mathbf{\hat{3}}=\mathbf{\hat{2}} \otimes \mathbf{\hat{3}'}  =\mathbf{3}\oplus\mathbf{3'}\,,\\
\nonumber&\mathbf{3} \otimes \mathbf{3}=\mathbf{3'} \otimes \mathbf{3'}=\mathbf{\hat{3}} \otimes \mathbf{\hat{3}}=\mathbf{\hat{3}'} \otimes \mathbf{\hat{3}'}=\mathbf{1}\oplus\mathbf{2}\oplus\mathbf{3}\oplus\mathbf{3'}, \\
\nonumber&\mathbf{3} \otimes \mathbf{3'}=\mathbf{\hat{3}} \otimes \mathbf{\hat{3}'}=\mathbf{1'}\oplus\mathbf{2}\oplus\mathbf{3}\oplus\mathbf{3'}, \\
\nonumber&\mathbf{3} \otimes \mathbf{\hat{3}}=\mathbf{3'} \otimes \mathbf{\hat{3}'}=\mathbf{\hat{1}}\oplus\mathbf{\hat{2}}\oplus\mathbf{\hat{3}}\oplus\mathbf{\hat{3}'}, \\
&\mathbf{3} \otimes \mathbf{\hat{3}'}=\mathbf{3'} \otimes \mathbf{\hat{3}}=\mathbf{\hat{1}'}\oplus\mathbf{\hat{2}}\oplus\mathbf{\hat{3}}\oplus\mathbf{\hat{3}'}\,.
\end{align}
We list the Clebsch-Gordan coefficients in the basis defined above in table~\ref{tab:S4xZ2_CG}. We use $\alpha_i$ to denote the elements of the first representation of the product and $\beta_i$ to denote those of the second representation.

\begin{table}[ht!]
\centering
\resizebox{1.0\textwidth}{!}{
\begin{tabular}{|c|c|c|c|c|c|c|c|c|c|c|c|c|c|}\hline\hline
\multicolumn{3}{|c}{~~~$\mathbf{1} \otimes \mathbf{2}= \mathbf{\hat{1}} \otimes \mathbf{\hat{2}} = \mathbf{2}~~~$} & \multicolumn{3}{|c}{~$\mathbf{1} \otimes \mathbf{\hat{2}}= \mathbf{\hat{1}} \otimes \mathbf{2} = \mathbf{\hat{2}}~$} & \multicolumn{3}{|c}{~~~$\mathbf{1'} \otimes \mathbf{2}= \mathbf{\hat{1}'} \otimes \mathbf{\hat{2}} = \mathbf{2}~~~$}&\multicolumn{3}{|c|}{$~~\mathbf{1'} \otimes \mathbf{\hat{2}} = \mathbf{\hat{1}'} \otimes \mathbf{2} = \mathbf{\hat{2}}~~$} \rule[-0.5ex]{-4pt}{3ex} \\ \hline
\multicolumn{3}{|c}{  $\mathbf{2}\sim\begin{pmatrix}
 \alpha\beta_1 \\
 \alpha\beta_2 \\
\end{pmatrix} $} &
\multicolumn{3}{|c}{$\mathbf{\hat{2}}\sim \begin{pmatrix}
 \alpha\beta_1 \\
 \alpha\beta_2 \\
\end{pmatrix} $} &
\multicolumn{3}{|c}{$\mathbf{2}\sim\begin{pmatrix}
 -\alpha\beta_1 \\
 \alpha\beta_2 \\
\end{pmatrix} $} &
\multicolumn{3}{|c|}{$\mathbf{\hat{2}}\sim\begin{pmatrix}
 -\alpha\beta_1 \\
 \alpha\beta_2 \\
\end{pmatrix} $} \rule[-3ex]{-4pt}{7ex} \\ \hline\hline

\multicolumn{6}{|c}{~~$\mathbf{1} \otimes \mathbf{3} = \mathbf{1'} \otimes \mathbf{3'} = \mathbf{\hat{1}} \otimes \mathbf{\hat{3}} = \mathbf{\hat{1}'} \otimes \mathbf{\hat{3}'}=\mathbf{3}~~$} & \multicolumn{6}{|c|}{~~$\mathbf{1} \otimes \mathbf{3'} = \mathbf{1'} \otimes \mathbf{3} = \mathbf{\hat{1}} \otimes \mathbf{\hat{3}'} = \mathbf{\hat{1}'} \otimes \mathbf{\hat{3}}=\mathbf{3'}~~$} \rule[-0.5ex]{-4pt}{3ex}  \\ \hline
\multicolumn{6}{|c}{$\mathbf{3}\sim\begin{pmatrix}
 \alpha\beta_1 \\
 \alpha\beta_2 \\
 \alpha\beta_3 \\
\end{pmatrix} $}&
\multicolumn{6}{|c|}{$ \mathbf{3'}\sim\begin{pmatrix}
 \alpha\beta_1 \\
 \alpha\beta_2 \\
 \alpha\beta_3 \\
\end{pmatrix} $} \rule[-4.5ex]{0pt}{10ex}\\ \hline
  \multicolumn{6}{|c}{~~$\mathbf{1} \otimes \mathbf{\hat{3}} = \mathbf{1'} \otimes \mathbf{\hat{3}'} = \mathbf{\hat{1}} \otimes \mathbf{3} = \mathbf{\hat{1}'} \otimes \mathbf{3'} = \mathbf{\hat{3}}~~$} & \multicolumn{6}{|c|}{~~$\mathbf{1} \otimes \mathbf{\hat{3}'} = \mathbf{1'} \otimes \mathbf{\hat{3}} = \mathbf{\hat{1}} \otimes \mathbf{3'} = \mathbf{\hat{1}'} \otimes \mathbf{3} = \mathbf{\hat{3}'}~~$} \rule[-0.5ex]{-4pt}{3ex} \\ \hline
\multicolumn{6}{|c}{$\mathbf{\hat{3}}\sim\begin{pmatrix}
 \alpha\beta_1 \\
 \alpha\beta_2 \\
 \alpha\beta_3 \\
\end{pmatrix}$ } &
\multicolumn{6}{|c|}{$\mathbf{\hat{3}'}\sim\begin{pmatrix}
 \alpha\beta_1 \\
\alpha\beta_2 \\
 \alpha\beta_3 \\
\end{pmatrix}$ } \rule[-4.5ex]{-4pt}{10ex} \\ \hline\hline

\multicolumn{4}{|c}{~~~~~~~~~$\mathbf{2} \otimes \mathbf{2} = \mathbf{1_s} \oplus \mathbf{1'_a} \oplus \mathbf{2_s}$~~~~~~~~~} & \multicolumn{4}{|c}{~~~~~~~~~~$\mathbf{2} \otimes \mathbf{\hat{2}} = \mathbf{\hat{1}} \oplus \mathbf{\hat{1}'} \oplus \mathbf{\hat{2}}$~~~~~~~~~~~} & \multicolumn{4}{|c|}{$\mathbf{\hat{2}} \otimes \mathbf{\hat{2}} = \mathbf{1_s} \oplus \mathbf{1'_a} \oplus \mathbf{2_s}$ } \rule[-0.5ex]{-4pt}{3ex}  \\ \hline
 \multicolumn{4}{|c}{  $ \begin{array}{l}
 \mathbf{1_s}\sim \alpha_1 \beta_2+\alpha_2 \beta_1 \\
 \mathbf{1'_a}\sim \alpha_1 \beta_2-\alpha_2 \beta_1 \\
 \mathbf{2_s}\sim\begin{pmatrix}
 \alpha_2 \beta_2 \\
 \alpha_1 \beta_1 \\
\end{pmatrix} \\
\end{array} $ } &
\multicolumn{4}{|c}{  $\begin{array}{l}
 \mathbf{\hat{1}}\sim \alpha_1 \beta_2+\alpha_2 \beta_1 \\
 \mathbf{\hat{1}'}\sim \alpha_1 \beta_2-\alpha_2 \beta_1 \\
 \mathbf{\hat{2}}\sim\begin{pmatrix}
 \alpha_2 \beta_2 \\
 \alpha_1 \beta_1 \\
\end{pmatrix} \\
\end{array} $} &
\multicolumn{4}{|c|}{ $ \begin{array}{l}
 \mathbf{1_s}\sim \alpha_1 \beta_2+\alpha_2 \beta_1 \\
 \mathbf{1'_a}\sim \alpha_1 \beta_2-\alpha_2 \beta_1 \\
 \mathbf{2_s}\sim\begin{pmatrix}
 \alpha_2 \beta_2 \\
 \alpha_1 \beta_1 \\
\end{pmatrix} \\
\end{array} $} \rule[-6ex]{-4pt}{13ex} \\ \hline\hline

\multicolumn{3}{|c}{~$\mathbf{2} \otimes \mathbf{3} = \mathbf{\hat{2}} \otimes \mathbf{\hat{3}} = \mathbf{3} \oplus \mathbf{3'} $} & \multicolumn{3}{|c}{$\mathbf{2} \otimes \mathbf{3'} = \mathbf{\hat{2}} \otimes \mathbf{\hat{3}'} =\mathbf{3} \oplus \mathbf{3'}~$} & \multicolumn{3}{|c}{~$\mathbf{2} \otimes \mathbf{\hat{3}} = \mathbf{\hat{2}} \otimes \mathbf{3} = \mathbf{\hat{3}} \oplus \mathbf{\hat{3}'}~$}& \multicolumn{3}{|c|}{~$\mathbf{2} \otimes \mathbf{\hat{3}'}= \mathbf{\hat{2}} \otimes \mathbf{3'} = \mathbf{\hat{3}} \oplus \mathbf{\hat{3}'}~$} \rule[-0.5ex]{-4pt}{3ex} \\ \hline
 \multicolumn{3}{|c}{ $\begin{array}{l}
 \mathbf{3}\sim\begin{pmatrix}
 \alpha_2 \beta_3 + \alpha_1 \beta_2  \\
 \alpha_2 \beta_1 + \alpha_1 \beta_3  \\
 \alpha_2 \beta_2 + \alpha_1 \beta_1  \\
\end{pmatrix}  \\
 \mathbf{3'}\sim\begin{pmatrix}
 \alpha_2 \beta_3 - \alpha_1 \beta_2  \\
 \alpha_2 \beta_1 - \alpha_1 \beta_3  \\
 \alpha_2 \beta_2 - \alpha_1 \beta_1  \\
\end{pmatrix} \\
\end{array} $} &
\multicolumn{3}{|c}{ $\begin{array}{l}
 \mathbf{3}\sim\begin{pmatrix}
 \alpha_2 \beta_3 - \alpha_1 \beta_2  \\
 \alpha_2 \beta_1 - \alpha_1 \beta_3  \\
 \alpha_2 \beta_2 - \alpha_1 \beta_1  \\
\end{pmatrix} \\
 \mathbf{3'}\sim\begin{pmatrix}
 \alpha_2 \beta_3 + \alpha_1 \beta_2  \\
 \alpha_2 \beta_1 + \alpha_1 \beta_3  \\
 \alpha_2 \beta_2 + \alpha_1 \beta_1  \\
\end{pmatrix} \\
\end{array} $} &
\multicolumn{3}{|c}{ $\begin{array}{l}
 \mathbf{\hat{3}}\sim\begin{pmatrix}
 \alpha_2 \beta_3 + \alpha_1 \beta_2  \\
 \alpha_2 \beta_1 + \alpha_1 \beta_3  \\
 \alpha_2 \beta_2 + \alpha_1 \beta_1  \\
\end{pmatrix} \\
 \mathbf{\hat{3}'}\sim\begin{pmatrix}
 \alpha_2 \beta_3 - \alpha_1 \beta_2  \\
 \alpha_2 \beta_1 - \alpha_1 \beta_3  \\
 \alpha_2 \beta_2 - \alpha_1 \beta_1  \\
\end{pmatrix} \\
\end{array} $} &
 \multicolumn{3}{|c|}{ $\begin{array}{l}
 \mathbf{\hat{3}}\sim\begin{pmatrix}
 \alpha_2 \beta_3 - \alpha_1 \beta_2  \\
 \alpha_2 \beta_1 - \alpha_1 \beta_3  \\
 \alpha_2 \beta_2 - \alpha_1 \beta_1  \\
\end{pmatrix} \\
 \mathbf{\hat{3}'}\sim\begin{pmatrix}
 \alpha_2 \beta_3 + \alpha_1 \beta_2  \\
 \alpha_2 \beta_1 + \alpha_1 \beta_3  \\
 \alpha_2 \beta_2 + \alpha_1 \beta_1  \\
\end{pmatrix} \\
\end{array} $} \rule[-8.5ex]{-4pt}{18ex} \\ \hline\hline

\multicolumn{6}{|c}{  $\mathbf{3} \otimes \mathbf{3} = \mathbf{3'} \otimes \mathbf{3'} = \mathbf{\hat{3}} \otimes \mathbf{\hat{3}} = \mathbf{\hat{3}'} \otimes \mathbf{\hat{3}'} = \mathbf{1} \oplus \mathbf{2} \oplus \mathbf{3} \oplus \mathbf{3'} $} & \multicolumn{6}{|c|}{ $\mathbf{3} \otimes \mathbf{3'} = \mathbf{\hat{3}} \otimes \mathbf{\hat{3}'} = \mathbf{1'} \oplus \mathbf{2} \oplus \mathbf{3} \oplus \mathbf{3'} $} \rule[-0.5ex]{-4pt}{3ex} \\ \hline
 \multicolumn{6}{|c}{ $\begin{array}{l}
 \mathbf{1}\sim \alpha_1 \beta_1+\alpha_2 \beta_3+\alpha_3 \beta_2 \\
 \mathbf{2}\sim\begin{pmatrix}
 \alpha_2 \beta_2+\alpha_1 \beta_3+\alpha_3 \beta_1 \\
 \alpha_3 \beta_3+\alpha_1 \beta_2+\alpha_2 \beta_1 \\
\end{pmatrix} \\
 \mathbf{3}\sim \begin{pmatrix}
 \alpha_3 \beta_2-\alpha_2 \beta_3 \\
 \alpha_2 \beta_1-\alpha_1 \beta_2 \\
 \alpha_1 \beta_3-\alpha_3 \beta_1 \\
\end{pmatrix}  \\
 \mathbf{3'}\sim \begin{pmatrix}
 2 \alpha_1 \beta_1-\alpha_2 \beta_3-\alpha_3 \beta_2 \\
 2 \alpha_3 \beta_3-\alpha_1 \beta_2-\alpha_2 \beta_1 \\
 2 \alpha_2 \beta_2-\alpha_1 \beta_3-\alpha_3 \beta_1 \\
\end{pmatrix} \\
  \end{array} $} &
 \multicolumn{6}{|c|}{ $  \begin{array}{l}
 \mathbf{1'}\sim \alpha_1 \beta_1+\alpha_2 \beta_3+\alpha_3 \beta_2\\
 \mathbf{2}\sim\begin{pmatrix}
 -(\alpha_2 \beta_2+\alpha_1 \beta_3+\alpha_3 \beta_1) \\
 \alpha_3 \beta_3+\alpha_1 \beta_2+\alpha_2 \beta_1 \\
\end{pmatrix} \\
 \mathbf{3}\sim\begin{pmatrix}
 2 \alpha_1 \beta_1-\alpha_2 \beta_3-\alpha_3 \beta_2 \\
 2 \alpha_3 \beta_3-\alpha_1 \beta_2-\alpha_2 \beta_1 \\
 2 \alpha_2 \beta_2-\alpha_1 \beta_3-\alpha_3 \beta_1 \\
\end{pmatrix} \\
 \mathbf{3'}\sim\begin{pmatrix}
 \alpha_3 \beta_2-\alpha_2 \beta_3 \\
 \alpha_2 \beta_1-\alpha_1 \beta_2 \\
 \alpha_1 \beta_3-\alpha_3 \beta_1 \\
\end{pmatrix} \\
\end{array} $} \rule[-12.5ex]{-4pt}{26ex} \\ \hline
 \multicolumn{6}{|c}{ $ \mathbf{3} \otimes \mathbf{\hat{3}} = \mathbf{3'} \otimes \mathbf{\hat{3}'} = \mathbf{\hat{1}} \oplus \mathbf{\hat{2}} \oplus \mathbf{\hat{3}} \oplus \mathbf{\hat{3}'} $} & \multicolumn{6}{|c|}{ $\mathbf{3} \otimes \mathbf{\hat{3}'} = \mathbf{3'} \otimes \mathbf{\hat{3}} = \mathbf{\hat{1}'} \oplus \mathbf{\hat{2}} \oplus \mathbf{\hat{3}} \oplus \mathbf{\hat{3}'} $} \rule[-0.5ex]{0pt}{3ex}\\ \hline
\multicolumn{6}{|c}{ $\begin{array}{l}
 \mathbf{\hat{1}}\sim \alpha_1 \beta_1+\alpha_2 \beta_3+\alpha_3 \beta_2 \\
 \mathbf{\hat{2}}\sim\begin{pmatrix}
 \alpha_2 \beta_2+\alpha_1 \beta_3+\alpha_3 \beta_1 \\
 \alpha_3 \beta_3+\alpha_1 \beta_2+\alpha_2 \beta_1 \\
\end{pmatrix} \\
 \mathbf{\hat{3}}\sim\begin{pmatrix}
 \alpha_3 \beta_2-\alpha_2 \beta_3 \\
 \alpha_2 \beta_1-\alpha_1 \beta_2 \\
 \alpha_1 \beta_3-\alpha_3 \beta_1 \\
\end{pmatrix} \\
 \mathbf{\hat{3}'}\sim\begin{pmatrix}
 2 \alpha_1 \beta_1-\alpha_2 \beta_3-\alpha_3 \beta_2 \\
 2 \alpha_3 \beta_3-\alpha_1 \beta_2-\alpha_2 \beta_1 \\
 2 \alpha_2 \beta_2-\alpha_1 \beta_3-\alpha_3 \beta_1 \\
\end{pmatrix} \\
\end{array} $} &
 \multicolumn{6}{|c|}{ $\begin{array}{l}
 \mathbf{\hat{1}'}\sim \alpha_1 \beta_1+\alpha_2 \beta_3+\alpha_3 \beta_2 \\
 \mathbf{\hat{2}}\sim\begin{pmatrix}
 -(\alpha_2 \beta_2+\alpha_1 \beta_3+\alpha_3 \beta_1) \\
 \alpha_3 \beta_3+\alpha_1 \beta_2+\alpha_2 \beta_1 \\
\end{pmatrix} \\
 \mathbf{\hat{3}}\sim\begin{pmatrix}
 2 \alpha_1 \beta_1-\alpha_2 \beta_3-\alpha_3 \beta_2 \\
 2 \alpha_3 \beta_3-\alpha_1 \beta_2-\alpha_2 \beta_1 \\
 2 \alpha_2 \beta_2-\alpha_1 \beta_3-\alpha_3 \beta_1 \\
\end{pmatrix} \\
 \mathbf{\hat{3}'}\sim\begin{pmatrix}
 \alpha_3 \beta_2-\alpha_2 \beta_3 \\
 \alpha_2 \beta_1-\alpha_1 \beta_2 \\
 \alpha_1 \beta_3-\alpha_3 \beta_1 \\
\end{pmatrix} \\
\end{array} $} \rule[-12.5ex]{-4pt}{26ex} \\ \hline\hline
\end{tabular} }
\caption{\label{tab:S4xZ2_CG}The Kronecker products and Clebsch-Gordan coefficients of the $S_4\times Z_2$ group.}
\end{table}

\section{Siegel modular forms of genus $g=2$ at level $n=2$ }
\label{A2}
Theta constant plays an essential role in the construction of classical Siegel modular forms on $Sp(2g, \mathbb{Z})$~\cite{Igusa1st}. A Theta constant on the Siegel space $\mathcal{H}_g$ is given by~\cite{thesis_Fiorentino,Piazza:2008qv}:
\begin{equation}
\label{eq:theta1}\theta[{}^a_b](\tau) := \sum_{m\in\mathbb{Z}^g}
\,e^{\pi i[(m+a/2)\tau\,(m+a/2)^t +(m+a/2)\,b^t]}\,,
\end{equation}
where $a=(a_1,a_2,\dots,a_g), b=(b_1,b_2,\dots,b_g)$ are row vectors with $a_i, b_i=0,1$ and $\sum a_ib_i\equiv 0$ mod $2$, and $m$ is a $g-$dimensional row vector with integral elements. The matrix
$\Delta\equiv\left[
\begin{matrix}   a_1\, a_2\,\dots\,a_g \\  b_1\,b_2\,\dots\,b_g \\ \end{matrix}\right] = \left[
\begin{matrix}   a \\  b \\ \end{matrix}\right]$ is called a characteristic. The action of siegel modular group $\Gamma_g$ on the characteristic is:
\begin{equation}
\gamma\cdot \left[\begin{matrix} a\\b \end{matrix}\right]=\left[\begin{matrix} c\\d \end{matrix}\right] \,, \quad  \begin{pmatrix} c^t \\ d^t  \end{pmatrix} = \Big[\begin{pmatrix}
D & -C \\
-B & A \\ \end{pmatrix} \begin{pmatrix} a^t \\ b^t  \end{pmatrix} + \begin{pmatrix} \texttt{diag}^t(C D^t) \\ \texttt{diag}^t(A B^t) \end{pmatrix} \Big]\,(\texttt{mod}\, 2)~~~,
\end{equation}
where the symbol $\texttt{diag}$ represents a row vector whose elements are the diagonal entries of the matrix, i.e. $\texttt{diag}(A)=(A_{11}, A_{22},\dots)$. The Theta constant fulfills the following transformation rule~\cite{thesis_Fiorentino}:
\begin{equation}
\label{eq:transf}
\theta[\gamma \cdot\Delta](\gamma \tau)\,=\,\kappa(\gamma)\chi_{\Delta}(\gamma)\det(C\tau+D)^{1/2}\theta[\Delta](\tau), ~~\gamma \in \Gamma_g,~~\tau \in \mathcal{H}_g\,,
\end{equation}
with
\begin{equation}
\chi_\Delta(\gamma)=e^{2\pi i \phi_\Delta(\gamma)},~~\phi_\Delta(\gamma)=-\frac{1}{8}(a B^t D a^t + b A^t C b^t -2a B^t C b^t ) +\frac{1}{4}(a D^t -b C^t) \texttt{diag}^t(A B^t)\,.
\end{equation}
The factor $\kappa(\gamma)$ is an eight-root of unity that only depends on $\gamma$, and has the following properties:
\begin{eqnarray}
\nonumber&&\kappa^8(\gamma) =1\,,\quad \forall \gamma \in \Gamma_g\,,\\
\nonumber&&\kappa^4(\gamma) =e^{\pi i \text{Tr}(B^t C)}\,,\quad \forall \gamma =\begin{pmatrix}
A & B \\
C & D
\end{pmatrix} \in \Gamma_g\,,\\
&&\kappa^2(\gamma) =\kappa^2((\gamma^{-1})^t)=\det(D)\,,\quad \forall \gamma =\begin{pmatrix}
A & B \\
0 & D
\end{pmatrix} \in \Gamma_g\,.
\end{eqnarray}
The second order Theta constant is frequently used to construct classical Siegel modular forms, and it is defined as:
\begin{equation}
\label{eq:theta2}\Theta[\sigma](\tau)= \theta[{}^\sigma_0](2\tau)= \sum_{m\in\mathbb{Z}^g}
\,e^{2\pi i(m+\sigma/2)\tau\,(m+\sigma/2)^t}\,.
\end{equation}
It satisfies the following identities:
\begin{eqnarray}
\nonumber&&\Theta[\sigma](\tau+B)=e^{\pi i[\sigma \texttt{diag}^t(B)-\frac{1}{2}\sigma B \sigma^t]}\Theta[\sigma](\tau)\,, \\
\label{eq:trans_M}&&\Theta[\sigma](-\tau^{-1})= \kappa(J)^{-1}\det(\tau/2)^{\frac{1}{2}}\sum_{\rho \in \mathbb{Z}_2^g}(-1)^{\rho \sigma^t}\Theta[\rho](\tau)\,.
\end{eqnarray}
\subsection{Modular forms at genus 2}
The Siegel modular group $\Gamma_2=Sp(4, \mathbb{Z})$ can be generated by four elements $T_1$, $T_2$, $T_3$ and $S$ in eq.~\eqref{eq:generators_SP4} which act on the modulus $\tau$ as
\begin{equation}
\tau\xrightarrow[]{T_i}\tau+B_i~~,~~~~~~~~~ \tau\xrightarrow[]{S}-\tau^{-1}\,.
\end{equation}
At weight $k=2$ and level $n=2$, the Seigel modular forms space $\mathcal{M}_2(\Gamma_2(2))$ is spanned by five linearly independent polynomials $p_0$, $p_1$, $p_2$, $p_3$ and $p_4$~\cite{Cacciatori:2007vk,Piazza:2008qv}:
\begin{align}
\label{eq:MF_g=2_n=2}
\nonumber&p_0=\Theta[00]^4(\tau)+\Theta[01]^4(\tau)+\Theta[10]^4(\tau)+\Theta[11]^4(\tau) \,,\\
\nonumber&p_1=2\left(\Theta[00]^2(\tau)\Theta[01]^2(\tau)+\Theta[10]^2(\tau)\Theta[11]^2(\tau)\right)\,,\\
\nonumber&p_2=2\left(\Theta[00]^2(\tau)\Theta[10]^2(\tau)+\Theta[01]^2(\tau)\Theta[11]^2(\tau)\right)\,, \\
\nonumber&p_3=2\left(\Theta[00]^2(\tau)\Theta[11]^2(\tau)+\Theta[01]^2(\tau)\Theta[10]^2(\tau)\right)\,, \\
&p_4=4 \Theta[00](\tau)\Theta[01](\tau)\Theta[10](\tau)\Theta[11](\tau)\, .
\end{align}
Using eq.~\eqref{eq:trans_M}, we can find the transformation rules of the above five Siegel modular form polynomials under the $T_i$ and $S$:
\begin{align}
\nonumber& T_1: p_0 \rightarrow p_0,~~ p_1\rightarrow p_1, ~~p_2 \rightarrow -p_2,~~p_3 \rightarrow -p_3,~~p_4 \rightarrow -p_4\,,\\
\nonumber&T_2: p_0\rightarrow p_0,~~ p_1 \rightarrow -p_1,~~ p_2 \rightarrow p_2,~~p_3 \rightarrow -p_3,~~p_4 \rightarrow -p_4\, \\
\nonumber& T_3: p_0 \rightarrow p_0,~~ p_1 \rightarrow p_1,~~p_2 \rightarrow p_2,~~p_3 \rightarrow p_3,~~p_4 \rightarrow -p_4\,,\\
\nonumber&S:p_0 \rightarrow \frac{1}{4}\det(\tau)^2[p_0 +3p_1 +3p_2 +3p_3 +6p_4]\,,\\
\nonumber&S: p_1\rightarrow  \frac{1}{4}\det(\tau)^2[p_0 -p_1 +3p_2 -p_3 -2p_4] ,\\
\nonumber&S:p_2 \rightarrow \frac{1}{4}\det(\tau)^2[p_0 +3p_1 -p_2 -p_3 -2p_4] ,\\
\nonumber&S: p_3 \rightarrow \frac{1}{4}\det(\tau)^2[p_0 -p_1 -p_2 +3p_3 -2p_4] ,\\
&S: p_4 \rightarrow \frac{1}{4}\det(\tau)^2[p_0 -p_1 -p_2 -p_3 +2p_4] \,.
\end{align}
The five modular forms $p_i$ can be arranged into a quintet $Y^T_{\mathbf{5}}=(Y_1, Y_2, Y_3, Y_4, Y_5)$ which form a five-dimensional representation $\mathbf{5}$ of $\Gamma_{2, 2}=S_6$.
If we choose the unitary representation matrices in $\mathbf{5}$ as follows:
\begin{align}
\nonumber& \rho(T_1)=\begin{pmatrix}
1 & 0 & 0 & 0 & 0 \\
0 & 1 & 0 & 0 & 0 \\
0 & 0 & -1 & 0 & 0 \\
0 & 0 & 0 & -1 & 0 \\
0 & 0 & 0 & 0 & -1
\end{pmatrix}\,, \quad
\rho(T_2)=\frac{1}{2}\begin{pmatrix}
-1 & \sqrt{3} & 0 & 0 & 0 \\
\sqrt{3} & 1 & 0 & 0 & 0 \\
0 & 0 & -1 & \sqrt{3} & 0 \\
0 & 0 & \sqrt{3} & 1 & 0 \\
0 & 0 & 0 & 0 & -2
\end{pmatrix} \,,\\
&
\rho(T_3)=\frac{1}{3}\begin{pmatrix}
3 & 0 & 0 & 0 & 0 \\
0 & 3 & 0 & 0 & 0 \\
0 & 0 & 0 & \sqrt{3}  & \sqrt{6} \\
0 & 0 & \sqrt{3} & 2 & -\sqrt{2} \\
0 & 0 & \sqrt{6} & -\sqrt{2} & 1
\end{pmatrix} \,, \quad
\rho(S)=\frac{1}{12}\begin{pmatrix}
3 & 3\sqrt{3} & 3\sqrt{3} & 9 & 0 \\
3\sqrt{3}  & -3  & -3  & \sqrt{3}  & 4\sqrt{6}  \\
3\sqrt{3} & -3 & 9 & -3\sqrt{3} & 0 \\
9 & \sqrt{3} &-3\sqrt{3} & -1 & -4\sqrt{2} \\
0 & 4\sqrt{6} & 0 & -4\sqrt{2} & 4
\end{pmatrix}\,.
\end{align}
then, up to an overall normalization,  the quintet Siegel form $Y_{\mathbf{5}}$ is given by:
\begin{align}
\nonumber&Y_1(\tau)=p_0(\tau)+3p_1(\tau)\,,\\
\nonumber&Y_2(\tau)=\sqrt{3}\,[p_0(\tau)-p_1(\tau)]\,,\\
\nonumber&Y_3(\tau)=\sqrt{3}\,[p_2(\tau)+p_3(\tau)-2p_4(\tau)]\,,\\
\nonumber&Y_4(\tau)=3p_2(\tau)-p_3(\tau)+2p_4(\tau)\,,\\
&Y_5(\tau)=2\sqrt{2}\,[p_3(\tau)+p_4(\tau)]\,.
\end{align}

\subsection{Restriction to the modular subspace with $\tau_3=0$}
In this case, the elements of the modular subgroup $N(H)$ are of the form $\hat{\gamma}_{+}$ and $\hat{\gamma}_{-}$ in eq.~\eqref{NorTwo1}. Since the diagonal entries of $\hat{\gamma}_{-}$ are zero, $\hat{\gamma}_{-}$ doesn't belong to the principal congruence modular subgroup $N(H,2)$.
The elements of $N(H, 2)$ is of the form of $\hat{\gamma}_{+}$ with $a_{1,4}$, $d_{1,4}$ odd and $b_{1, 4}$, $c_{1, 4}$ even, and consequently it is isomorphic to $\Gamma_1(2)\times\Gamma_1(2)$. Taking $\tau_3=0$ in the expressions of $p_{0,1,2,3,4}$ in eq.~\eqref{eq:MF_g=2_n=2}, we can obtain the weight 2 modular forms of $N(H,2)$ as follow,
\begin{eqnarray}
\nonumber&&~~~~~~~~~~~~~~ p_0(\tau)=e_1(\tau_1)e_1(\tau_2),~~~~p_1(\tau)=e_1(\tau_1)e_2(\tau_2)\,,\\
\label{eq:SMF-tau3=0}&&p_2(\tau)=e_2(\tau_1)e_1(\tau_2),~~~~p_3(\tau)=e_2(\tau_1)e_2(\tau_2),~~~~
p_4(\tau)= e_2(\tau_1)e_2(\tau_2)\,,
\end{eqnarray}
where $e_1(\tau_i)=\Theta[0]^4(\tau_i)+\Theta[1]^4(\tau_i)$, $e_2(\tau_i)= 2\Theta[0]^2(\tau_i)\Theta[1]^2(\tau_i)$ are the weight 2 modular forms of $\Gamma_1(2)$, and they form a doublet of the finite Siegel modular group $\Gamma_{1,2}\cong S_3$. The $q-$expansions of $e_{1,2}(\tau)$ read as
\begin{align}
\nonumber& e_1(\tau)=1+24q+24q^2+96q^3+24q^4+144q^5+\dots\,,\\
&\sqrt{3}\,e_2(\tau)=8\sqrt{3}q^{1/2}(q+4q+6q^2+8q^3+13q^4+12q^5+\dots),~~~~q=e^{2\pi i\tau}\,,
\end{align}
which are exactly the same as those given in~\cite{Kobayashi:2018vbk} up to a overall constant. Moreover, we see $p_3(\tau)=p_4(\tau)$ in this modular subspace so that the Siegel modular form space $\mathcal{M}_2(N(H,2))$ is a four-dimensional subspace of $\mathcal{M}_2(\Gamma_2(2))$. Actually eq.~\eqref{eq:SMF-tau3=0} implies that $\mathcal{M}_2(N(H,2))$ is a tensor product of two $\mathcal{M}_2(\Gamma_1(2))$:
\begin{equation}
\begin{pmatrix}
p_0(\tau) \\p_1(\tau) \\p_2(\tau) \\p_3(\tau)
\end{pmatrix} = \begin{pmatrix}
e_1(\tau_1) e_1(\tau_2) \\e_1(\tau_1) e_2(\tau_2) \\e_2(\tau_1) e_1(\tau_2) \\e_2(\tau_1) e_2(\tau_2)
\end{pmatrix}=\begin{pmatrix}
e_1(\tau_1) \\e_2(\tau_1)
\end{pmatrix} \otimes \begin{pmatrix}
e_1(\tau_2) \\e_2(\tau_2)
\end{pmatrix}\,.
\end{equation}
It is quite straightforward to check $p_i(\hat{\gamma}_{+}\tau)=(c_{1}\tau_1 +d_{1})^{2}(c_{4}\tau_2 + d_{4})^{2}p_i(\tau)=\det(\hat{C}_{+}\tau+\hat{D}_{+})^2p_i(\tau)$, consequently $p_i(\tau)$ are really weight 2 Siegel modular forms of $N(H,2)$. Furthermore, we find that the above 4 linearly independent
Siegel modular forms $p_{0,1,2,3}$ furnish a four-dimensional irreducible representation of the finite Siegel modular subgroup $N_2(H)\cong(S_3\times S_3 )\rtimes C_2$.
\subsection{Restriction to the modular subspace with $\tau_1=\tau_2$}
When we restrict $\tau$ to $\Omega$ of eq.~\eqref{Omega}, the relation $p_1(\tau)=p_2(\tau)$ is fulfilled, thus the five linearly independent modular forms of weight 2  collapse to four. They can be organized into an invariant singlet and an irreducible triplet of the finite Siegel modular subgroup $N_2(H)\cong S_4\times Z_2$:
\begin{align}
\label{moularformsC2xS4}
\nonumber& \mathbf{3}': ~~~Y_{\mathbf{3}'}(\tau)= \begin{pmatrix}
p_0(\tau)+4p_1(\tau)-p_3(\tau) \\
p_0(\tau)-2p_1(\tau)-p_3(\tau)-2i\sqrt{3}p_4(\tau) \\
p_0(\tau)-2p_1(\tau)-p_3(\tau)+2i\sqrt{3}p_4(\tau)
\end{pmatrix}\equiv\begin{pmatrix}
Y_1(\tau) \\ Y_2(\tau) \\ Y_3(\tau)
\end{pmatrix} \,,\\
&\mathbf{1}: ~~~Y_{\mathbf{1}}(\tau)=p_0(\tau)+3p_3(\tau)\equiv Y_4(\tau)\,.
\end{align}
The Fourier expansions of $Y_{1,2,3,4}$ are:
\begin{align}
\nonumber
Y_1(\tau)&=1+32q_1^{\frac{1}{2}}-q_1(8q_3^{-1}+8q_3)+q_1^{\frac{3}{2}}(512+192q_3^{-1}+192q_3) +q_1^2(64+24q_3^{-2}+24q_3^2)\\
\nonumber&+q_1^{\frac{5}{2}}(1152+416q_3^{-2}+1024q_3^{-1}+1024q_3+416q_3^2)+ q_1^{3}(-32 q_3^{-3}-192q_3^{-1}+192q_3-32q_3^3) \\
\nonumber&+q_1^{\frac{7}{2}}(2048+448q_3^{-3}+1536q_3^{-2}+2496q_3^{-1}+2496q_3+1536q_3^2+448q_3^3)+\dots\\
\nonumber
Y_2(\tau)&=1-16q_1^{\frac{1}{2}}-q_1(8q_3^{-1}+64i\sqrt{3}q_3^{-\frac{1}{2}}+64i\sqrt{3}q_3^{\frac{1}{2}}+8q_3)-q_1^{\frac{3}{2}}(256+96q_3^{-1}+96q_3)+q_1^{2}(64 \\
\nonumber&+ 24q_3^{-2}-128i\sqrt{3}q_3^{-\frac{3}{2}} -384i\sqrt{3}q_3^{-\frac{1}{2}}-384i\sqrt{3}q_3^{\frac{1}{2}}-128i\sqrt{3}q_3^{\frac{3}{2}}+24q_3^2)-q_1^{\frac{5}{2}}(576 +208q_3^{-2}\\
\nonumber& +512q_3^{-1}+512q_3+208q_3^2) - q_1^{\frac{7}{2}}(1024+224q_3^{-3}+768q_3^{-2}+1248q_3^{-1}+1248q_3+768q_3^2\\
\nonumber&+224q_3^3)+\dots \\
\nonumber
Y_3(\tau)&=1-16q_1^{\frac{1}{2}}-q_1(8q_3^{-1}-64i\sqrt{3}q_3^{-\frac{1}{2}}-64i\sqrt{3}q_3^{\frac{1}{2}}+8q_3)-q_1^{\frac{3}{2}}(256+96q_3^{-1}+96q_3)+q_1^{2}(64 \\
\nonumber&+ 24q_3^{-2}+128i\sqrt{3}q_3^{-\frac{3}{2}} +384i\sqrt{3}q_3^{-\frac{1}{2}}+384i\sqrt{3}q_3^{\frac{1}{2}}+128i\sqrt{3}q_3^{\frac{3}{2}}+24q_3^2)-q_1^{\frac{5}{2}}(576 +208q_3^{-2}\\
\nonumber& +512q_3^{-1}+512q_3+208q_3^2) - q_1^{\frac{7}{2}}(1024+224q_3^{-3}+768q_3^{-2}+1248q_3^{-1}+1248q_3+768q_3^2\\
\nonumber &+224q_3^3)+\dots\,,\\
\nonumber
Y_4(\tau)&=1+q_1(192+24q_3^{-1}+24q_3)+q_1^2(576+24q_3^{-2}+768q_3^{-1}+768q_3+24q_3^2)+q_1^3(3072 \\
\nonumber&+96q_3^{-3}+1152q_3^{-2}+576q_3^{-1}+576q_3 +1152q_3^2+96q_3^3) + q_1^4(576+24 q_3^{-4}+1536q_3^{-3} \\
&+2304q_3^{-2}+4608q_3^{-1}+4608q_3+2304q_3^2+1536q_3^3+24q_3^4)+\dots \,,
\end{align}
with $q_1=e^{2\pi i\tau_1}$ and $q_3=e^{2\pi i\tau_3}$.

The higher weight Siegel modular forms can be constructed from the tensor product of the weight two modular forms $Y_{\mathbf{1}}(\tau)$ and $Y_{\mathbf{3}'}(\tau)$ in eq.~\eqref{moularformsC2xS4}. Using the Clebsch-Gordan coefficients in appendix~\ref{app:S4xZ2-group}, we find
\begin{align}
\nonumber & \mathbf{1}:~~~ \left\{
\begin{array}{l}
Y^{(4)}_{\mathbf{1}a}= Y_\mathbf{1} Y_\mathbf{1} = Y_4^2\,,\\
Y^{(4)}_{\mathbf{1}b}= (Y_\mathbf{3'} Y_\mathbf{3'})_\mathbf{1} = Y_1^2 + 2 Y_2Y_3\,,
\end{array}\right. \\
\nonumber&\mathbf{2}: ~~~Y^{(4)}_{\mathbf{2}}= (Y_\mathbf{3'} Y_\mathbf{3'})_\mathbf{2} = \begin{pmatrix}Y_2^2+2Y_1Y_3 \\ Y_3^2 +2Y_1Y_2 \end{pmatrix} \,, \\
\nonumber&\mathbf{3}: ~~~Y^{(4)}_{\mathbf{3}}= (Y_\mathbf{3'} Y_\mathbf{3'})_\mathbf{3} = (0,0,0)^T \,,\\
\label{eq:wt4SMF} &\mathbf{3}':~~~\left\{
\begin{array}{l}
Y^{(4)}_{\mathbf{3}'a}=  Y_\mathbf{1} Y_\mathbf{3'} = Y_4
(Y_1 , Y_2 , Y_3 )^T \,,\\
Y^{(4)}_{\mathbf{3}'b}= (Y_\mathbf{3'} Y_\mathbf{3'})_\mathbf{3'} = 2 \begin{pmatrix}
Y_1^2-Y_2Y_3 \\ Y_3^2-Y_1Y_2 \\ Y_2^2-Y_1Y_3
\end{pmatrix}\,.
\end{array}
\right.
\end{align}
Hence the weight four Siegel modular forms decompose as $\mathbf{1}\oplus\mathbf{1}\oplus\mathbf{2}\oplus\mathbf{3}'\oplus\mathbf{3}'$ under $S_4\times Z_2$. The weight 6 Siegel modular forms in the subspace with $\tau_1=\tau_2$ can be generated from the tensor products of weight 2 Siegel modular forms in eq.~\eqref{moularformsC2xS4} and  weight 4 Siegel modular forms in eq.~\eqref{eq:wt4SMF}:
\begin{align}
\nonumber&\mathbf{1}: ~~~
\left\{
\begin{array}{l}
Y^{(6)}_{\mathbf{1}a}= Y_\mathbf{1} Y^{(4)}_{\mathbf{1}a} = Y_4^3\,,\\
Y^{(6)}_{\mathbf{1}b}= Y_\mathbf{1} Y^{(4)}_{\mathbf{1}b} = Y_4Y_1^2 + 2 Y_2Y_3Y_4\,,\\
Y^{(6)}_{\mathbf{1}c}= (Y_\mathbf{3'} Y^{(4)}_{\mathbf{3}'b})_\mathbf{1} =2(Y_1^3+Y_2^3+Y_3^3-3Y_1Y_2Y_3)\,,\\
Y^{(6)}_{\mathbf{1}d}= (Y_\mathbf{3'} Y^{(4)}_{\mathbf{3}'a})_\mathbf{1} =Y_4Y_1^2 + 2 Y_2Y_3Y_4
\,,
\end{array}
\right.\\
\nonumber&\mathbf{2}: ~~~
\left\{
\begin{array}{l}
Y^{(6)}_{\mathbf{2}a}= (Y_\mathbf{3'} Y^{(4)}_{\mathbf{3}'a})_\mathbf{2} = Y_4 \begin{pmatrix}
Y_2^2+2Y_1Y_3 \\ Y_3^2+2Y_1Y_2
\end{pmatrix} \,,\\
Y^{(6)}_{\mathbf{2}b}= (Y_\mathbf{3'} Y^{(4)}_{\mathbf{3}'b})_\mathbf{2} = (0,0,0)^T \,,\\
Y^{(6)}_{\mathbf{2}c}= Y_\mathbf{1} Y^{(4)}_{\mathbf{2}} = Y_4 \begin{pmatrix}
Y_2^2+2Y_1Y_3 \\ Y_3^2+2Y_1Y_2
\end{pmatrix}\,,
\end{array}
\right.\\
\nonumber&\mathbf{3}:~~~
\left\{
\begin{array}{l}
Y^{(6)}_{\mathbf{3}a}=  (Y_\mathbf{3'} Y^{(4)}_{\mathbf{3}'b})_\mathbf{3} = 2 \begin{pmatrix}
Y_3^3-Y_2^3 \\ 2Y_1^2Y_2-Y_2^2Y_3-Y_3^2Y_1 \\ Y_2^2Y_1+Y_3^2Y_2-2Y_1^2Y_3
\end{pmatrix}  \,,\\
Y^{(6)}_{\mathbf{3}b}=  (Y_\mathbf{3'} Y^{(4)}_{\mathbf{3}'a})_\mathbf{3} = (0,0,0)^T\,,\\
Y^{(6)}_{\mathbf{3}c}=  (Y_\mathbf{3'} Y^{(4)}_{\mathbf{2}})_\mathbf{3} =  \begin{pmatrix}
Y_3^3-Y_2^3 \\ 2Y_1^2Y_2-Y_2^2Y_3-Y_3^2Y_1 \\ Y_2^2Y_1+Y_3^2Y_2-2Y_1^2Y_3
\end{pmatrix}  \,,
\end{array}
\right.
\end{align}

\begin{align}
\nonumber \mathbf{3}': ~~~
\left\{
\begin{array}{l}
Y^{(6)}_{\mathbf{3}'a}=  Y_\mathbf{1} Y^{(4)}_{\mathbf{3}'a} = Y_4^2(Y_1, Y_2, Y_3)^T  \,,\\
Y^{(6)}_{\mathbf{3}'b}=  Y_\mathbf{1} Y^{(4)}_{\mathbf{3}'b} = 2 Y_4 \begin{pmatrix}
Y_1^2-Y_2Y_3 \\ Y_3^2-Y_1Y_2 \\ Y_2^2-Y_1Y_3
\end{pmatrix}   \,,\\
Y^{(6)}_{\mathbf{3}'c}= ( Y_\mathbf{3}' Y^{(4)}_{\mathbf{3}'b} )_{\mathbf{3}'}= 2 \begin{pmatrix}
2Y_1^3- Y_2^3-Y_3^3 \\ 3Y_2^2Y_3-3Y_3^2Y_1 \\ 3Y_3^2Y_2-3Y_2^2Y_1
\end{pmatrix}   \,,\\
Y^{(6)}_{\mathbf{3}'d}= ( Y_\mathbf{3}' Y^{(4)}_{\mathbf{2}} )_{\mathbf{3}'}=  \begin{pmatrix}
Y_2^3+Y_3^3+ 4Y_1Y_2Y_3 \\ 3Y_3^2Y_1+2Y_1^2Y_2 +Y_2^2Y_3 \\ 3Y_2^2Y_1+2Y_1^2Y_3+Y_3^2Y_2
\end{pmatrix}   \,,\\
Y^{(6)}_{\mathbf{3}'e}= ( Y_\mathbf{3'} Y^{(4)}_{\mathbf{3}'a} )_{\mathbf{3}'}= 2 Y_4 \begin{pmatrix}
Y_1^2-Y_2Y_3 \\ Y_3^2-Y_1Y_2 \\ Y_2^2-Y_1Y_3
\end{pmatrix}   \,,\\
Y^{(6)}_{\mathbf{3}'f}=  Y_{\mathbf{3}'} Y^{(4)}_{\mathbf{1}a} = Y_4^2(Y_1, Y_2, Y_3)^T \,,\\
Y^{(6)}_{\mathbf{3}'g}=  Y_{\mathbf{3}'} Y^{(4)}_{\mathbf{1}b} = (Y_1^2+2Y_2Y_3)(Y_1, Y_2, Y_3)^T \,,
\end{array}
\right.
\end{align}
We see that these weight 6 Siegel modular forms are not all linearly independent from each other, and the following relations are satisfied
\begin{eqnarray}
\nonumber&&Y^{(6)}_{\mathbf{1}b} = Y^{(6)}_{\mathbf{1}d} ,~ Y^{(6)}_{\mathbf{2}a} = Y^{(6)}_{\mathbf{2}c},~Y^{(6)}_{\mathbf{3}a} = 2 Y^{(6)}_{\mathbf{3}c},\\
&& Y^{(6)}_{\mathbf{3}'e} = Y^{(6)}_{\mathbf{3}'b},~Y^{(6)}_{\mathbf{3}'f} = Y^{(6)}_{\mathbf{3}'a},~  Y^{(6)}_{\mathbf{3}'g}= [Y^{(6)}_{\mathbf{3}'c}+2Y^{(6)}_{\mathbf{3}'d}]/4\,.
\end{eqnarray}
Hence the weight 6 Siegel modular forms in the subspace of $\tau_1=\tau_2$
has dimension 20, they can be decomposed into $\mathbf{1}\oplus\mathbf{1}\oplus\mathbf{1}\oplus\mathbf{2}\oplus\mathbf{3}\oplus\mathbf{3}'\oplus\mathbf{3}'\oplus\mathbf{3}'\oplus\mathbf{3}'$ under $S_4\times Z_2$, and the basis vectors can be chosen to be
\begin{align}
Y^{(6)}_{\mathbf{1}a},~ Y^{(6)}_{\mathbf{1}b},~ Y^{(6)}_{\mathbf{1}c},~ Y^{(6)}_{\mathbf{2}a},Y^{(6)}_{\mathbf{3}a},~ Y^{(6)}_{\mathbf{3}'a},~Y^{(6)}_{\mathbf{3}'b},~Y^{(6)}_{\mathbf{3}'c},~Y^{(6)}_{\mathbf{3}'d} \,.
\end{align}

\subsection{Restriction to the modular subspace with dimension one}
As in the case of 2-dimensional modular subspace, the expressions of the Siegel modular forms in the 1-dimensional modular subspace can be straightforwardly obtained, and they can be arranged into multiplets of the corresponding finite Siegel modular subgroup $N_2(H)$. Furthermore, we find that the following nontrivial relations between the original Siegel modular forms $p_{0,1,2,3,4}$ are satisfied:
\begin{align}
\label{eq:relations_MF_d=1}
\nonumber
&\tau=\begin{pmatrix} i & 0\\ 0 & \tau_1 \end{pmatrix}\,:
\,~~~~\qquad p_0(\tau) = 3 p_2(\tau),\quad p_1(\tau) = 3 p_3(\tau) \quad\text{and}\quad p_3(\tau) = p_4(\tau) \,, \\
\nonumber
&\tau=\begin{pmatrix} \omega & 0\\ 0 & \tau_1 \end{pmatrix}\,:
\,~~~~\qquad p_0(\tau) = i\sqrt{3}\, p_2(\tau),\quad p_1(\tau) = i\sqrt{3}\, p_3(\tau) \quad\text{and}\quad p_3(\tau) = p_4(\tau) \,, \\
\nonumber
&\tau=\begin{pmatrix} \tau_1 & 0\\ 0 & \tau_1 \end{pmatrix}\,:
\,\,~~~\qquad p_1(\tau) = p_2(\tau) \quad\text{and}\quad p_3(\tau) = p_4(\tau) \,, \\
\nonumber
&\tau=\begin{pmatrix} \tau_1 & 1/2\\ 1/2 & \tau_1 \end{pmatrix}\,:
\,\qquad p_1(\tau) = p_2(\tau) \quad\text{and}\quad p_4(\tau) = 0 \,, \\
&\tau=\begin{pmatrix} \tau_1 & \tau_1/2 \\ \tau_1/2 & \tau_1 \end{pmatrix}\,:
\,~\quad p_1(\tau) = p_2(\tau) = p_3(\tau)  \,.
\end{align}

\end{appendices}
\section*{Acknowledgements}
We thank Luca Martucci for a useful discussion. This project has received support in part by the European Union's Horizon 2020 research and innovation programme under the Marie Sklodowska-Curie grant
agreement N$^\circ$~674896 and 690575 and by the National Natural Science Foundation of China under Grant Nos 11975224, 11835013, 11947301. The research of F.~F.~was supported in part by the INFN. F.~F.~thanks the University of Science and Technology of China (USTC) in Hefei for hospitality in July 2019, when this project started. GJD thanks the  Department of Physics and Astronomy, University of Padova for hospitality in January 2020.

\providecommand{\href}[2]{#2}\begingroup\raggedright\endgroup

\end{document}